\documentclass[twocolumn,english,prb,floatfix,superscriptaddress,twocolumn,amsmath,amssymb,reprint]{revtex4}

\usepackage[T1]{fontenc}
\usepackage[latin9]{inputenc}
\usepackage{color}
\definecolor{shadecolor}{rgb}{1, 0, 0}
\usepackage{verbatim}
\usepackage{float}
\usepackage{framed}
\usepackage{amsmath}
\usepackage{amssymb}
\usepackage{graphicx}
\usepackage{multirow}
\usepackage{mathrsfs}
\usepackage{relsize}

\def\sumslashD{\mathop{\sum \kern-1.4em -\kern 0.5em}}
\def\sumslash{\mathop{\sum \kern-1.2em -\kern 0.5em}}
\def\intslash{\mathop{\int \kern-0.9em -\kern 0.5em}}
\def\intslashD{\mathop{\int \kern-1.1em -\kern 0.5em}}

\makeatletter
\PassOptionsToPackage{caption=false}{subfig}
\usepackage{hyperref}
\hypersetup{
breaklinks=true,
colorlinks=true,
citecolor=blue,
linkcolor=blue,
filecolor=blue,
urlcolor=blue
}
\IfFileExists{lmodern.sty}{\usepackage{lmodern}}{}
\makeatother

\usepackage{babel}

\begin{document}

\title{Theory of phonon instabilities in Weyl semimetals at high magnetic fields}
  
\author{Sarbajaya Kundu}
\email{sarbajay.kundu@ufl.edu}
\affiliation{D\'epartement de physique, Institut quantique and Regroupement Qu\'eb\'ecois sur les Mat\'eriaux de Pointe,  Universit\'{e} de Sherbrooke, Sherbrooke, Qu\'{e}bec J1K 2R1, Canada}
\affiliation{Department of Physics, University of Florida, Gainesville, Florida 32611, USA}
\author{Claude Bourbonnais}
\affiliation{D\'epartement de physique, Institut quantique and Regroupement Qu\'eb\'ecois sur les Mat\'eriaux de Pointe,  Universit\'{e} de Sherbrooke, Sherbrooke, Qu\'{e}bec J1K 2R1, Canada}
\author{Ion Garate}
\affiliation{D\'epartement de physique, Institut quantique and Regroupement Qu\'eb\'ecois sur les Mat\'eriaux de Pointe,  Universit\'{e} de Sherbrooke, Sherbrooke, Qu\'{e}bec J1K 2R1, Canada}
\date{\today}                                       

\begin{abstract}
The behavior of three-dimensional (3D) semimetals under strong 
magnetic fields is a topic of recurring interest in condensed
matter physics. Recently, the advent of Weyl and Dirac semimetals has brought about an interesting platform
for potentially uncovering phases of matter that combine nontrivial band topology and interactions.
While electronic instabilities of such semimetals at strong magnetic fields have
been explored theoretically and experimentally,
the role of electron-phonon interactions therein has been largely neglected. 
In this paper, we study the interplay of electron-electron and electron-phonon interactions
in a minimal two-node model of Weyl semimetal.
Using a Kadanoff-Wilson renormalization
group approach, we analyze lattice (Peierls) instabilities emerging from chiral and nonchiral Landau levels as a function of the magnetic field. 
We consider both the adiabatic and the nonadiabatic phonon regimes, in the presence or in the absence of improper symmetries that relate Weyl nodes of opposite chirality. 
We find that (i) the Cooper channel, often neglected in recent studies, can prevent purely electronic instabilities while enabling lattice instabilities that are not Bardeen-Cooper-Schrieffer-like; (ii) breaking the improper symmetry that relates the two Weyl nodes suppresses the Cooper channel, thereby increasing the critical temperature for the lattice instability; (iii) in the adiabatic phonon regime, lattice instabilities can preempt purely electronic instabilities; 
(iv) pseudoscalar phonons are more prone to undergo a Peierls instability than scalar phonons. 
In short, our study emphasizes the importance of taking electron-phonon interactions into account for a complete understanding of interacting phases of matter in Dirac and Weyl semimetals at high magnetic fields.



  
\end{abstract}

\maketitle

\section{Introduction}

In recent times, the competing effects of electronic correlations and strong spin-orbit
coupling have generated considerable interest in the condensed matter
community \cite{doi:10.1146/annurev-conmatphys-031115-011319,PhysRevLett.109.066401,Schaffer_2016}.
Topological materials \cite{doi:10.1146/annurev-conmatphys-031214-014501,doi:10.1146/annurev-conmatphys-062910-140432,doi:10.7566/JPSJ.82.102001,Hasan_2015,RevModPhys.82.3045,RevModPhys.83.1057,Yan_2012}
provide an excellent platform for the potential realization of exotic
electronic phases and novel phenomena resulting from such
an interplay. 

A particular example is provided by Weyl semimetals (WSMs) \cite{doi:10.1146/annurev-conmatphys-031016-025458,doi:10.1146/annurev-conmatphys-031113-133841,RevModPhys.90.015001},
which are topologically nontrivial gapless systems featuring pairs of non-degenerate
bands, touching each other at isolated points in the band structure,
with a low-energy description in terms of massless, linearly dispersing
Weyl fermions. Electronic interaction effects in these systems have been extensively explored, using various approaches \cite{PhysRevB.95.201102,PhysRevLett.113.136402,PhysRevLett.109.196403,PhysRevB.98.241102,PhysRevB.96.195160,PhysRevB.94.241102,PhysRevB.94.125135,PhysRevB.90.035126,PhysRevB.89.235109,PhysRevB.89.014506,PhysRevB.96.165203,PhysRevB.100.245137,Yi2017possible}. 

More generally, the behavior of three-dimensional
semimetals under a large applied magnetic field has presented
a problem of recurring interest for the community\cite{Arnold2017,Yakovenko93,Kumar_2010,LeBeouf2017,PhysRevB.79.241101,PhysRevLett.110.266601,PhysRevLett.54.1182,TAKAHASHI1994384,Yaguchi_2009}.
For such systems, in the so-called quantum limit, where the Fermi level
intersects only the lowest Landau band, the nesting of Fermi surfaces
is greatly enhanced, stabilizing field-induced symmetry-breaking phases,
such as density wave orders \cite{Arnold2017,PhysRevB.79.241101,Yaguchi_2009,TAKAHASHI1994384}.
The quantum limit can be achieved relatively easily in Dirac/Weyl semimetals,
owing to their low carrier density, and the extra valley or orbital
degrees of freedom may further enrich the types of phases that can
be realized. 

On the experimental side, high-field measurements of Dirac/Weyl semimetals have also given rise to anomalous features \cite{Zhang2016,Wang2020,Tang2019,Ramshaw2018,Liu2016} that hint at the possibility of field-induced phase transitions in the quantum limit, thus contributing to the interest in this area. Such anomalies
are often attributed to interaction-induced instabilities occurring
at high fields\cite{Zhang2016,Ramshaw2018,Tang2019}. Besides, the high-field regime provides a setting that can be useful for potentially
isolating decisive experimental signatures of Weyl fermions\cite{Hui2019,Kim2017,Ramshaw2018,rinkel2019,PhysRevB.86.115133,PhysRevX.4.031035,Zhang2019}. 

From a purely theoretical point of view, much of the existing effort has
been directed specifically towards examining the effect of electronic interactions
in three-dimensional semimetals at high fields\cite{Roy2015, Miransky2015, Shindou2019,Song2017,Trescher2017,Xiao2016,Yang2011,TAKAHASHI1994384}.
In this context, the appearance of a chiral symmetry-breaking, fully
gapped charge-density wave (CDW) order, even for sufficiently weak
repulsive electron-electron interactions, for topological Dirac and Weyl semimetals has been discussed in the literature \cite{Roy2015,Song2017,Trescher2017,Xiao2016,Yang2011}. 
Such studies
have employed both analytical and numerical approaches, but have generally
focused their attention solely on the quantum limit. 

On the other hand, while phonons in Weyl semimetals have gained considerable attention\cite{Song2016, Hui2019,rinkel2019,Moller2017,nguyen2020,PhysRevB.101.085202,PhysRevB.95.165114,PhysRevB.97.195106,PhysRevLett.110.046402,PhysRevLett.119.107401,PhysRevLett.125.146402, Wang2020} due to their interplay with the nontrivial electronic band topology,
the role of phonons in the magnetic field-induced {\em instabilities} of Weyl semimetals
has barely been explored.

The need for including electron-phonon interactions in theoretical
treatments of such systems has been raised in part by recent high-field measurements
in the Weyl semimetal TaAs \cite{Ramshaw2018}, where a strong increase
in the ultrasonic attenuation at low temperatures has been observed.
Likewise, it has been recently suggested that electron-phonon interactions can be important for the field-induced ordered states reported in graphite\cite{LeBeouf2017} and in the Dirac semimetal ZrTe$_{5}$\cite{Zhao2020, Qin2020}, although the discussion on the presence or absence of order in ZrTe$_5$ is still ongoing\cite{Okada1982, Tian2021, Galeski2021, Ehmcke2021}.

In this paper, we aim to contribute along the preceding line of research by studying possible instabilities of a Weyl semimetal at high magnetic fields,
through the interplay of electron-electron and electron-phonon interactions.
The primary objective is to theoretically predict the high-field lattice instabilities
for a minimal (two-node) Weyl semimetal model, as a function of the magnitude
and direction of the magnetic field, in the presence and absence of improper symmetries relating the two Weyl nodes, and for both the adiabatic and nonadiabatic phonon regimes.

To that end, we adopt the Kadanoff-Wilson renormalization group (RG) approach\cite{Bourbon89,Bourbon91},
which has the advantage of treating different electronic and lattice instabilities on an equal footing and allows to
include multiple energy scales with ease. 
It follows from our analysis that electron-phonon interactions tend to augment the charge-density wave fluctuations in the system and that, under certain conditions, lattice instabilities preempt those of purely electronic origin.
It is, therefore, important to take into account the effect of electron-phonon interactions while studying potential instabilities in Weyl semimetals at high magnetic fields. 
Our analysis also evidences that particle-particle scattering (a.k.a. the "Cooper channel"), which has been often neglected before, plays an important role when the two Weyl nodes are related by a mirror plane. Indeed, the Cooper channel interferes destructively with the particle-hole scattering (a.k.a. the "Peierls channel"), thereby preventing purely electronic instabilities while enabling lattice instabilities.
Another important finding from our study is that breaking the mirror symmetry (which can be done for example by rotating the external magnetic field) suppresses the Cooper channel, thereby increasing the critical temperature for the lattice instability.


The paper is organized as follows. In Sec. \ref{sec:prel}, we begin by recalling the minimal
model of two Weyl nodes in a quantizing magnetic field and in the 
presence of electron-phonon and electron-electron interactions. Then, we briefly review the Kadanoff-Wilson renormalization group
method. 

In Sec. \ref{sec:cll}, we write down the RG equations when the Fermi level intersects
only the chiral Landau levels.
We derive the condition for phonon softening from the renormalization
of the phonon part of the action, and use it to obtain the Peierls
transition temperature, in hitherto unexplored scenarios and as a function of different tuning parameters.
For instance, we take into account the effect of an asymmetry in
the position of the Weyl nodes of opposite chiralities, as well as
in their velocities.
We also extend our analysis to the nonadiabatic regime, where the bare phonon energy
exceeds the energy scale of the Peierls transition. 
We consider competing logarithmic divergences in the Cooper and Peierls
channels whenever relevant, and find that the results are sensitively
dependent on the competition between the two contributions.

In Sec. \ref{sec:noncll}, we describe the
corresponding results for the case when the Fermi level also intersects
the first nonchiral Landau level. Here, we distinguish between the
behavior of the scalar and pseudoscalar phonon modes, and we find that the latter are qualitatively more prone to undergo an instability.

In Sec. \ref{sec:real}, we briefly extrapolate our results to real Weyl semimetals and comment on their applicability to Dirac semimetals.
In Sec. \ref{sec:conc}, we summarize our results and state the main conclusions. 
We also compare our work with recent theories\cite{Zhao2020,Qin2020} of lattice instabilities in ZrTe$_5$, as well as with theories of magnetic catalysis\cite{Miransky2015}.
Finally, the appendices contain details of the RG calculation.

\section{Model}
\label{sec:prel}

For most of this work, we limit ourselves to a minimal model of two untilted Weyl nodes of opposite chirality.
In Sec. \ref{sec:real}, we discuss briefly how the results from our toy model might extrapolate to models with various pairs of Weyl nodes, with a tilt.

\subsection{Free Weyl fermions in a magnetic field}

We consider a Weyl semimetal of spatial dimensions $L_x\times L_y\times L_z$ under a magnetic field ${\bf B}=B\hat{\bf z}$, with periodic boundary conditions in the $z$ direction. 
The Landau levels originating from a Weyl fermion of a given chirality $\tau$ (with $\tau=\pm 1$) are characterized by the following good quantum numbers: the Landau level index $n$ ( $n\in\mathbb{Z}$, with $n>0$ for conduction-band Landau levels, $n<0$ for valence-band Landau levels, $n=0$ for the chiral Landau level), the guiding center $X$, and the wave vector $k$ along the direction of the external magnetic field, measured from the position of the Weyl node.
Note that $k$ is assumed to be bound by a cutoff,  within which the Weyl fermion approximation is valid.
This cutoff is smaller than the internodal distance.
Below, we will often denote the collection of good quantum numbers with the letter $\alpha$, i.e. $\alpha=(n,X,k,\tau)$.
In addition, we disregard the Zeeman effect, whose effect in our model is limited to a shift in the Weyl nodes' location.

The eigenenergies are independent of $X$ and given by 
\begin{align}
\label{eq:LL_energies}
  \epsilon_{n X k \tau} &=\left\{\begin{array}{cc} \tau b_0 + \epsilon_{n k\tau} & \text{   $(n\neq 0)$}\\
   \tau b_0 +\hbar v_\tau \tau k & \text{  $(n=0)$}\end{array}\right.,
\end{align}
where $\epsilon_{n k\tau} = \hbar v_\tau \tau {\rm sign}(n)\sqrt{k^2+ 2 |n|/l_B^2}$, 
$v_\tau$ is the node-dependent magnitude of the Dirac velocity (not to be confused with the $v_{n k \tau}$ coefficient below), and $2 b_0$ is the energy shift between the two nodes of opposite chirality. 
In the presence of an improper symmetry that relates the two nodes (e.g. a mirror plane), $b_0=0$ and $v_+ = v_-$.
Below, we will study the general case in which both $b_0$ and $v_+-v_-$ maybe nonzero.

The eigenspinors, written in the pseudospin basis $\{\sigma\} = \{\uparrow,\downarrow\}$ that describes the two degenerate bands at the Weyl node in the absence of magnetic fields, are
\begin{equation}
\label{eq:spinor_cll}
\langle{\bf r}|\Psi_{0 X k \tau}\rangle= \frac{e^{i (k+k_\tau) z}}{\sqrt{L_z}}
      \left(\begin{array}{c} 0\\
        h_{0 X}(x,y)
      \end{array}\right)
\end{equation}
for the chiral ($n=0$) Landau level, and 
\begin{equation}
\label{eq:spinor_ncll}
\langle{\bf r}|\Psi_{n X k \tau}\rangle= \frac{e^{i (k+k_\tau) z}}{\sqrt{L_z}}
      \left(\begin{array}{c} u_{n k \tau} h_{|n|-1, X}(x,y)\\
        v_{n k\tau} h_{|n|, X}(x,y)
      \end{array}\right)
\end{equation}
for the nonchiral  ($n\neq 0$) Landau levels, 
where $k_\tau$ is the projection of the momentum of the Weyl node of chirality $\tau$ along the magnetic field, 
\begin{align}
  u_{n k\tau} &= \frac{1}{\sqrt{2}} \tau {\rm sign}(n) \sqrt{1+\frac{\hbar v_\tau k}{\epsilon_{n k \tau}}}\\
  v_{n k\tau} &= \frac{1}{\sqrt{2}} \sqrt{1-\frac{\hbar v_\tau k}{\epsilon_{n k \tau}}}
  \end{align}
  are position-independent coefficients of the Landau-level spinors,
  \begin{align}
  h_{n X}(x,y) &= \frac{(-1)^{n}}{\sqrt{L_y}} e^{-i X y/l_B^2} \varphi_{n}(x-X)\\
  \varphi_n(x) &=\left(\frac{1}{\pi l_B^2}\right)^{1/4}\frac{1}{\sqrt{2^n n!}} H_n\left(\frac{x}{l_B}\right) e^{-x^2/2 l_B^2}
\end{align}
are Landau-level wave functions and $H_n(x)$ are the Hermite polynomials.
It is useful to note that
\begin{align}
 &\langle h_{|n|, X} | h_{|n'|, X'}\rangle\equiv \int dx dy\, h^*_{|n| X} h_{|n'| X'}= \delta_{X X'} \delta_{|n|  |n'|},\nonumber\\
  & u_{n k\tau}^2 + v_{n k \tau}^2 =1.
\end{align}

Using the shorthand notation, the free fermion Hamiltonian can be written as
\begin{equation}
{\cal H}_e^{(0)} = \sum_\alpha \epsilon_\alpha c^\dagger_\alpha c_\alpha,
\end{equation}
where $c^\dagger_\alpha\equiv c^\dagger_{n X k \tau}$ is a (dimensionless) operator that creates an electron in state $|\Psi_{n X k \tau}\rangle$.

For the moment, we neglect the single-particle hybridization gap due to magnetic tunneling between chiral Landau-levels of opposite chirality. 
This gap can become measurable at high magnetic fields\cite{Zhang2017, Ramshaw2018}, provided that (i) the magnetic field is perpendicular to the wave vector that separates Weyl nodes in momentum space ; (ii)  the magnetic length is comparable to or shorter than the inverse of the distance in momentum space that separates two Weyl nodes at zero field.
The hybridization gap depends strongly on the orientation of the magnetic field, and it is believed to be relatively negligible when the magnetic field is parallel to the wave vector connecting the two Weyl nodes\cite{Kim2017, Chan2017, Saykin2018}. 
In Sec.~\ref{sec:real}, we will briefly comment on the implications of the hybridization gap in our theory.

\subsection{Electron-phonon interaction}

The electron-phonon interaction in the band eigenstate basis is given by
\begin{widetext}
  \begin{equation}
  {\cal H}_{\rm ep} =\sum_{\alpha \alpha'} \sum_{{\bf q}, j}  \tilde{g}^{ep}_{\alpha\alpha', j}({\bf q})   c^\dagger_{\alpha} c_{\alpha'} (a^\dagger_{{\bf q},j}+a_{-{\bf q},j}),
\end{equation}
where $a^\dagger_{{\bf q},j}$ is a (dimensionless) operator that creates a phonon mode $j$ with momentum ${\bf q}$, and
\begin{align}
  \label{eq:e_ph_me}
  &\tilde{g}_{\alpha\alpha', j}^{ep}({\bf q})\equiv \tilde{g}_{\tau\tau',j}({\bf q})\langle \Psi_{n X, k, \tau} |e^{i {\bf q}\cdot{\bf r}}| \Psi_{n' X' k' \tau'}\rangle\nonumber\\
  &=\tilde{g}_{\tau\tau',j}({\bf q})\delta_{k'+k_{\tau'}, k+k_\tau-q}\delta_{X',X+q_y l_B^2}\left[u_{n k\tau} u_{n' k' \tau'}  \langle h_{|n|-1, X}|e^{i {\bf q}_\perp\cdot{\bf r}_\perp} | h_{|n'|-1, X'}\rangle
    + v_{n k\tau} v_{n' k' \tau'} \langle h_{|n|, X} |e^{i {\bf q}_\perp\cdot{\bf r}_\perp}| h_{|n'|, X'}\rangle\right]\nonumber\\
  &=\tilde{g}_{\tau\tau',j}({\bf q})\delta_{k'+k_{\tau'}, k+k_\tau-q}\delta_{X',X+q_y l_B^2} e^{i q_x(X+X')/2}\left[u_{n k\tau} u_{n' k' \tau'}  F_{|n|-1, |n'|-1}({\bf q}_\perp)+ v_{n k \tau} v_{n' k' \tau'} F_{|n|,|n'|}({\bf q}_\perp)\right] 
\end{align}
is the electron-phonon matrix element (with dimensions of energy) in the low-energy Hilbert space spanned by Weyl fermions.
For brevity, Eq. (\ref{eq:e_ph_me}) has been written using the eigenstates of the nonchiral Landau levels (Eq.~(\ref{eq:spinor_ncll})). 
Yet, it can be easily adapted to include chiral Landau levels via $u_{0,k,\tau}\to 0$ and $v_{0,k,\tau}\to 1$.
In the numerical estimates performed below, we will use\cite{Kaasbjerg2012, Mahan2000b} 
\begin{equation}
\label{eq:ep_approx}
\tilde{g}_{\tau\tau',j}({\bf q})^2 \sim\frac{\hbar d_{{\rm op},j}^2}{2 \rho {\cal V} \omega_{0,j}({\bf q})}
\end{equation}  
for long-wavelength optical phonons, where ${\cal V}$ is the crystal volume, $\rho$ is the atomic mass density, $d_{{\rm op},j}$ is the optical deformation potential   (dimensions of energy divided by length) and $\omega_{0,j}({\bf q})$ is the unperturbed phonon frequency for mode $j$. For the case of long-wavelength acoustic phonons, we replace $d_{{\rm op},j} \to |{\bf q}| d_{{\rm ac},j}$, where $d_{{\rm ac},j}$ is the acoustic deformation potential (dimensions of energy).

In Eq.~(\ref{eq:e_ph_me}), we have defined the form factors\cite{MacDonald1986,Goerbig2011} 
\begin{equation}
\label{eq:form_factor}
  F_{n m}({\bf q}_\perp) \equiv \sqrt{\frac{{\rm min}(n,m)!}{{\rm max}(n,m)!}} \left(\frac{(\mp q_y - i q_x) l_B}{\sqrt{2}}\right)^{|n-m|} L_{{\rm min}(n,m)}^{|n-m|}\left(\frac{q_\perp^2 l_B^2}{2}\right) e^{-q_\perp^2 l_B^2/4},
\end{equation}
where $n,m$ are nonnegative integers, the $+$ and $-$ signs correspond to the cases $n>m$ and $n<m$, respectively, and the functions $L_n^\alpha$ are the generalized Laguerre polynomials.
Also, we have used ${\bf q}=q \hat{\bf z} + {\bf q}_\perp$, ${\bf r}= z \hat{\bf z} + {\bf r}_\perp$. 
We note that, for long-wavelength phonons ($q$ smaller than the internodal distance), Eq.~(\ref{eq:e_ph_me}) vanishes unless $\tau=\tau'$.
In addition, in the limit ${\bf q}_\perp\to 0$, Eq.~(\ref{eq:e_ph_me}) becomes
\begin{equation}
  \langle \Psi_{n X, k, \tau} |e^{i q z}| \Psi_{n' X' k' \tau'}\rangle = \delta_{k'+k_{\tau'}, k+k_\tau-q}\delta_{|n|, |n'|} \delta_{X, X'}
    \left(u_{n,k,\tau} u_{n', k', \tau'}  + v_{n,k,\tau} v_{n', k', \tau'}\right).
\end{equation}
In Eq. (\ref{eq:e_ph_me}), we have assumed (for simplicity) that the electron-phonon coupling is diagonal in the pseudospin basis.
Admittedly, for a generic nesting wave vector, there is no symmetry-based reason that would preclude the electron-phonon coupling from having pseudospin-dependent terms. However, the pseudospin structure of the electron-phonon coupling is not qualitatively crucial for our theory; what is crucial is that the coupling between Fermi points connected by the nesting wave vector be nonzero. Any pseudospin-dependence of such electron-phonon interaction will simply modify (by a numerical factor) the effective electron-phonon coupling appearing in our expressions for the critical temperature in Secs \ref{sec:cll} and \ref{sec:noncll}.

\subsection{Electron-electron interaction}

The electron-electron interaction in the band eigenstate basis is given by
\begin{align}
  {\cal H}_{ee} \simeq \frac{1}{2 {\cal V}} \sum_{\bf q} \sum_{\{\alpha\}}
  \tilde{g}_{\alpha_1\alpha_2}^{\alpha_3\alpha_4} ({\bf q})c^\dagger_{\alpha_1} c_{\alpha_3} c^\dagger_{\alpha_2}  c_{\alpha_4},
\end{align}
where we have neglected umklapp processes, 
\begin{equation}
\label{eq:e_e_me}
 \tilde{g}_{\alpha_1\alpha_2}^{\alpha_3\alpha_4} ({\bf q}) \equiv V({\bf q})  \langle\Psi_{n_1 X_1 k_1 \tau_1}|e^{i {\bf q}\cdot{\bf r}} |\Psi_{n_4 X_4 k_4 \tau_4}\rangle \langle\Psi_{n_2 X_2 k_2 \tau_2}|e^{-i {\bf q}\cdot{\bf r}} |\Psi_{n_3 X_3 k_3 \tau_3} \rangle
 \end{equation}
 \end{widetext}
 is the Coulomb matrix element, $V({\bf q})=e^2/(\epsilon_0\epsilon_\infty {\bf q}^2)$ is the Coulomb potential and $\epsilon_\infty$ is the cutoff-dependent contribution from "high-energy" electrons  (of energy exceeding the ultraviolet RG cutoff) to dielectric screening.
Screening effects originating from low-energy fermions will be discussed below.



\subsection{Partition function}

In what follows, we will apply the renormalization group approach\cite{Bourbon89} to the Peierls instability of the above model.
The starting point is the  functional integral representation of the partition function in the presence of a magnetic field along $z$,
\begin{equation}
\label{eq:Z}
Z=\int \mathfrak{D}\psi^\dagger \mathfrak{D}\psi\int D\phi 
\,e^{S[\psi^\dagger,\psi,\phi,h]/\hbar},
\end{equation}
 which is expressed in terms of a trace over the fermion ($\psi^{(\dagger)})$ and phonon ($\phi$) fields. Here, $S$ stands for the action 
\begin{align}
S[\psi^\dagger,\psi,\phi,h] &= \ S^{(0)}_e[\psi^\dagger,\psi]+ S_{ee}[\psi^\dagger,\psi]\nonumber\\
 &+  S^{(0)}_p[\phi]+S_{ep}[\psi^\dagger,\psi,\phi] + S_{h}[\psi^\dagger,\psi,h],
\end{align}
which comprises five parts.
First, the free electron action is 
 \begin{equation}
 \label{eq:freefer}
S_e^{(0)}[\psi^\dagger,\psi] =\sum_\alpha \sum_{\omega_n}\left(i\hbar\omega_n+\mu-\epsilon_\alpha\right)\psi^\dagger_\alpha (\omega_n)\psi_\alpha (\omega_n),
\end{equation}
where 
$\psi_\alpha$ and $\psi_{\alpha'}^\dagger$ are Grassmann fields (with dimensions $\sqrt{\rm time}$), $\omega_n=(2n+1)\pi k_BT/\hbar$ ($n\in\mathbb{Z}$) are fermionic Matsubara frequencies at temperature $T$ and $\mu$ is the chemical potential.
In Eq.~(\ref{eq:freefer}), we have used the convention
\begin{equation}
\psi_\alpha(t)=\sqrt{\frac{1}{\hbar\beta}} \sum_{\omega_n} e^{-i\omega_n t} \psi_\alpha(\omega_n),
\end{equation}
where $\beta=1/k_B T$.
Second, the free phonon action is
 \begin{equation}
 S_p^{(0)} [\phi]= -\sum_{{\bf q},j} \sum_{\omega_m}   [\mathscr{D}^0_j({\bf q},\omega_m)]^{-1}|\phi_j({\bf q},\omega_m)|^2,
\end{equation}
 where
\begin{equation}
\label{D}
  \mathscr{D}_j^0({\bf q},\omega_m) = [\omega_m^2+\omega_{0,j}({\bf q})^2]^{-1}
\end{equation}
is the free phonon propagator  and
$\omega_m=2\pi k_BT m/\hbar$ ($m\in\mathbb{Z}$) are bosonic Matsubara frequencies, $\omega_{0,j}(\bf q)$ is the bare phonon dispersion, and 
$\phi_j({\bf q},\omega_m)$ is the phonon displacement field for mode $j$  with wave vector  ${\bf q}$ (with dimensions of $\sqrt{\rm action} \times {\rm time}$). 
In second quantized form, 
\begin{equation}
\phi_j ({\bf q},t) \to \frac{1}{2}\sqrt{\frac{\hbar}{\omega_{0,j}({\bf q})}} \left (a_{{\bf q},j}(t)^\dagger + a_{-{\bf q}, j}(t)\right).
\end{equation}
Third, the action for the electron-phonon interaction is
 \begin{widetext}
\begin{equation}
S_{ep}[\psi^\dagger,\psi,\phi]=-\int_0^{\hbar\beta} dt {\cal H}_{\rm ep}(t)=-\sqrt{\frac{\pi v_F}{\beta \cal V}}\sum_{\omega_n,\omega_m} \sum_{{\bf q},j} \sum_{\alpha,\alpha'} z_{\alpha\alpha', j} g^{ep}_{\alpha\alpha', j}({\bf q})\psi^\dagger_{\alpha}(\omega_n+\omega_m) \psi_{\alpha'}(\omega_n) \phi_j({\bf q},\omega_m),
\end{equation}
where $z_{\alpha\alpha', j}$ stands as a renormalization factor of the electron-phonon coupling ($z_{\alpha\alpha', j}=1$ at the bare level)  and 
\begin{equation}
\label{eq:ep_conv}
g^{ep}_{\alpha \alpha', j}({\bf q})= \frac{2}{\hbar} \sqrt{\frac{\omega_{0,j}({\bf q}) {\cal V}}{\pi v_F}}\tilde{g}^{ep}_{\alpha\alpha',j}({\bf q})
\end{equation}
is the electron-phonon coupling with dimensions of velocity ($\tilde{g}_{\alpha\alpha', j}^{ep}$ was defined in Eqs.~(\ref{eq:e_ph_me}) and~(\ref{eq:ep_approx})).
The action for electron-electron interactions is
 \begin{equation}
 S_{ee}[\psi^\dagger,\psi] =-\int_0^{\hbar\beta} dt {\cal H}_{\rm ee}(t)=- \frac{\pi v_F}{\beta \cal V} \sum_{\{\alpha\},\{\omega_n\}} \sum_{\bf q}g_{\alpha_1\alpha_2}^{\alpha_3\alpha_4}({\bf q}) \psi^\dagger_{\alpha_1}(\omega_{n_1}) \psi^\dagger_{\alpha_2}(\omega_{n_2})\psi_{\alpha_3}(\omega_{n_3})\psi_{\alpha_4}(\omega_{n_1}+\omega_{n_2}-\omega_{n_3}),
  \end{equation}
where
\begin{equation}
\label{eq:ee_conv}
g_{\alpha_1\alpha_2}^{\alpha_3\alpha_4}({\bf q})=\frac{\tilde{g}_{\alpha_1\alpha_2}^{\alpha_3\alpha_4}({\bf q})}{2\pi \hbar v_F} 
\end{equation}
is the electron-electron coupling parameter with dimensions of length squared ($\tilde{g}_{\alpha_1\alpha_2}^{\alpha_3\alpha_4}$ was defined in Eq.~(\ref{eq:e_e_me})). 

Finally, the source field contribution for susceptibilities is given by
\begin{equation}
\label{sh0}
S_{h}[\psi^\dagger,\psi,h] = \sqrt{\frac{\pi v_F}{\beta \cal V}} \sum_{\nu}\sum_{{\bf q},\omega_m} \sum_{\sigma,\sigma'} z_{\nu;\sigma,\sigma'} \Big\{O^\dagger_{\nu; \sigma,\sigma'} (\omega_m) h_{\nu; \sigma,\sigma'}({\bf q},\omega_m) + O_{\nu; \sigma,\sigma'} (\omega_m) h^*_{\nu; \sigma,\sigma'}({\bf q},\omega_m)\Big\},
\end{equation}
where $\sigma$ and $\sigma'$ are pseudospin labels, $h_{\nu;\sigma,\sigma'}$ are the source fields, and the composite fields for charge density-wave ($\nu={\rm CDW}$) and Cooper pairing ($\nu={\rm SC}$)  take their respective forms:
\begin{equation}
\label{Ocdw}
O_{{\rm CDW};\sigma,\sigma'} (\omega_m) = \sum_{\alpha,\alpha'}\sum_{\omega_n} f_{\alpha,\alpha'; \sigma,\sigma'}^{\rm CDW}\psi^\dagger_{\alpha}(\omega_n+\omega_m) \psi_{\alpha'}(\omega_n)
\end{equation} 
and 
\begin{equation}
\label{Osc}
O_{{\rm SC};\sigma,\sigma'} (\omega_m) = \sum_{\alpha,\alpha'}\sum_{\omega_n} f_{\alpha,\alpha'; \sigma,\sigma'}^{\rm SC}\psi_{\alpha}(\omega_n+\omega_m) \psi_{\alpha'}(\omega_n).
\end{equation} 
Here, $z_{\nu;\sigma,\sigma'}$ stands for the renormalization factor for vertex part of the susceptibility,  and $f_{\alpha,\alpha';\sigma,\sigma'}^\nu$ is the related form factor (more on this below).  
For a CDW susceptibility that is pseudospin-independent, we take $h_{{\rm CDW};\sigma,\sigma'} \propto  \delta_{\sigma,\sigma'}$; for a pseudospin-singlet Cooper pairing susceptibility, we take $h_{{\rm SC};\sigma,\sigma'} \propto  \delta_{\sigma,-\sigma'}$, and so on.

\subsection{Method}
\label{subsec:method}

Starting from Eq.~(\ref{eq:Z}), we perform successive partial integrations of fermion fields located in the outer energy shell of width $\Lambda_n(l) dl$ on both sides of the Fermi level. For  each outer shell integration  a complete trace over Matsubara frequencies is carried out  from which the temperature dependence of the flow is obtained. 
Here, $\Lambda_n(l)=\Lambda_n e^{-l}$ is the distance in energy from the Fermi level to the outer shell at step $l$ of the integration ($l>0$), $2\Lambda_n$ is the initial bandwidth cutoff  of the Landau band $n$, and $dl\ll 1$ is the integration step.
The fermion fields in the outer shell are denoted as $\{\bar{\psi}, \bar{\psi}^\dagger\}$.
The recursive procedure is carried out perturbatively, keeping fixed the inner ($<$) shell variables and  using $S^{(0)}_e[\bar{\psi}^\dagger,\bar{\psi}]$ as the outer shell free  fermion part.  This yields
\begin{align}
\label{RGTransf}
Z &\sim \int D\phi\int_< \mathfrak{D}\psi^\dagger\, \mathfrak{D}\psi \,e^{S[\psi^\dagger,\psi,\phi,h]_l/\hbar} \int \mathfrak{D}\bar{\psi}^\dagger \mathfrak{D}\bar{\psi} \,e^{S[\psi^\dagger,\psi,\bar{\psi}^\dagger,\bar{\psi},\phi,h]_{dl}/\hbar}\nonumber\\
& \propto \int D\phi\int_< \mathfrak{D}\psi^\dagger \mathfrak{D}\psi \,e^{S[\psi^\dagger,\psi,\phi,h]_{l+dl}/\hbar}.
\end{align}
 Using the linked cluster theorem, the outer shell integration at the one-loop level yields the recursion relations
 \begin{align}
  \label{Sp0}
  S_p^{(0)}[\phi]_{l+dl} &= S_p^{(0)}[\phi]_l+\frac{1}{2}\langle S_{e p}^2\rangle_{_{dl}} + {\cal O}(\phi^4) \\
   \label{zeph0}
 S_{e p}[\psi^\dagger,\psi,\phi]_{l+dl} &= S_{e p}[\psi^\dagger,\psi,\phi]_l+\langle S_{e p} S_{ee}\rangle_{_{dl}}+\ldots \\
  \label{g1g2}
 S_{ee}[\psi^\dagger,\psi]_{l+dl} &= S_{ee}[\psi^\dagger,\psi]_l + \frac{1}{2} \langle S_{ee}^2\rangle_{_{dl}}+ \ldots\\
 \label{sh}
  S_{h}[\psi^\dagger,\psi,h]_{l+dl} &= S_{h}[\psi^\dagger,\psi,h]_l+\langle S_{h} S_{ee}\rangle_{_{dl}}+\ldots
 \end{align}
which are expressed in terms of free fermion averages $\langle \ldots \rangle_{_{dl}}$ in the outer energy shell.
Equation (\ref{Sp0}) gives the recursion relation for the renormalization of the purely phonon part of the action. At the harmonic $\phi^2$ level it leads to the phonon the self-energy correction  $\delta \scalebox{1.55}{$\pi$} $ to the phonon propagator via  $\mathscr{D}_{j,l}^{-1} = \mathscr{D}_{j,0}^{-1} -  \scalebox{1.55}{$\pi$}_{j,l} $, as shown in Fig.~\ref{fig_RG1}(a). For simplicity, we neglect the hybridization between different phonon modes induced by the electron-phonon coupling; this is a reasonable approximation for phonon modes that are well-separated in energy from the rest.
Accordingly, $\scalebox{1.55}{$\pi$}$ and $\mathscr{D}$ have a single phonon mode index. At the one-loop level, we do not consider electronic self-energy corrections   to $S_e^{(0)}$ since for effective momentum independent couplings these only affect the chemical potential which can rescaled back to its original value at each step of the RG at fixed band filling. 
Equation (\ref{zeph0}),  depicted by Fig.~\ref{fig_RG1}(b), refers to the renormalization of the electron-phonon vertex part $z_{\alpha,\alpha',j}$ due to electron-electron interactions and  Eq. (\ref{g1g2}), depicted in Fig.~\ref{fig_RG1}(c), gives the  recursion relation for a combination of electron-electron couplings $g_{\alpha_1,\alpha_2}^{\alpha_3,\alpha_4} $ involved in the renormalization of  $z_{\alpha\alpha',j}$ and source field vertex  $z_\nu$ of Fig.~\ref{fig_RG1}(d).
 \begin{figure}[h]
  \begin{center}
    \includegraphics[scale=0.48]{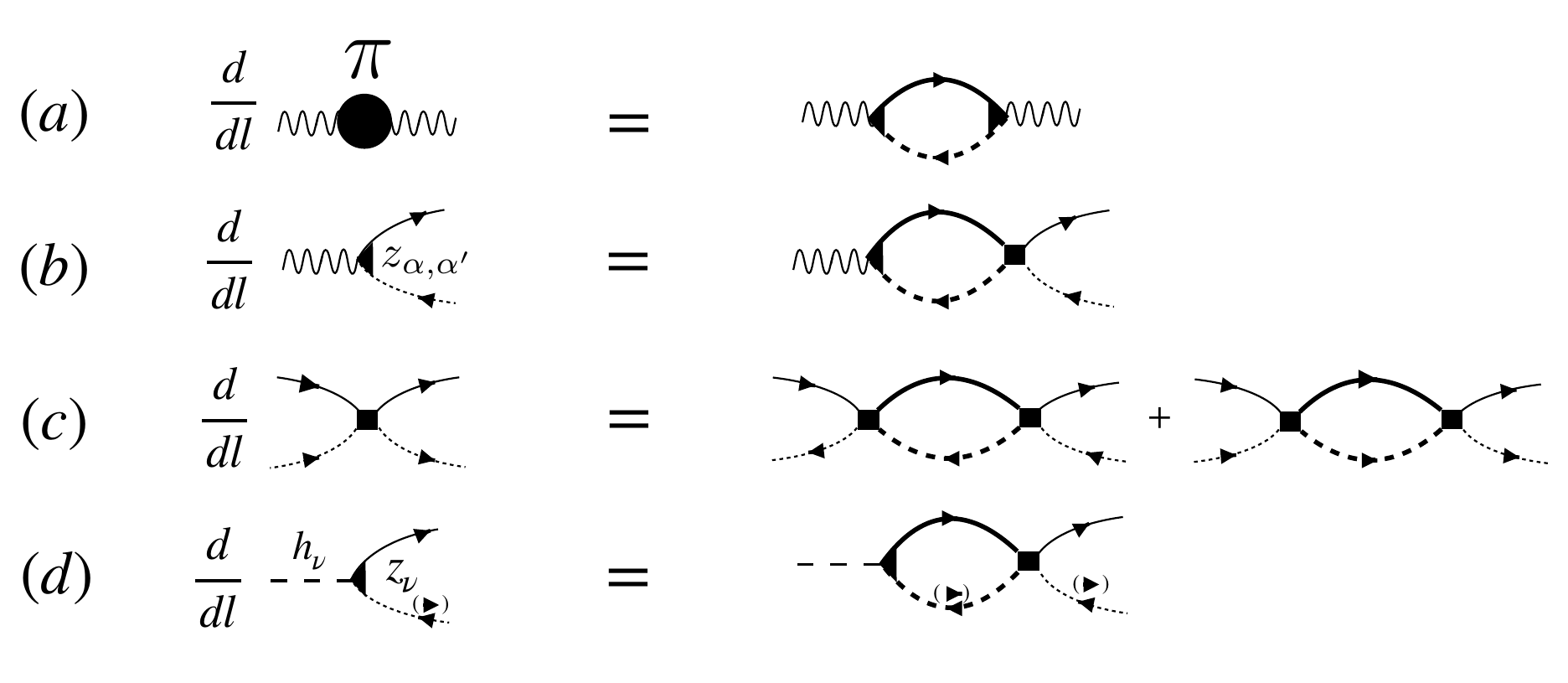}
\caption{Renormalization group flow equations for (a) the phonon self-energy; (b) electron-phonon vertex part;  (c) electron-electron scattering amplitudes (full square); (d) and source field vertex for CDW (SC) susceptibility. The full and dotted thick (thin) loop fermion lines refer to  outer (inner) energy shell fermions near the right and left Fermi points of the Landau band $n$ (we consider $|n|=0$ in Sec. \ref{sec:cll} and $|n|=0, 1$ in Sec. \ref{sec:noncll}).  }
 \label{fig_RG1}
  \end{center}
\end{figure} 
 
 \end{widetext}
 
\section{Peierls instability from chiral Landau levels}
\label{sec:cll}

We begin by considering the theory of the Peierls instability in the quantum limit, where the Fermi energy intersects only the chiral Landau levels. 

\subsection{Electron-phonon and electron-electron couplings}

\begin{figure}[t]
  \begin{center}
    \includegraphics[width=0.9\columnwidth]{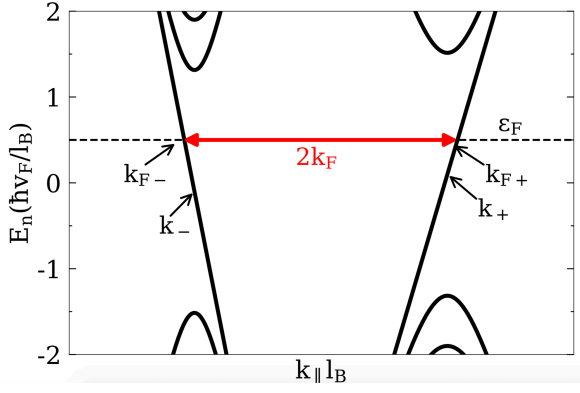}
\caption{Landau level spectrum of two Weyl nodes of opposite chirality (Eq.~(\ref{eq:LL_energies})) in the quantum limit, where the Fermi energy $\epsilon_F$ intersects only chiral Landau levels. The parameter values in Eq. (\ref{eq:LL_energies}) are $v_+=v_F$, $v_-=1.5 v_F$, $b_0=0.1 \hbar v_F/l_B$. The distance between the two Fermi points is denoted as $2 k_F=k_{F+} -  k_{F-}$, where $k_{F+}$ and $k_{F-}$ are the Fermi wave vectors measured from the center of the Brillouin zone (along the direction of the magnetic field). 
The locations of the Weyl nodes, $k_+$ and $k_-$, are also indicated and measured from the center of the Brillouin zone (along the direction of the magnetic field).
}
 \label{fig:ql}
  \end{center}
\end{figure}

Figure~\ref{fig:ql} shows the low-energy electronic structure in the quantum limit. 
The distance between the two Fermi points along the direction of the magnetic field is denoted as $2 k_F= k_{F+} - k_{F-} $ (see Fig.~\ref{fig:ql}).
According to this definition, the Fermi wave vector depends on the distance between the Weyl nodes (note that a different definition for $k_F$ will be used when we treat nonchiral Landau levels in Sec. \ref{sec:noncll}).
We take the magnitudes of the Dirac velocities to be $v_\tau = v_F$ for $\tau=1$ and $v_\tau=v_F + \delta v$ for $\tau=-1$, where $\delta v \geq -v_F$ (the latter condition is due to the fact that the two nodes have opposite chirality). 

Assuming that the total carrier density per unit volume $n_e$ is independent of the magnetic field, the $B-$dependence of $k_F$ in the quantum limit is obtained from 
\begin{equation}
2 k_F = |4\pi^2 l_B^2 n_e+b|, 
\end{equation}
where $b=|k_+-k_-|$ is the momentum separation between the two Weyl nodes along the direction of the magnetic field.
Here, $n_e$ is defined by counting the number of electronic states between $k_\tau$ and the Fermi energy for each node $\tau$, and then summing over the two nodes.
For an overall electron-doped system ($n_e>0$), $2 k_F > b$ and $k_F$ decreases with $B$. 
For an overall hole-doped system ($n_e<0$), ${\rm min}({k_F})=0$.
Regardless of the doping, $2 k_F \to b$ as $B\to\infty$.
Thus, at very high magnetic field, $k_F$ can be made small by orienting the magnetic field perpendicular to the direction of separation between the two nodes (because $b=0$ in that case).

We will be interested in phonons whose wave vectors connect the two Fermi points, as these are subject to a Peierls instability.
We believe such internodal electron-phonon scattering is generically allowed, despite assertions to the contrary in the literature\cite{nguyen2020}.
Indeed, from a group theory point of view, there is no selection rule that forbids electron-phonon scattering between nodes of opposite chirality. 
In other words, because the little group of Weyl nodes located at arbitrary points in the Brillouin zone comprises only the identity operation, generic phonon modes of the crystal can scatter electrons between two nodes of opposite chirality.

From Eq.~(\ref{eq:e_ph_me}), the bare electron-phonon matrix element for the chiral Landau level can be written as
\begin{align}
\label{eq:cll_me}
g_{\alpha\alpha',j}^{ep} = & g_{\tau\tau',j}  \langle\Psi_{0 X, k, \tau} |e^{i {\bf q}\cdot{\bf r}}| \Psi_{0 X' k' \tau'}\rangle\nonumber\\
=& g_{\tau\tau',j}  \delta_{k'+k_{\tau'}, k+k_\tau-q}\delta_{X',X+q_y l_B^2}\nonumber\\
& \times e^{i q_x(X+X')/2} e^{-q_\perp^2 l_B^2/4},
\end{align}
where $g_{\tau\tau',j}$ is related to $\tilde{g}_{\tau\tau',j}$ in Eq.~(\ref{eq:ep_approx}) via the relation (\ref{eq:ep_conv}).
When $q_\perp l_B\gtrsim 1$, the electron-phonon matrix element is exponentially suppressed. 
For a phonon wave vector connecting the two Fermi points, we set $\tau=-\tau'$ in Eq.~(\ref{eq:cll_me}).
Hereafter, we label $g_{\tau,-\tau,j}\equiv g_{x,j}$ and we neglect the momentum-dependence of $g_{x,j}$. 
The latter approximation is justified because we will be interested in phonon momenta close to $2 k_F \hat{\bf z}$ and, in such vicinity, the ${\bf q}$-dependence of $g_{x,j}$ will be smooth and qualitatively unimportant for our theory. 

The matrix elements for the vertex part of electronic CDW and SC susceptibilities lead to the form factors entering in Eqs. (\ref{Ocdw}) and (\ref{Osc}),
\begin{align}
\label{fcdw}
   f_{\alpha\alpha';\sigma,\sigma'}^{\rm CDW} & = \delta_{\sigma,\downarrow}\delta_{\sigma',\downarrow}\delta_{k'+k_{\tau'}, k+k_\tau-q}\delta_{X',X+q_y l_B^2}\nonumber\\
 &\times e^{i q_x (X+X')/2} e^{-q_\perp^2 l_B^2/4} \\
 \label{fsc}
  f_{\alpha\alpha';\sigma,\sigma'}^{\rm SC}  &  = \delta_{\sigma,\downarrow}\delta_{\sigma',\downarrow}\delta_{k'+k_{\tau'}, -k-k_\tau-q}\delta_{X',-X+q_y l_B^2}\nonumber\\
 & \times e^{i q_x (X+X')/2} e^{-\left[q_x^2+\left(q_y-2X/l_B^2\right)^2\right]l_B^2/4}, 
\end{align}
where we have used the fact that the $n=0$ Landau levels are pseudospin-polarized (recall Eq.~(\ref{eq:spinor_cll})). This allows us to omit the $\sigma$ and $\sigma'$ labels from the CDW and SC renormalization factors, denoting them simply as $z_{\rm CDW}$ and $z_{\rm SC}$. 

Similarly, the Coulomb matrix elements for the electron-electron interactions (see Eqs. (\ref{eq:e_e_me}) and (\ref{eq:ee_conv})) read
\begin{align}
\label{eq:cll_me2}
g_{\alpha_1\alpha_2}^{\alpha_3\alpha_4} ({\bf q}) =&  \frac{V({\bf q})}{2\pi\hbar v_F} e^{i q_x(X_1+X_4)/2} e^{-i q_x(X_2+X_3)/2} e^{-q_\perp^2 l_B^2/2}\nonumber\\
&\times \delta_{X_4,X_1+q_y l_B^2} \delta_{X_3,X_2-q_y l_B^2} \nonumber\\
&\times \delta_{k_4+k_{\tau_4}, k_1+k_{\tau_1}-q} \delta_{k_3+k_{\tau_3}, k_2+k_{\tau_2}+q}.
 \end{align}
In what follows, we separate $g_{\alpha_1\alpha_2}^{\alpha_3\alpha_4}$ in two parts, which are the most important marginal couplings involved in the lattice instabilities.
The first part is the long-wavelength part, known as the forward Coulomb scattering. 
For this scattering, the momentum transfer ${\bf q}$ is small compared to the internodal distance. 
Thus, we take $\tau_1=\tau_4$ and $\tau_2=\tau_3$ in Eq.~(\ref{eq:cll_me2}).
Moreover, the divergence of the Coulomb potential at ${\bf q}\to 0$ is removed by replacing 
\begin{equation}
\label{eq:Vsc}
V({\bf q}) \to \lim_{q\to 0}V({\bf q})/\epsilon({\bf q},0)\equiv g_2 \pi \hbar v_F,
\end{equation}
where 
\begin{equation}
\epsilon({\bf q},\omega) = 1-V({\bf q}) \Pi({\bf q},\omega)
\end{equation}
is the contribution of Weyl fermions to the dielectric function,
\begin{equation}
\Pi({\bf q},\omega) = \frac{1}{{\cal V}}\sum_{\alpha,\alpha'}|\langle \Psi_{\alpha}|e^{i{\bf q}\cdot{\bf r}}|\Psi_{\alpha'}\rangle|^2 \frac{f_\alpha-f_{\alpha'}}{E_\alpha-E_{\alpha'}+\hbar\omega+i 0}
\end{equation}
is the electronic polarization function in the random phase approximation (RPA) and $f_\alpha$ is the Fermi distribution.
The reason for using static screening in Eq.~(\ref{eq:Vsc}) is that frequency-dependence is irrelevant in the RG sense. 
Then, keeping only the chiral Landau level contribution to the zero-temperature polarization function \cite{rinkel2019}, which is justified for a UV cutoff $\Lambda_0$ that does not exceed the distance in energy between the $n=0$ and $n=1$ Landau levels, we get
\begin{equation}
\label{g20}
g_2= 2 \pi l_B^2 \frac{v_F+\delta v}{v_F + \delta v/2}.
\end{equation}
We emphasize that this RPA expression for $g_2$ is only the bare or initial value of the coupling in the RG flow, i.e. $g_2\equiv g_2(l=0)$.
We can rewrite Eq.~(\ref{g20}) as $g_2 \pi \hbar v_F = e^2/(\epsilon_\infty q_{TF}^2)$,
where 
\begin{equation}
q_{TF}^2 = \frac{e^2}{\epsilon_\infty} \nu(\epsilon_F)
\end{equation}
is the square of the Thomas-Fermi screening wave vector and
\begin{equation}
\nu(\epsilon_F) = \frac{1}{4\pi^2 l_B^2 \hbar v_F}\frac{2+\delta v/v_F}{1+\delta v/v_F}
\end{equation}
 is the density of states at the Fermi level.
For $|\delta v|/v_F\ll 1$,  $(q_{TF} l_B)^2$ is equal to an effective "fine structure constant" $e^2/(2\pi^2 \hbar v_F \epsilon_\infty)$, which is smaller than unity in typical Weyl semimetals. 
Thus, we expect $q_{TF} l_B \lesssim 1$ in common Weyl semimetals in the quantum limit.

The second part of $g_{\alpha_1\alpha_2}^{\alpha_3\alpha_4}$ is the "short"-wavelength part, known as the backward Coulomb scattering.
Here, the momentum transfer ${\bf q}$ connects the two nodes of opposite chirality. 
Thus, we take $\tau_1=-\tau_4$, $\tau_2=-\tau_3$, $\tau_4=-\tau_3$ in Eq.~(\ref{eq:cll_me2}) and we replace
\begin{equation}
V({\bf q}) \to V (2 k_F \hat{\bf z}+{\bf q}_\perp).
\end{equation}
Then, we define 
\begin{align}
\label{g10}
g_1&=\frac{1}{\pi \hbar v_F}\frac{\int_0^{1/l_B} dq_\perp\,q_\perp  V(2 k_F\hat{\bf z}+{\bf q}_\perp)}{\int_0^{1/l_B} dq_\perp\, q_\perp} \nonumber\\
&= \frac{e^2 l_B^2}{\pi\hbar  v_F \epsilon_\infty} \ln\left(1+\frac{1}{4 k_F^2 l_B^2}\right).
\end{align}
Like for $g_2$ in Eq.~(\ref{g20}), the value of $g_1$ in Eq.~(\ref{g10}) is only the initial (bare) value of the coupling in the RG flow. 

Two remarks are in order here. First, our expression for the bare $g_1$ is unscreened. This is well justified when $k_F l_B\gtrsim 1$. In such regime, $(2 k_F)^2 \gg q_{TF}^2$ and the non-logarithmic part of the RPA screening (coming from the "fast" electrons) is qualitatively unimportant.
Regarding the logarithmically divergent part of the RPA screening at $q=2 k_F$, it is not to be included in the bare $g_1$, but rather it is treated below in the course of the RG flow.
Second, the averaging over the transverse momentum ${\bf q}_\perp$ with a cutoff $1/l_B$ in Eq.~(\ref{g10}) is motivated by simplicity (so that the effective electron-electron coupling is independent of wave vector) and by the fact that the Coulomb matrix element in Eq.~(\ref{eq:cll_me2}) is exponentially suppressed for $q_\perp l_B\gtrsim 1$.
This averaging procedure is justified provided that $k_F l_B$ is not small. When $k_F l_B\to 0$, the backward Coulomb scattering becomes singular and this singularity has to be treated on an equal  footing as the one in the forward Coulomb scattering. The study of the regime $k_F l_B \ll 1$ goes beyond the scope of our paper.


The relative strength between short- and long-range Coulomb interactions plays an important role in the Peierls instability. 
As we show below, a larger value of $g_1/g_2$ results in a lower critical temperature for the Peierls instability.
From Eqs.~(\ref{g20}) and (\ref{g10}), the ratio between the bare couplings is
\begin{equation}
\frac{g_1}{g_2} = \frac{e^2}{2\pi^2 \hbar v_F \epsilon_\infty} \frac{1+ \delta v/2 v_F}{1 +\delta v/v_F}\ln\left(1+\frac{1}{4 k_F^2 l_B^2}\right),
\end{equation}
which depends on the magnetic field explicitly via $l_B$ and implicitly via $k_F$. 
When $k_F l_B\gg 1$, we have $g_1/g_2 \sim q_{TF}^2/k_F^2$ (note that  $k_F l_B\gg 1$ is {\em not} incompatible with being in the quantum limit, provided that the Weyl nodes are well-separated in momentum space).
When $k_F l_B\lesssim 1$, $g_1/g_2 \sim (q_{TF} l_B)^2 \ln[1+1/(2 k_F l_B)^2]$ 
 increases with $B$.
In summary, for a weakly interacting Weyl semimetal  (small fine structure constant), 
one expects $g_1/g_2\ll 1$.
However, for more strongly interacting systems with $k_F l_B\lesssim 1$, it is possible that $g_1 \simeq g_2$.



\subsection{RG equations for the adiabatic Peierls instability}

The outer shell corrections entering the RG recursive relations (Eqs.~(\ref{Sp0}), (\ref{zeph0}) and (\ref{g1g2})) and calculated in Appendix A are all evaluated in the limit of small $q_\perp $ up to the cutoff $1/l_B$, above which all the couplings are 
exponentially suppressed (see  Eqs. (\ref{eq:cll_me}) and (\ref{eq:cll_me2})). Within the magnetic  length scale $l_B$, the couplings can  then be considered as essentially local in the transverse directions, which yield one-dimensional-like loop corrections for  the flow equations.   

From  Eq. (\ref{zeph0}) and the results of Appendix A, the RG transformation  for the internode electron-phonon vertex part $z_{\tau,-\tau}$ (which renormalizes the bare coupling $g_{x,j}$ introduced below Eq.~(\ref{eq:cll_me})) 
 is purely electronic and independent of the phonon mode $j$; it leads  to the scaling equation
\begin{equation}
\label{zpm}
\frac{d}{dl} \ln z_{\tau,-\tau}= {\alpha_0\over 2} { (g_2-g_1)}\lambda_P,
\end{equation}
where 
\begin{equation}
\label{alpha0}
\alpha_0=\frac{1}{2\pi l_B^2 (1+ \delta v/2v_F)}
\end{equation}
is a factor that originates from the three-dimensionality of the system, and $\lambda_P(l,T,\delta v)$ is the loop cutoff function of the Peierls scattering channel resulting from an exact summation over intermediate Matsubara fermion frequencies ($0\le\lambda_P\le 1$, see Eq. (\ref{IPS})).


The flow of the electron-phonon vertex then depends on the combination of electron-electron couplings  $g_2-g_1$ governed by Eq. (\ref{g1g2}).  
The derivations in Appendix A at the one-loop level lead to the scaling equations 
\begin{align}
\label{g1g2ChiralA}
 {dg_1\over dl}  &  = -{1\over 2}g_1^2 \alpha_0\lambda_P + g_2g_1\alpha_0(\lambda_P + \lambda_C) \\
 {dg_2\over dl}    &  = {1\over 2}g_1^2\alpha_0 \lambda_C  + {1\over 2}g_2^2\alpha_0(\lambda_P + \lambda_C),
 \label{g1g2ChiralB}
\end{align}
which can be merged to give
\begin{equation}
\label{g2mg1}
{d(g_2-g_1)\over dl}  =  {1\over 2} \alpha_0 (g_2-g_1)^2 (\lambda_P+ \lambda_C),
\end{equation}
as a  relevant combination  of couplings in presence of  effectively spinless fermions shown in Fig. \ref{fig_RG1} (c). In our local transverse scheme for the couplings, this equation is similar to the one obtained by Yakovenko for the problem of 3D electron gas in the presence of strong magnetic field\cite{Yakovenko93}.    Here  $\lambda_C(l,T,\Delta,\delta v)$  is the loop cutoff function of the Cooper scattering channel, as obtained from the summation over Matsubara fermion frequencies ($-1\leq\lambda_C\leq 0$, see Eq. (\ref{Cooper})). This $\lambda_C$ is opposite in sign to $\lambda_P$ and  it is also cut off by the energy scale 
\begin{equation}
\label{eq:Delta_main}
\Delta\equiv \hbar v_F(k_{F+}+k_{F-})
\end{equation}
associated to mirror symmetry-breaking.
In the RG equations (\ref{zpm}), (\ref{g1g2ChiralA}) and (\ref{g2mg1}), the variables $g_1$, $g_2$, $\lambda_C$ and $\lambda_P$ vary with $l$ (they are not to be confused with their bare values). In what follows we shall proceed to the   integration of the flow equations as a function of $l$. At sufficiently  large $l $, the functions $\lambda_{P,C}$ will  cut-off  the flow providing in turn  either a $T$ or a  $\Delta$ dependence   to various quantities depending whether  $k_BT< \Delta$ or $k_BT> \Delta$.  From the $T$ dependence, for instance, one can then extract   the temperature scale of various instabilities of the model.  

When the two Weyl nodes are related by an improper symmetry operation that remains unbroken in the presence of the external  magnetic field, we get $\Delta=0$. 
Such is the case, for instance, if at zero magnetic field the two nodes were related to one another by a mirror plane and the applied magnetic field is oriented perpendicular to that plane.
If the magnetic field is tilted away from the normal to the mirror plane, one generically has $\Delta\neq 0$.
Thus, the energy scale $\Delta$ is partially tunable with the magnitude and orientation of the magnetic field. 
Of course, if the two nodes are unrelated by an improper symmetry at zero magnetic field, then $\Delta\neq 0$ irrespective of the magnetic field.


When ${\bf q} = 2 k_F \hat{\bf z}$, the  RG transformation for the phonon propagator of mode $j$ 
leads to the phonon self-energy corrections  $\delta \scalebox{1.55}{$\pi$}_j $  shown in  Fig.~\ref{fig_RG1}~(a)  and   in Eq. (\ref{Sp0}). In the adiabatic case, these corrections are purely electronic in character. At the one-loop level at step $\ell$, one has
\begin{align}
\label{Kohn_chiral}
\mathscr{D}_{j,l}^{-1} = &\, \mathscr{D}_{j,0}^{-1} -  \scalebox{1.55}{$\pi$}_{j,l}\cr
=  & \, \omega_m^2 + \omega_{0,j}^2(\mathbf{q})\left[1-  \frac{\alpha_0}{2} g_{x,j}^{\prime2}\chi(l)\right],
\end{align}
where $g_{x,j}'\equiv g_{x,j}/\omega_{0,j}$ and
\begin{equation}
\label{Ki0}
\chi(l) =  \int_0^l z_{\tau,-\tau}^2(l')\lambda_P dl'
\end{equation}
is the charge density-wave susceptibility of the electron system along $\hat{\bf z}$ and near $2k_F$.

Equation~(\ref{Kohn_chiral}) describes the softening of phonons due to their coupling to Weyl fermions. 
The vanishing of the term inside the square bracket sets the temperature scale for the Peierls instability.
For pedagogical reasons, we compare Eq.~(\ref{Kohn_chiral})  with the standard RPA expression for the renormalized phonon frequency (see e.g. Refs. [\onlinecite{mahan2000, rinkel2019}]).
The connection emerges if we set the bare $g_2$ to zero and if we neglect $\lambda_C$ in our RG equations.
Under those conditions,  Eq.~(\ref{Ki0}) becomes
\begin{equation}
\label{eq:chi_RPA}
\chi(l)=\frac{\int_0^l \lambda_P dl'}{1+ \frac{\alpha_0 g_1}{2} \int_0^l \lambda_P dl'},
\end{equation}
where $g_1$ here is the bare coupling for the backward Coulomb scattering.
If we substitute Eq.~(\ref{eq:chi_RPA}) in Eq.~(\ref{Kohn_chiral}), we recover the usual RPA result for the renormalized phonon frequency, with $\alpha_0\int_0^l \lambda_P dl'/2$ playing the role of the electronic polarization bubble and  $1+ g_1\alpha_0 \int_0^l \lambda_P dl'/2$ describing the contribution from low-energy electrons to the dielectric function.
Thus, we learn that the widely used RPA expressions are valid only when the forward Coulomb scattering and the scattering amplitude in the Cooper channel are negligible. 
Yet, as we see below, both $g_2$ and $\lambda_C$ play an important role in the Peierls instabilities of Weyl semimetals at high magnetic fields.


\subsubsection{Adiabatic regime with mirror symmetry}
\label{sec:cll_mirror}

Let us suppose that the two Weyl nodes of opposite chirality are related by a mirror plane at zero magnetic field. 
This is a common circumstance in a variety of real Weyl semimetals. 
In the presence of a magnetic field, mirror symmetry is preserved provided that the field is oriented perpendicular to the mirror plane. Then, one has $\Delta=0$, $\delta v=0$ and  $\lambda_P(l,T) =-\lambda_C(l,T)$, the latter of which become cut off only by the temperature. It follows that interference between both (Peierls and Cooper) scattering channels is maximum, so that 
\begin{align}
\label{g1g2Chiral}
\frac{d g_1}{dl}&=-{\alpha_0\over 2} g_1^2\lambda_P\cr
\frac{d g_2}{dl}&= {\alpha_0\over 2} g_1^2\lambda_C =\frac{d g_1}{dl}.
\end{align}
The solution for $g_1(l)$ at low temperature (where $\lambda_{P}\approx1$) is given by
\begin{equation}
\label{g1l}
g_1(l) = {g_1 \over 1+ {1\over 2} \alpha_0 g_1 l},
\end{equation}
indicating that repulsive backward scattering decays with  $l$ and is marginally irrelevant. From Eqs. (\ref{g1g2Chiral}) or (\ref{g2mg1}), we see that for the mirror symmetric situation,
$g_1-g_2$ is an invariant of the RG flow. This is a consequence of the fact that, since fermions in chiral Landau levels are effectively spinless, the $g_1$ term can be transformed  by exchange to a $g_2$ process, which conserves particles on each node  and satisfies electron-hole symmetry at $\Delta=0$ \cite{Fowler76}.  

The invariant combination $g_2(l)-g_1(l)=g_2-g_1$   remains marginal and  nonuniversal. One can then write down the solution for the electron-phonon vertex part (Eq. (\ref{zpm})) as 
\begin{equation}
\label{zeph}
z_{\tau,-\tau} = e^{{1\over 2}\gamma \int_0^l \lambda_P(l') dl'},
\end{equation}
where 
 \begin{equation}
\label{eq:gamma}
\gamma= {\alpha_0(g_2-g_1)}
\end{equation}
is a nonuniversal power law exponent dependent on Coulomb interactions and the magnetic field. 
At ${\bf q}=2k_F \hat{\bf z}$, $\lambda_P=1$ up to the temperature cutoff value 
\begin{equation}
l_T\equiv \ln (\Lambda_0/k_B T),
\end{equation}
 and $\lambda_P\simeq 0$ for $l>l_T$.
Then, from Eq.~(\ref{Kohn_chiral}), the vanishing of the renormalized phonon frequency 
in the static ($\omega_m=0$)  limit leads to the Peierls critical temperature scale for the $j$ mode,
\begin{equation}
\label{Tc0}
T_{c,0}^j={\Lambda_0\over k_B}\left( \alpha_0 g_{x,j}^{\prime2}\over 2\gamma +  \alpha_0 g_{x,j}^{\prime2}\right)^{1/\gamma}.
\end{equation}
In practice, the Peierls instability is set by the mode $j$ for which $T_{c,0}^j$ is maximum.
The power law dependence for the critical temperature originates from the competition between the Peierls and the Cooper scattering channels, and differs qualitatively from the exponential, BCS-like forms proposed in earlier mean-field treatments \cite{Yang2011, Zhang2016, Qin2020}.
The latter studies have implicitly ignored the possibility of a Cooper channel contribution. We emphasize that the Cooper channel is important even though the two Weyl nodes are not time-reversed partners of one another;
the reason is that the mirror plane relating the two Weyl nodes passes from the $\Gamma$ point, so that the two nodes have opposite momenta along the direction of the magnetic field. 
It is as though the presence of the magnetic field rendered an effectively one-dimensional problem with the $k\to -k$~ symmetry that is required for a full Cooper channel contribution.

Let us comment on some qualitative aspects of Eq.~(\ref{Tc0}). When $g_2> g_1$, Peierls instability occurs at any nonzero value of $g_{x}^{\prime2}$ and charge-density wave correlations of the electron system enhance the  Peierls temperature with respect to the BCS-like limit  $k_BT^j_{c,0} = \Lambda_0 \exp[-2/ (\alpha_0 g_{x,j}^{\prime2})]$, which we recover when electron-electron interactions are vanishingly small ($\gamma\to 0^+$). 
When $g_2< g_1$ (i.e. $\gamma <0$), 
a Peierls instability can still occur but only if the electron-phonon coupling strength exceeds a threshold value ($\alpha_0 g_{x,j}^{\prime2} > -2\gamma$). 
In this sense, the phonon softening is qualitatively weaker if $g_1>g_2$ than if $g_1<g_2$.
As mentioned above, $g_1>g_2$ takes place if $q_{TF}> {\rm max}(k_F, 1/l_B)$.
The fact that the screening due to the short-range ($2 k_F$) part of the Coulomb interaction tends to suppress the Peierls instability when $q_{TF} > k_F$ is consistent with the literature in quasi one-dimensional organic conductors at zero magnetic field \cite{barisic83}.

\begin{figure}[h]
  \begin{center}
    \includegraphics[width=1.0\columnwidth]{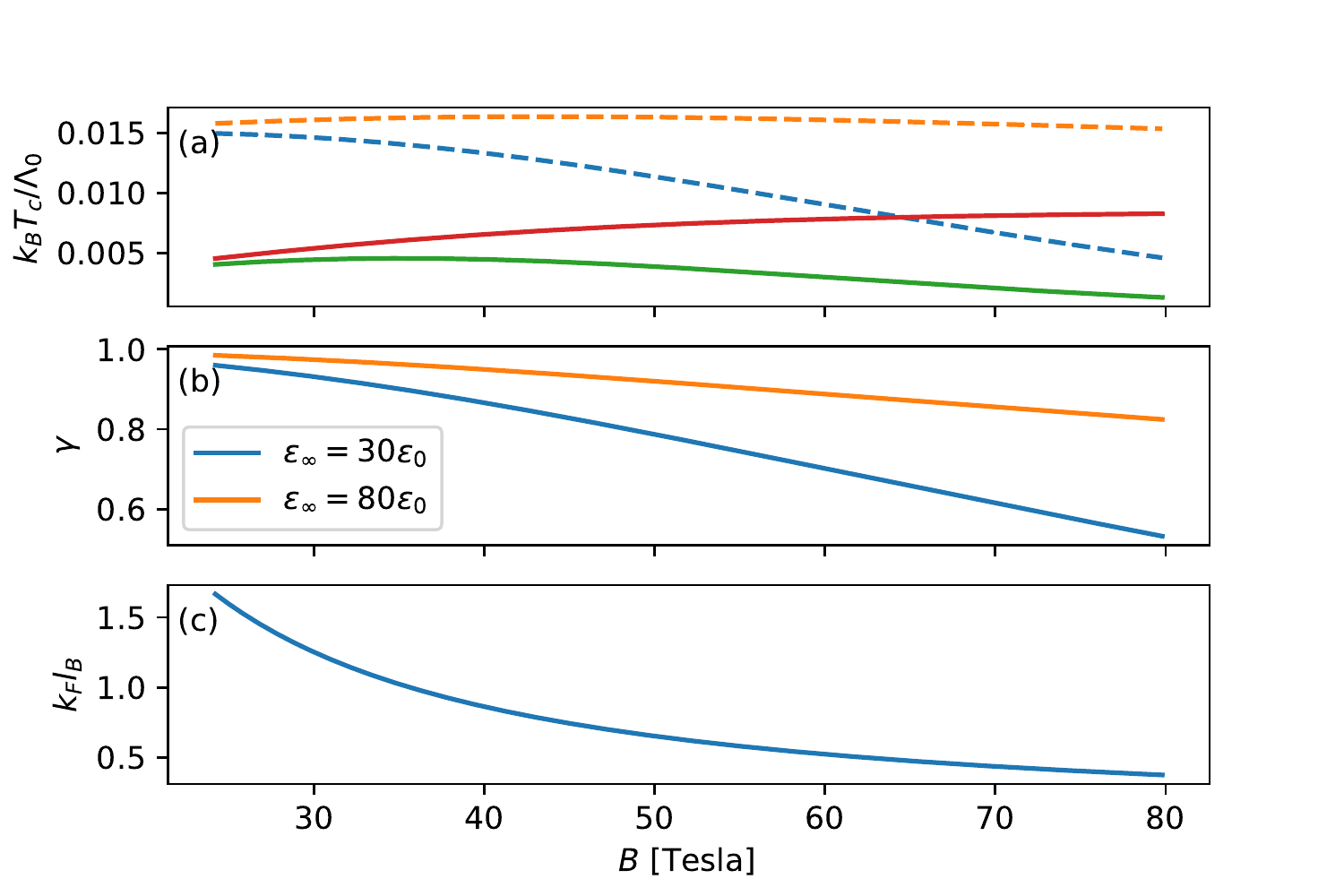}
\caption{(a) Estimate of the Peierls transition temperature in the adiabatic phonon approximation as a function of the magnetic field, with mirror symmetry (solid lines) and without mirror symmetry (dashed lines) for $\Lambda_0=50\, {\rm meV}$ (considering a field-dependent cutoff does not introduce qualitative changes to our results). For the curves with the dashed lines, we take $\Delta=5\, {\rm meV}$ as the mirror symmetry-breaking energy scale. Only the magnetic field values for which the system is in the quantum limit are displayed.
Two different values of $\epsilon_\infty$ are considered: $30$ (blue and green curves), $80$ (orange and red curves).
The transition temperature $T_{c}$ is enhanced by mirror-symmetry breaking. Also, the transition temperature is higher for larger $\epsilon_\infty$.
(b) The electron-electron interaction parameter $\gamma$ (Eq.~(\ref{eq:gamma})) as a function of the magnetic field, for two different values of $\epsilon_\infty$.
(c) $k_F l_B$ as a function of the magnetic field. The parameter values are indicated in the main text.
}
 \label{fig:Tc}
  \end{center}
\end{figure}

Next, we carry out a numerical estimate of Eq.~(\ref{Tc0}) (see Fig. \ref{fig:Tc} for the variation of $T^j_{c,0}$ as a function of the field value $B$).
Combining Eqs.~(\ref{eq:ep_approx}) and (\ref{eq:ep_conv}), we have 
\begin{equation}
\alpha_0 g^{\prime2}_{x,j}\simeq \frac{e B d_j^2}{\pi^2   v \rho \hbar^2 \omega_{0,j}({\bf q})^2}.
\end{equation}
We consider optical phonons with $d_j= 5\, {\rm eV}/\AA$ and  $\hbar\omega_{0,j}({\bf q}) = 10\, {\rm meV}$.
We also take $\rho=10^4\, {\rm kg/m}^3$, $v_\tau=10^5\, {\rm m/s}$, $b=0.01 \AA^{-1}$ and $n_e=5\times 10^{17}\, {\rm cm}^{-3}$.
We take two possible values for $\epsilon_\infty$, $30$ and $80$.

In Fig. \ref{fig:Tc}, the dependence of $T^j_{c,0}$ on $B$ is nonmonotonic. 
For low-to-moderate magnetic fields, $g_1/g_2\ll 1$ and hence $\gamma\simeq 1$ (far from the BCS-like regime).
In this regime, $\gamma\gg \alpha_0 g^{\prime2}_{x,j}$ and $k_B T^j_{c,0}/\Lambda_0 \simeq  \alpha_0g^{\prime2}_{x,j}/2$ grows linearly with $B$.
As the magnetic field is made stronger, $g_1/g_2$ increases and as a result $\gamma$ decreases.
At high enough magnetic field, this eventually leads to a suppression of $T_c$ due to $g_1$.
The magnitude of the maximum critical temperature
depends sensitively on the parameters of the model. 
For instance, if $d_j=2 {\rm eV}/\AA$, everything else being equal, the maximum of $T^j_{c,0}$ is suppressed by almost an order of magnitude. 
For a given material, higher carrier density implies a higher value of the maximal $T^j_{c,0}$.
In Fig. \ref{fig:Tc}, the maximum of $T^j_{c,0}$ attains $\sim 1\%$ of the cutoff energy scale $\Lambda_0$. 
 The value for $\Lambda_0$ is roughly set by the distance in energy between the Fermi level and the bottom of the $n=1$ nonchiral Landau level band.  We estimate $\Lambda_0\sim \hbar v_F/l_B \gtrsim 20\, {\rm meV}$ for the quantum limit, which is inferior to the values used in the recent literature \cite{Zhang2016}.
Based on these numbers, we conclude that the emergence of a measurable Peierls instability in a mirror-symmetric Weyl semimetal at high magnetic field is far from obvious. We must emphasize, however, that while these numbers are reasonable across a wide family of materials, reliable numerical estimates for $T_{c}$ in individual Weyl semimetal materials are beyond the scope of our work. For one thing, the value of the electron-phonon coupling is not known reliably in these materials. Likewise, the presence of multiple pairs of nodes in several real Weyl semimetals introduces additional complications in making quantitative predictions about field-induced instabilities (see Sec.\ref{sec:real} for further comments in this regard).

\subsubsection{Adiabatic regime with broken mirror symmetry}
\label{sec:cll_no_mirror}

We now consider the more general case with broken mirror symmetry.
Nonzero $\delta v$ and $\Delta=\hbar v_F (k_{F+}+k_{F-})$
are inevitable if the two Weyl nodes of opposite chirality are unrelated by an improper symmetry. 
In the instance in which the two nodes are related by a mirror plane at zero magnetic field, such mirror symmetry can be effectively broken by orienting the magnetic field away from the normal  to the mirror plane. 
In such case, $\Delta$ can be large (of the order of $\hbar v_F (k_++k_-))$ when the magnetic field is parallel to the mirror plane. 

From  Eq.~(\ref{g2mg1}), the solution for $g_2(l)-g_1(l)$ can be obtained,
\begin{equation}
\label{C1}
g_2(l)-g_1(l) = {g_2-g_1\over 1- {1\over 2}\alpha_0(g_2-g_1) \int_0^l (\lambda_P + \lambda_C)dl'}.
\end{equation}
A finite $\delta v$ alone does not affect the logarithmic singularity of the scattering channels, but leads to a density of states renormalization by the factor $(1+ \delta v/2v_F)^{-1}$, which we have included in the definition for $\alpha_0$ (Eq.~\ref{alpha0}). 
In contrast, $\Delta$ acts as a pair breaking perturbation for the logarithmic singularity of the Cooper scattering channel, which is thereby suppressed at sufficiently low energy.  Electron-hole symmetry at $2k_F$ (nesting), however, is preserved and  the related  logarithmic singularity  in the Peierls channel therefore remains essentially unaffected at all $l< l_T$. 

From Eq. (\ref{Cooper}), one can distinguish  three regimes for the cutoff functions $\lambda_C$ and $\lambda_P$ as a function of  $l$.  
First, at  $l <l_\Delta= \ln (\Lambda_0/\Delta)$,  the influence of $\Delta$ is small and one still  has $\lambda_C\approx -\lambda_P=-1$. Therefore, the flow equations (\ref{g1g2Chiral}-\ref{zeph})  remain valid together with the expression for the Peierls critical temperature (Eq. (\ref{Tc0})), with the proviso that $k_BT^j_{c,0}> \Delta$.
 
The second regime, where  $l> l_\Delta$, becomes applicable if $\Delta$ exceeds the Peierls transition temperature for the mirror-symmetric case (Eq. (\ref{Tc0})).
In this regime, the Cooper loop is cut off by $\Delta$ and becomes vanishingly small ($\lambda_C\approx 0$). Accordingly,  the expression for $T^j_{c,0}$ is modified. The flow of interactions (Eq.~(\ref{C1})) for $\lambda_P\approx 1$ up to $l_T$ will develop  a simple pole  singularity at the  characteristic temperature
   \begin{equation}
\label{ }
T^*_0 = {\Delta\over k_B } e^{-2/\alpha_0(g_2-g_1)},
\end{equation} 
which corresponds to the scale of an intrinsic  instability of the electron gas against a $(0,0,2k_F)$ charge-density-wave formation. This can be checked by looking at the vertex part $z_{\rm CDW}$ for the electronic  CDW susceptibility. From (\ref{sh0}), (\ref{sh}) and (\ref{fcdw}), its flow equation reads
\begin{equation}
\label{zcdw}
\frac{d}{dl} \ln z_{\rm CDW}= {\alpha_0\over 2} { (g_2-g_1)}\lambda_P,
\end{equation}
which unsurprisingly coincides with (\ref{zpm}) for $z_{\tau,-\tau}$. Using (\ref{C1}), one has 
\begin{equation}
\label{zdelta}
z_{\rm CDW}(l) = {z_{\rm CDW}(\Delta)\over [1-{1\over 2}\alpha_0(g_2-g_1) (l-l_\Delta)]},
\end{equation}
where $z_{\rm CDW}(\Delta)=z_{\tau,-\tau}(\Delta)$, as  given by Eq.~(\ref{zeph}) when evaluated at $l=l_\Delta$. The expression for  $z_{\rm CDW}(l_T)$ and thus the CDW susceptibility (\ref{Ki0}) develops a simple pole singularity at $T^*_0$. 

 However, in the presence of a nonzero electron-phonon coupling, this instability is preempted by the Peierls transition.  
According  to Eq.~(\ref{C1}), the phonon softening condition (Eq.~(\ref{Kohn_chiral})) then gives a Peierls transition temperature 
\begin{equation}
\label{Tc0Delta}
T^j_{c,0}= {\Delta \over k_B} \exp\left( -{2- \alpha_0g_{x,j}^{\prime 2}\chi(\Delta)\over \gamma [1 -{1\over2}\alpha_0\chi(\Delta) g_{x,j}^{\prime 2}] +\alpha_0z_{\tau,-\tau}^2(\Delta)g_{x,j}^{\prime 2}}\right),
\end{equation}
 where $\chi(\Delta)$ is the contribution to the charge density-wave electronic susceptibility (\ref{Ki0}) up to $l_\Delta$. These correlations enhance the transition temperature with respect to the case without electron-phonon interactions.
We remind the reader that Eq.~(\ref{Tc0Delta}) holds only if $k_B T^j_{c,0}\leq \Delta$; in the opposite regime, one can ignore mirror symmetry-breaking and use Eq.~(\ref{Tc0}).
We also note that $T_{c,0}^j$ exceeds the temperature scale $T_0^*$ of the purely electronic instability, with $T_{c,0}^j\to T_0^*$ when $g_{x,j}\to 0$.

Finally, the third regime takes place when $\Delta>\Lambda_0$, i.e. when mirror symmetry is strongly broken. In this case, $l_\Delta\to 0$ and the Peierls transition temperature is still given by Eq.~(\ref{Tc0Delta}), but with $\chi(\Delta)\to 0$ in the exponent and $\Delta\to\Lambda_0$ in front of the exponential, i.e. 
\begin{equation}
\label{Tc0Delta2}
T^j_{c,0}= {\Lambda_0\over k_B} \exp\left( -\frac{2}{\gamma  +\alpha_0 g_{x,j}^{\prime 2}}\right).
\end{equation}

Figure \ref{fig:Tc}(a)  shows the evolution of the transition temperature as a function of the applied magnetic field for $\Lambda_0=50~{\rm meV}$ (considering a field-dependent cutoff does not introduce qualitative changes to our results) and ${\Delta=5~{\rm meV}}$ (assumed for simplicity to be field-independent), for the same parameters as the in the mirror-symmetric case. 
Interestingly, the transition temperature is significantly enhanced when mirror symmetry is broken. 
As shown by Eq. (\ref{zdelta}), the presence at finite $l$ of a simple pole singularity  magnifies  the charge density-wave response of the electron system with respect to the power law  (\ref{zeph}) for $\Delta=0$. This  is responsible for such an enhancement in  $T^j_{c,0}$.  
  

In sum, there is a qualitative difference for the Peierls instabilities of the chiral Landau levels depending on whether or not they are related by an improper symmetry. 
In the presence of mirror symmetry, the instability vanishes in the absence of electron-phonon interaction. This is because of the destructive interference between the Cooper and the Peierls channels. 
When the mirror symmetry is broken, the Cooper channel is suppressed and this opens the possibility for having a purely electronic instability even when the electron-phonon coupling is negligible. 
However, the presence of electron-phonon coupling makes the instability more pronounced, and thus a lattice instability preempts the purely electronic instability. 

\subsection{Nonadiabatic Peierls instability}
\label{sec:nonad}

Thus far, our results were obtained in the so-called adiabatic regime, where $\hbar\omega_{0,j}(2 k_F \hat{\bf z}) \ll  k_BT^j_{c,0}$ and the electronic degrees of freedom are `faster' than the phonons.
In this case, the effective electron-electron interaction induced by the exchange of phonons is strongly retarded and its contribution to  higher order vertex corrections in  the phonon self-energy  (Eq. (\ref{Kohn_chiral})), which would result from the contraction of  the generated  anharmonic terms in the phonon action,  is negligible. This leaves us with the diagram in Fig. \ref{fig_RG1}(a) of the adiabatic approximation, in which vertex corrections are purely electronic in character.

However, it is apparent from the numerical estimates of Fig.\ref{fig:Tc} that the scenario $\hbar\omega_{0,j}(2 k_F \hat{\bf z}) \gg k_BT^j_{c,0}$
can be commonly realized in practice in Weyl semimetals.
This defines the nonadiabatic regime where phonon-mediated interactions between electrons tend to be nonretarded. Such interactions make a significant contribution to the electronic vertex in Fig. \ref{fig_RG1}(b-c) and 
contribute to the phonon self-energy depicted in Fig. \ref{fig_RG1}(a). It follows that the criterion of instability against the formation of a superstructure, as derived in Eq. (\ref{Kohn_chiral}) for phonons, no longer holds and must be modified. Therefore, at energy scales below $\hbar\omega_{0,j}(2 k_F \hat{\bf z})$, we focus on the electronic degrees of freedom and derive 
the condition for the instability of the electron system against  the formation of a $2k_F \hat{\bf z}$ charge-density wave, using Eqs. (\ref{g1g2ChiralA}) and (\ref{g1g2ChiralB}).

We shall tackle this problem using a two-cutoff RG approach \cite{Caron84}.  First, we integrate  the flow equations of the previous section in the adiabatic sector, where $\Lambda_0(l) \ge   \hbar\bar{\omega}_{0}$, here defined for simplicity as the minimum energy of the set of phonon modes $\{\hbar\omega_{0,j}\}$. 
The scaled cutoff energy  reaches the characteristic phonon energy at $l^*$, i.e. $\Lambda_0(l^*)\equiv  \hbar\bar{\omega}_{0}$. 
For $l>l^*$, all phonon modes enter the nonadiabatic regime and  we carry out the partial trace integration $\int D\phi\  e^{(S^{(0)}_{p,l^*}+ S^{}_{ep,l^*})/\hbar}$ over phonon degrees of freedom  in the partition function (Eq. (\ref{RGTransf})). By dropping  anharmonic terms generated by the RG transformation, the integration is gaussian and yields attractive, phonon-mediated interactions between electrons. The   backscattering amplitude mediated by phonons takes the form
\begin{equation}
\label{g1ph}
\mathfrak{g}_1 = - {1\over 2}z_{\tau,-\tau}^2(l^*)\sum_j\mathscr{D}_{j,l^*}(0,2k_F \hat{\bf z})g_{x,j}^{2}.
\end{equation}  
This coupling can be considered as essentially nonretarded, i.e. independent of frequency. 
As such, its amplitude will simply add to the purely repulsive $g_1(l^*)$ part obtained  from Eq. (\ref{g1l}), so that the  initial condition at  $l^*$ for (\ref{g1g2Chiral})  becomes $g_1^*= g_1(l^*) + \mathfrak{g}_1$. With sizable electronic vertex corrections $z_{\tau,-\tau}(l^*)$ and  Peierls fluctuations, which imply a large $\mathscr{D}_{j,l^*}(0,2k_F \hat{\bf z})$,  the likelihood for a total attractive backscattering amplitude $g_1^*$ is high. 

The integration of phonon modes at small ${\bf q}$ will also give rise to an effective attractive forward scattering amplitude
 \begin{equation}
\label{g2ph}
\mathfrak{g}_2 = - {1\over 2}\sum_j\mathscr{D}_{j,0}(0,q\hat{\bf z}\sim0)g_{\tau,\tau,j}g_{-\tau,-\tau,j},
\end{equation} 
which will add to $g_2(l^*)$ to give $g_2^*= g_2(l^*) + \mathfrak{g}_2$ as the boundary condition for the forward scattering at $l^*$. However, in the absence of electronic enhancement ($z_{\tau,\tau}=1$) and a lack of ${\bf q} \sim 0$ phonon softening, this attractive contribution is expected to be small, so that  $g_2^*$ remains repulsive.

Given these corrections to the electronic interactions, we will investigate the possibility for an electronic instability in the Weyl semimetal model, in the nonadiabatic regime.

\subsubsection{Nonadiabatic regime with mirror symmetry}
\label{sec:nonad_mir}

We first consider the possibility for instabilities when the two Weyl nodes are related by perfect mirror symmetry ($\Delta=0,\delta v=0)$. In these conditions, $g_1$ and $g_2$ at $l> l^*$ are still governed by the flow of Eq.~({\ref{g2mg1}), for which the interference between the Cooper and the Peierls scattering channels is maximum ($\lambda_C=-\lambda_P=-1$). The invariant $g_2(l) -g_1(l)=g_2^*-g_1^*$ is now fixed by the boundary conditions at $l^*$.  Therefore, the vertex part (\ref{zcdw}) of the $2k_F$ charge-density-wave susceptibility  at $l>l^*$ takes the scaling form 
\begin{equation}
\label{zna}
z_{\rm CDW}(l) = z_{\rm CDW}(l^*) e^{\gamma^*(l-l^*)/2},
\end{equation}
where $\gamma^*= \alpha_0 (g_2^*-g_1^*)$ and $z_{\rm CDW}(l^*)$ corresponds to the same expression given in Eq. (\ref{zeph}). 
The $2k_F\hat{\bf z}$ charge-density-wave susceptibility, Eq. (\ref{Ki0}), when evaluated at  the temperature cutoff value $l_T$, has the  power law form
\begin{equation}
\label{ }
\chi(T) = \chi(l^*) + {{z^{2}_{\rm CDW}(l^*)}\over \gamma^{*}}\left( \left({  \hbar\bar{\omega}_{0}\over k_BT}\right)^{\gamma^*} -1\right),
\end{equation}
where $\chi(l^*)$ is the contribution to the susceptibility (\ref{Ki0}) up to $l^*$. In the nonadiabatic regime, the chiral mirror symmetric Landau level thus evolves towards a Luttinger liquid behavior with no  instability at finite temperature. 

As  a function of phonon frequency, it follows that the system will crossover from a classical  $2k_F\hat{\bf z}$ Peierls superstructure at small  $ \hbar\bar{\omega}_{0}\ll k_B T_{c,0}^j$ to a Luttinger liquid at high phonon frequency $ \hbar\bar{\omega}_{0}\gg k_B T_{c,0}^j$. Such a quantum-classical crossover  is reminiscent  to the one encountered in  the model of spinless electrons coupled to optical phonons  in one dimension.  The zero temperature Peierls state is  known to be suppressed by quantum lattice fluctuations at large phonon frequency \cite{Hirsch83,Caron84}. Regarding  the strength of correlations,   attractive corrections coming from phonons yield $\gamma^*>\gamma $ at any field values, so that charge density wave correlations are   enhanced by phonons  in the  nonadiabatic domain.  

One may also wonder whether  other types of correlations, such as the superconducting ones, may become singular in the nonadiabatic domain.  It turns out that the  phonon renormalization of $g_1^*$ to the attractive sector will make this possibility unlikely. This is especially  the case for the equal-pseudospin Cooper pairing susceptibility. From Eqs.~(\ref{sh0}), (\ref{Osc}) and (\ref{sh}), its vertex part obeys the flow equation of Fig.~\ref{fig_RG1} (e), namely
\begin{equation}
\label{ }
{d\ln z_{\rm SC}\over dl} = -{\alpha_0\over 2} { (g_1-g_2)}\lambda_C.
\end{equation}
Its solution takes  the scaling form $z_{\rm SC} \sim e^{\gamma_{\rm SC} l^*/2}e^{\gamma_{\rm SC}^*(l-l^*)/2}$, but with a negative power law exponent $\gamma_{\rm SC}^*=-\gamma^*$ that strongly suppresses  superconducting correlations  in the nonadiabatic domain.

\subsubsection{Nonadiabatic regime with broken mirror symmetry}
\label{sec:nonad_nomir}

In the absence of mirror symmetry ($\Delta\neq 0$ and $\delta v\neq 0$), one can distinguish two regimes of nonadiabaticity. The first one is characterized by   $l^*<l_\Delta$, i.e. $ \hbar \bar{\omega}_{0} > \Delta$.
In this regime,  the nonadiabatic conditions described in Sec. \ref{sec:nonad_mir} come first, where the Peierls scale $T_{c,0}^j$ becomes  irrelevant and  Luttinger  liquid conditions for electrons prevail in the interval $l_\Delta> l>l^*$. Within our sharp cutoff boundary approach,   the logarithmic singularity of the Cooper scattering channel is suppressed  at $l > l_\Delta$, where $\lambda_C\approx0$, and  the couplings in Eq. (\ref{C1}) develop a simple pole singularity:
\begin{equation}
\label{ }
g_2(l)-g_1(l) = {g^*_2-g^*_1\over 1- {1\over 2}\alpha_0(g^*_2-g^*_1) (l-l_\Delta)},
\end{equation}
where  $g_{1,2}^*=g_{1,2}(l^*) + \mathfrak{g}_{1,2}$ are evaluated at $l^*$. 
The singularity  influences in turn the vertex part of the charge density-wave susceptibility, 
\begin{equation}
z_{\tau,-\tau}(l) = {z_{\tau,-\tau}(l_\Delta)\over 1- {1\over 2}\alpha_0(g^*_2-g^*_1) (l-l_\Delta)},
\end{equation}
where $z_{\tau,-\tau}(l_\Delta)$ is given by Eq. (\ref{zna}).  The Luttinger liquid  is therefore  unstable against the formation of a charge-density-wave state. Using $l_T=\ln (\Lambda_0/k_BT)$,  the temperature scale of the instability  is given by
\begin{equation}
\label{Tc0nonadia}
T^*_0 = {\Delta\over k_B } e^{-2/\alpha_0(g^*_2-g^*_1)}.
\end{equation}

Next, we examine the second regime, where $l^*>l_\Delta$, i.e. $ \hbar \bar{\omega}_{0} < \Delta$. In this regime, symmetry breaking effects  described in Sec.~\ref{sec:cll_no_mirror}~first set in, before the onset of nonadiabatic regime at $l^*$. In this case the combination of couplings at $l>l^*$ shows a  simple pole structure
\begin{equation}
\label{ }
g_2(l)-g_1(l) = {g^*_2-g^*_1\over 1- {1\over 2}\alpha_0(g^*_2-g^*_1) (l-l^*)}.
\end{equation}
Here $g^*_2-g^*_1$ is given by the solution of  Eq. (\ref{C1}), first up to  $l_\Delta$ and then to $l^*$ where the phonon induced contributions $\mathfrak{g}_{1,2}$ are added to $g_{1,2}(l^*)$. According to Eq. (\ref{zpm})  the  vertex part of the charge-density-wave susceptibility will have the  form 
 \begin{equation}
\label{ }
z_{\tau,-\tau}(l) = {z_{\tau,-\tau}(l^*) \over 1- {1\over 2}\alpha_0(g^*_2-g^*_1) (l-l^*)},
\end{equation}
where $z_{\tau,-\tau}(l^*)$ is given by (\ref{zdelta}). The temperature scale of the instability   against charge-density-wave superstructure becomes
\begin{equation}
\label{Tc0nonadiadelta}
T^*_0 = { \hbar \bar{\omega}_{0} \over k_B } e^{-2/\alpha_0(g^*_2-g^*_1)}.
\end{equation}
One can conclude that mirror symmetry-breaking with finite  $\Delta$ invariably makes the Luttinger liquid  of the nonadiabatic regime unstable against the formation of a charge-density-wave state. 

The nonadiabatic regime has also been addressed in Ref. [\onlinecite{Shindou2019}], albeit for a different model of semimetal with spin-polarized electron and hole bands in the quantum limit. In that paper, an attractive electron-electron interaction mediated by a screened electron-phonon coupling have been considered, and phases such as an excitonic insulator, charge Wigner crystal and charge-density wave have been predicted using a parquet renormalization group analysis. 

\section{Peierls instability from nonchiral Landau levels}
\label{sec:noncll}

In this section, we consider the case in which the Fermi level crosses not only the chiral Landau levels, but also the $|n|=1$ nonchiral Landau levels.
For now, we will be interested in the Peierls instability that results from electronic transitions taking place inside the $|n|=1$ Landau level.
In the end of the section, we will compare this instability to the one taking place in the $n=0$ Landau level. 
We will neglect the possibility of instabilities arising from transitions connecting chiral and nonchiral Landau levels.
For such transitions, the form factors in Eq.~(\ref{eq:form_factor}) vanish when ${\bf q}$ is parallel to the magnetic field.

\subsection{Electron-phonon and electron-electron couplings}

\begin{figure}[h]
  \begin{center}
    \includegraphics[width=0.9\columnwidth]{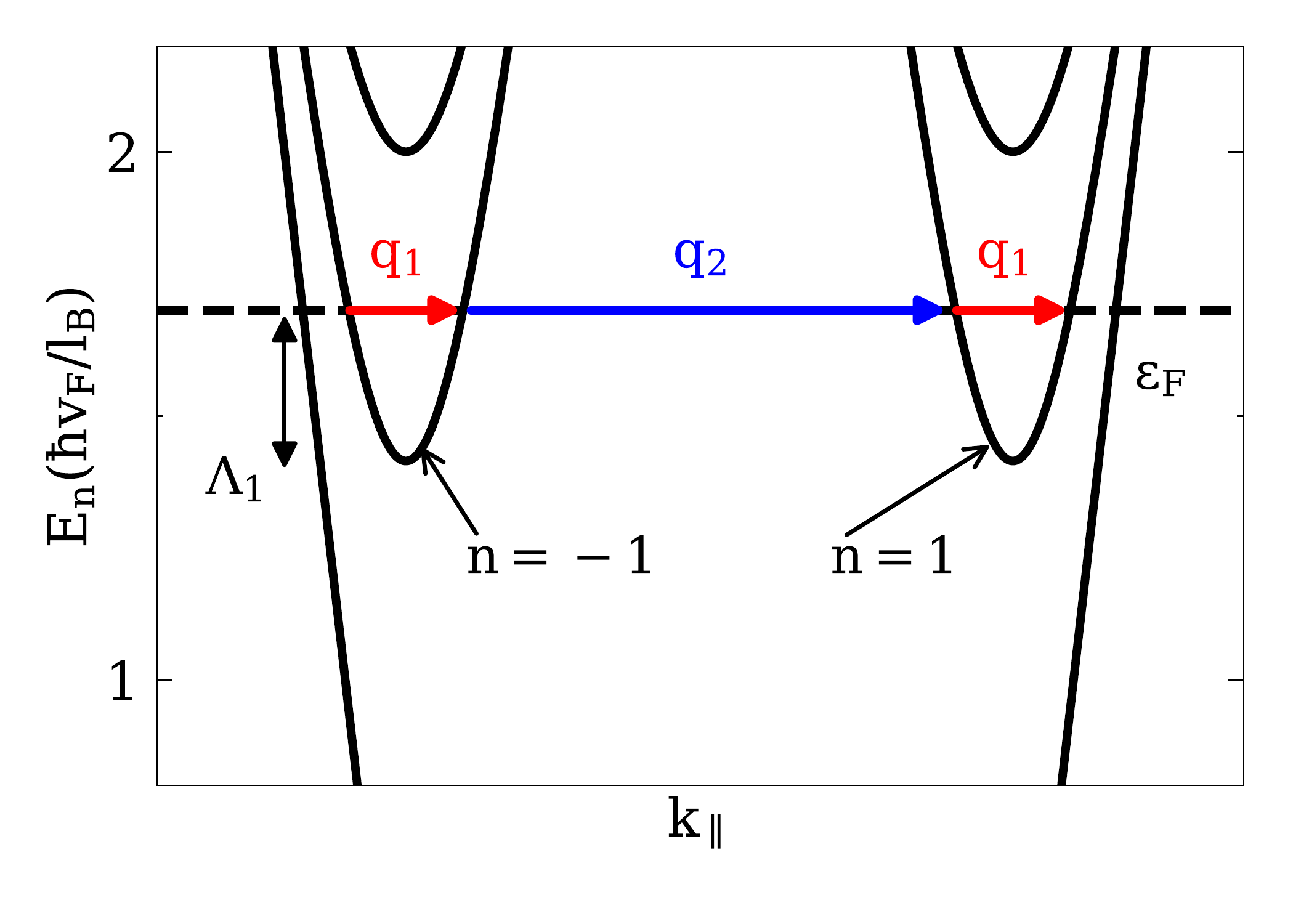}                                                                                         
\caption{Landau level spectrum of two Weyl nodes of opposite chirality, near the quantum limit, where $\rm{k}_{||}$ refers to the component of the momentum along the direction parallel to the magnetic field.  The Fermi energy $\epsilon_F$ intersects the first nonchiral Landau level on each node ($n=1$ for the positive chirality node, and $n=-1$ for the negative chirality node). 
The wave vectors ${\bf q}_1$, ${\bf q}_2$, ${\bf q}_3={\bf q}_1+{\bf q}_2$ and ${\bf q}_4=2 {\bf q}_1 + {\bf q}_2$  connect different Fermi points within the $|n|=1$ set of Landau levels.
The energy $\Lambda_1$ corresponds to the distance from the bottom of the $|n|=1$ Landau levels to $\epsilon_F$; it plays the role of an ultraviolet cutoff.
In this figure, it is assumed that the two nodes of opposite chirality are related by a mirror plane. 
Otherwise, the vector ${\bf q}_1$ would have a different length for the two nodes.}
 \label{fig:nonc}
  \end{center}
\end{figure} 

We begin by assuming that the two Weyl nodes of opposite chirality are related by a mirror plane.
Then, we linearize the dispersion for the $|n|=1$ Landau levels in the vicinity of the Fermi points.
Within the subspace of $|n|=1$ Landau levels, several different phonon wave vectors $\pm {\bf q}_i$ ($i=1,2,3,4$) connect different Fermi points (see Fig. \ref{fig:nonc}).
The wave vectors at which instabilities are more likely are $\pm {\bf q}_1$, $\pm {\bf q}_2$ and $\pm {\bf q}_4=\pm (2 {\bf q}_1+{\bf q}_2)$, because they connect electronic branches that are perfectly nested.
On the one hand, the theory of phonon instabilities at $\pm {\bf q}_2$ and $\pm {\bf q}_4$, which connect nodes of opposite chirality,  is very similar to the one discussed in the preceding subsection for chiral Landau levels. 
The only significant difference is that the ultraviolet cutoff appearing in the expression for the Peierls transition temperature is now given by $\Lambda_1$, instead of $\Lambda_0$, and that the matrix elements of the different couplings are multiplied by a factor $\lambda$ defined below in Eq.~(\ref{eq:lambda}). 
On the other hand, the instability at wave vector $\pm {\bf q}_1$, which connects two $|n|=1$ branches at the same Weyl node, presents some conceptual novelties. 
Thus, hereafter we concentrate on this wave vector, whose magnitude we denote as $2 k_F$.
The maximum value of $k_F$ we consider is $\sqrt{2}/l_B$ (beyond this value, the Fermi level intersects the $|n|=2$ Landau levels).

Let us label $L$ (left) and $R$ (right) the two Fermi points separated by $q_1$ in a node of a given chirality $\tau$.
The electron-phonon matrix element connecting $L$ to $R$ reads
\begin{align}
\label{eq:gLR}
&g_{LR,\tau}^{ph}(\mathbf{q})  =g^{ep}_{\tau}\langle\Psi_{n,X,L,\tau}|e^{i {\bf q}\cdot{\bf r}} |\Psi_{n,X^{\prime},R,\tau}\rangle\nonumber\\
 & \simeq \lambda\, g^{ep}_{\tau} \delta_{q, -2 k_F}\delta_{X^{\prime},X+q_{y}l_{B}^{2}}
e^{iq_{x} (X+X^\prime)/2} e^{-q_{\bot}^{2}l_{B}^{2}/4},
\end{align}
where we have used $n=1$ ($n=-1$) for $\tau=1$ ($\tau=-1$), and we have defined 
\begin{equation}
\label{eq:lambda}
\lambda=\frac{1}{\sqrt{(k_{F}l_{B})^{2}/2+1}}.
\end{equation}
One can similarly obtain an expression for $g_{RL,\tau}^{ph}(\mathbf{q})$; the only difference with respect to  Eq.~(\ref{eq:gLR}) is that $\delta_{q, -2 k_F}$ is replaced by $\delta_{q, 2 k_F}$.
In the derivation of Eq.~(\ref{eq:gLR}), we have approximated $\exp(-q_\perp^2 l_B^2/4) \left[1 + O(q_\perp^2 l_B^2)\right]\simeq \exp(-q_\perp^2 l_B^2/4) $.
This approximation is motivated by the fact that, for $q_{\bot}l_{B}\geq1$, the electron-phonon matrix element is exponentially suppressed. 

Comparing Eq.~(\ref{eq:gLR}) with Eq.~(\ref{eq:cll_me}), we recognize that the electron-phonon matrix elements for chiral and nonchiral Landau levels differ by a factor $\lambda$.
This difference emerges from the pseudospin textures of the respective Landau levels. 
For chiral Landau levels, the expectation value of the pseudospin is either perfectly aligned or antialigned with the $z$ axis (depending on the sign on $B$), irrespective of the wave vector and the chirality  (cf. Eq.~(\ref{eq:spinor_cll})).
Hence, the overlap factor for the spinors of the two chiral Landau levels of opposite chirality is unity.
In contrast, for the $|n|=1$ level, the expectation value of the pseudospin rotates in momentum space: it points along $x$ at the bottom of the $|n|=1$ band, and gradually tilts towards $z$ as $k$ departs from the energy minimum (cf. Eq.~(\ref{eq:spinor_ncll})). 
Moreover, the left- and right-moving branches in a nonchiral Landau level have opposite orientations of the pseudospin along $z$.
This leads to an overlap factor $\lambda<1$ between the branches connected by the wave vector ${\bf q}_1$.
The overlap factor is gradually suppressed for larger $q_1 l_B$.

In practice,  $q_1=2 k_F$ is a very small fraction of the Brillouin zone. Therefore, the ${\bf q}$-dependence of $g^{ep}_\tau$ in Eq. (\ref{eq:gLR}) can be neglected and its transformation properties under symmetry operations can be determined from the irreducible representations at zero wave vector.
In the presence of mirror symmetry, one may have $q=0$ phonons whose deformation potentials are even or odd under the mirror operation.
We denote these phonons as scalar and pseudoscalar, respectively. 
In reality, scalar and pseudoscalar phonons are invariant under all proper symmetry operations; however, for our model of two Weyl nodes, only the mirror operation is relevant. 
Consequently, we have
\begin{align}
\label{geph}
g^{ep}_{1} &=g^{ep}_{-1}\equiv g_{s} \text{ (scalar phonons)}\nonumber\\
g^{ep}_{1} &=-g^{ep}_{-1}\equiv g_{ps} \text{ (pseudoscalar phonons)}.
\end{align}


The Coulomb matrix elements can be calculated in the same way as for the chiral Landau levels.
The long-wavelength (${\bf q}\simeq 0$) part of the Coulomb interaction (counterpart of Eq.~(\ref{g20})) is now given by 
\begin{equation}
\label{eq:g2_nc}
g_{2}=\frac{2\pi l_{B}^{2}}{1+2\sqrt{1+\frac{2}{k_F^2 l_B^2}}},
\end{equation}
which incorporates the static screening from electrons in the $|n|=0$ and $|n|=1$ levels, at zero temperature (we assume $\Lambda_1\gg k_B T$).

Likewise, the short-wavelength ($q\simeq q_1 =2 k_F$) part corresponding to interbranch momentum transfer 
within a node reads
\begin{equation}
g_{1} =\frac{e^{2}\lambda^{2} l_{B}^{2}}{\pi \hbar v_{F}\epsilon_{\infty}}\ln\left(1+\frac{1}{4k_{F}^{2}l_{B}^{2}}\right),
\end{equation}
where we have averaged over $q_\perp$ like in Eq. (\ref{g10}), assuming that $k_F l_B$ is not small. 
Note that this expression is identical to that of the previous section, except for the $\lambda^2$ factor coming from the momentum-space pseudospin texture of the nonchiral Landau levels.
Then, 
\begin{align}
\frac{g_1}{g_2} &= \frac{e^2}{\pi^2 \hbar v_F \epsilon_\infty} \lambda' \ln\left(1+\frac{1}{4 k_F^2 l_B^2}\right)\nonumber\\
\lambda' &\equiv \frac{1}{2 + k_F^2 l_B^2} \left(1+ 2 \sqrt{1+ \frac{2}{k_F^2 l_B^2}}\right).
\end{align}
Once again, compared to the discussion of Sec. \ref{sec:cll} on chiral Landau levels, a dimensionless factor $\lambda'$ appears in the ratio $g_1/g_2$ for nonchiral Landau levels. 

\subsection{RG equations and phonon spectrum}

Like in Sec.~\ref{sec:cll}, the outer shell corrections entering the RG recursive relations (Eqs.~(\ref{Sp0}), (\ref{zeph0}) and (\ref{g1g2})) are all evaluated in the limit of small $q_\perp $ up to the cutoff $1/l_B$, above which all the couplings are 
exponentially suppressed. Within the magnetic  length scale $l_B$, the couplings can  then be considered as essentially local in the transverse directions, which yield one-dimensional-like loop corrections for  the flow equations.   
From the results of Appendix B, the RG transformations for the electron-phonon vertices $z_j$ of Fig.~\ref{fig_RG1} (b), corresponding to scalar ($j=s$) and pseudoscalar ($j={ps}$) modes, are given by 
\begin{align}
\label{eq:rg_ep_nc}
\frac{d}{dl} \ln z_j &=\frac{\gamma_j(l)}{2}\lambda_P,
\end{align}
where 
\begin{align}
\gamma_s(l) &= \alpha_1(g_{2}(l)-2g_{1}(l))\nonumber\\
\gamma_{ps}(l) &=\alpha_1 g_{2}(l)
\end{align}
and $\alpha_{1}=1/(2\pi l_{B}^{2})$. Here the cutoff function $\lambda_P$ of  the Peierls loop for the non chiral Landau is given by Eq.~(\ref{eq:ipnc}) of Appendix B. 
Remarkably, the RG flow and the corresponding phonon softening for the {\em pseudoscalar} phonons is independent of the coupling $g_{1}$. As explained in Appendix B, this is only true for the case where the two Weyl nodes are related by mirror symmetry.
The physical meaning behind the difference in scalar and pseudoscalar phonon modes originates from the fact that the latter do not create any net charge fluctuations and hence they are not screened by $g_1$. 
In other words, $g_1$ couples to total charge fluctuations, which are absent in the case of pseudoscalar phonons. 
In contrast, $g_2$ (which is associated to vertex corrections) couples to all phonons.

As for the electron-electron interactions, the results of Appendix B lead to the following RG equations:
\begin{align}
\label{eq:rg_ee_nc}
\frac{dg_{1}}{dl}&=(-g_{1}^{2}+g_{1}g_{2})\alpha_{1}\lambda_P\nonumber\\
\frac{dg_{2}}{dl}&=\frac{g_{2}^{2}}{2}\alpha_{1}\lambda_P.
\end{align}
Unlike in Sec. \ref{sec:cll}, the one-loop corrections to both the couplings $g_{1}$ and $g_{2}$ arise from the Peierls channel alone. The contribution from the Cooper channel, neglected here, is no longer logarithmically divergent  because the two electronic branches connected by $q_1$ are not symmetric with respect to center of the Brillouin zone.
The last diagram of Fig.~\ref{fig_RG1} (c) is therefore (approximately) absent.

The solutions of Eq.~(\ref{eq:rg_ee_nc}) can be written as 
\begin{align}
\label{eq:ee_nc_sol}
\gamma_{j}(l) &=\frac{\gamma_{j}}{1-\frac{\gamma_{j}}{2}\int_0^l\lambda_P dl'},
\end{align}
where $\gamma_j$ is the bare ($l=0$) value of $\gamma_j(l)$.
 In the absence of a logarithmically  singular Cooper channel, there is no invariant combination of couplings  in the course of the RG flow. While $\gamma_{ps}(l)$ is marginally relevant,  $\gamma_{s}(l)$ may be marginally irrelevant or relevant depending upon the relative values of $g_{2}(l)$ and $2g_{1}(l)$. 
 
Using Eq.~(\ref{eq:ee_nc_sol}), Eq.~(\ref{eq:rg_ep_nc}) can be solved to yield
\begin{align}
\label{eq:ep_nc_sol}
z_{j} &=\frac{1}{1-\frac{\gamma_{j}}{2}\int_0^l\lambda_P dl'}.
\end{align}
Thus, the simple poles in the electron-electron couplings translate into corresponding poles in the electron-phonon vertices $z_{s}$ and $z_{ps}$, which essentially demonstrates the coupling of different charge-density wave correlations to the scalar and pseudoscalar phonon modes. 
The charge density wave emerging at the instability of pseudoscalar phonons is special in that the total charge density is uniform, while the {\em difference} in the charge density between the two Weyl nodes of opposite chirality oscillates in space with a period of $\pi/k_F$.

The self-energy corrections for the scalar and pseudoscalar phonon propagators shown in Fig.~\ref{fig_RG1} (a) are given by
\begin{align}
\label{eq:phprop_nc}
\mathscr{D}_l^{-1} = &\, \mathscr{D}_0^{-1} -  \scalebox{1.55}{$\pi$}_l\cr
=  & \, \omega_m^2 + \omega_{0,j}^2(\mathbf{q})\left[1-  {\alpha_1} g_{j}^{\prime2}\chi_j(l)\right],
\end{align}
where $\omega_{0,j}$  is the bare frequency for the scalar {($ j =s$)} and pseudoscalar   {($ j =ps$)} phonons, $g_{j}^{\prime2}= g_{j}^{2}\lambda^2/\omega_{0,j}^2$,  and
\begin{equation}
\label{Ki}
\chi_{j}(l) =  \int_0^l z_{j}^2(l') \lambda_P dl'
\end{equation}
refers to the corresponding electronic susceptibility involved in the phonon softening.
Since we are interested in ${\bf q} = 2 k_F \hat{\bf z}$, we take $\lambda_P \approx 1$ for  $l< l_T= \ln(\Lambda_1/k_B T)$ and $\lambda_P\simeq 0$ for $l>l_T$ in Eq.~(\ref{Ki}). 

Equation (\ref{eq:phprop_nc}) allows us to study the adiabatic phonon softening caused by electrons. 
When $\alpha_1 g_{j}^{\prime2}\chi_{j}(l_T)=1$ at $\omega_m=0$, the system undergoes a Peierls instability. 
The transition temperature for the instability at $2 k_F\hat{\bf z}$ is given by
\begin{align}
\label{eq:Tc_nc}
T_{c,1}^j &=\Lambda_{1}\exp\left(-\frac{2}{2\alpha_{1} g_{j}^{\prime2}+\gamma_{j}}\right) \text{  ($j=s,ps$)},
\end{align}
where the subscript $1$ is to remind that the instability considered here emerges from within the $|n|=1$ nonchiral Landau levels and their coupling to phonons.
The Peierls transition temperature is of the BCS (exponential) form, in agreement to the phenomenological expression used in Ref. [\onlinecite{Liu2016}], although renormalized by electron-electron interactions.
This is unlike in the chiral Landau levels related by a mirror plane (or by another improper symmetry), where the Cooper scattering channel is present, and consequently, a power law behavior of the transition temperature is realized.

We note that Eq.~(\ref{eq:Tc_nc}) is valid only when $2\alpha_{1} g_{j}^{\prime2}+\gamma_{j}>0$.
If $2\alpha_{1} g_{j}^{\prime2}+\gamma_{j}<0$, $T_{c,1}^j=0$ and there is no Peierls instability.
For scalar phonons, $\gamma_s$ becomes negative at high magnetic fields, which leads to $2\alpha_{1} g_{s}^{\prime2}+\gamma_{s}<0$.
In contrast, for pseudoscalar phonons, $\gamma_{ps}$ is positive regardless of the magnetic field (because $g_2>0$) and therefore $2\alpha_{1} g_{ps}^{\prime2}+\gamma_{ps}$ is always positive.
Thus, pseudoscalar phonons remain susceptible to a Peierls instability even when the scalar phonons are not.

The qualitative difference in the softening between scalar and pseudoscalar phonons was first noticed in Ref.~[\onlinecite{rinkel2019}], in the context of the random phase approximation.
This approximation amounts to neglecting the bare value of $g_2$ in Eqs. (\ref{eq:rg_ep_nc}) and (\ref{eq:rg_ee_nc}).
Under such condition, our theory would predict 
\begin{align}
\label{eq:chi_RPA_nc}
\chi_s(l) &= \frac{\int_0^l \lambda_P dl'}{1+g_1 \alpha_1 \int_0^l \lambda_P dl'}\nonumber\\
\chi_{ps}(l) &= \int_0^l \lambda_P dl'.
\end{align}
Hence, the Peierls instability (the divergence of $\chi(l)$ at $l\to\infty$, which takes place when $\lambda_P\simeq 1$) would be suppressed for scalar phonons due to RPA screening, but not for pseudoscalar phonons. 
According to  Eqs.~(\ref{eq:chi_RPA_nc}) and (\ref{eq:phprop_nc}), pseudoscalar phonons would undergo a Peierls instability for an infinitesimally weak electron-phonon coupling, while it would take a minimum strength of electron-phonon coupling to have a Peierls instability of scalar phonons. 
This statement is in agreement with the results of Ref.~[\onlinecite{rinkel2019}].
However, our more complete theory, which goes beyond RPA, adds nuance to such results.
For instance, the $g_2$ contribution ignored in the RPA helps the Peierls instability of scalar phonons, which can now take place for infinitesimally weak electron-phonon interactions, provided that $g_2> 2 g_1$ (as shown in Eq.~(\ref{eq:Tc_nc})).

In Eq.~(\ref{eq:Tc_nc}), the cutoff energy scale $\Lambda_{1}$ decreases as the strength of the applied magnetic field increases (unlike $\Lambda_0$ in the preceding section), ultimately vanishing at the threshold of the quantum limit.
The exact functional dependence can be calculated analytically.
When $\Lambda_{1}$ is vanishingly small, there is no Peierls transition in the nonchiral Landau level. 

\begin{figure}[h]
  \begin{center}
    \includegraphics[width=1.0\columnwidth]{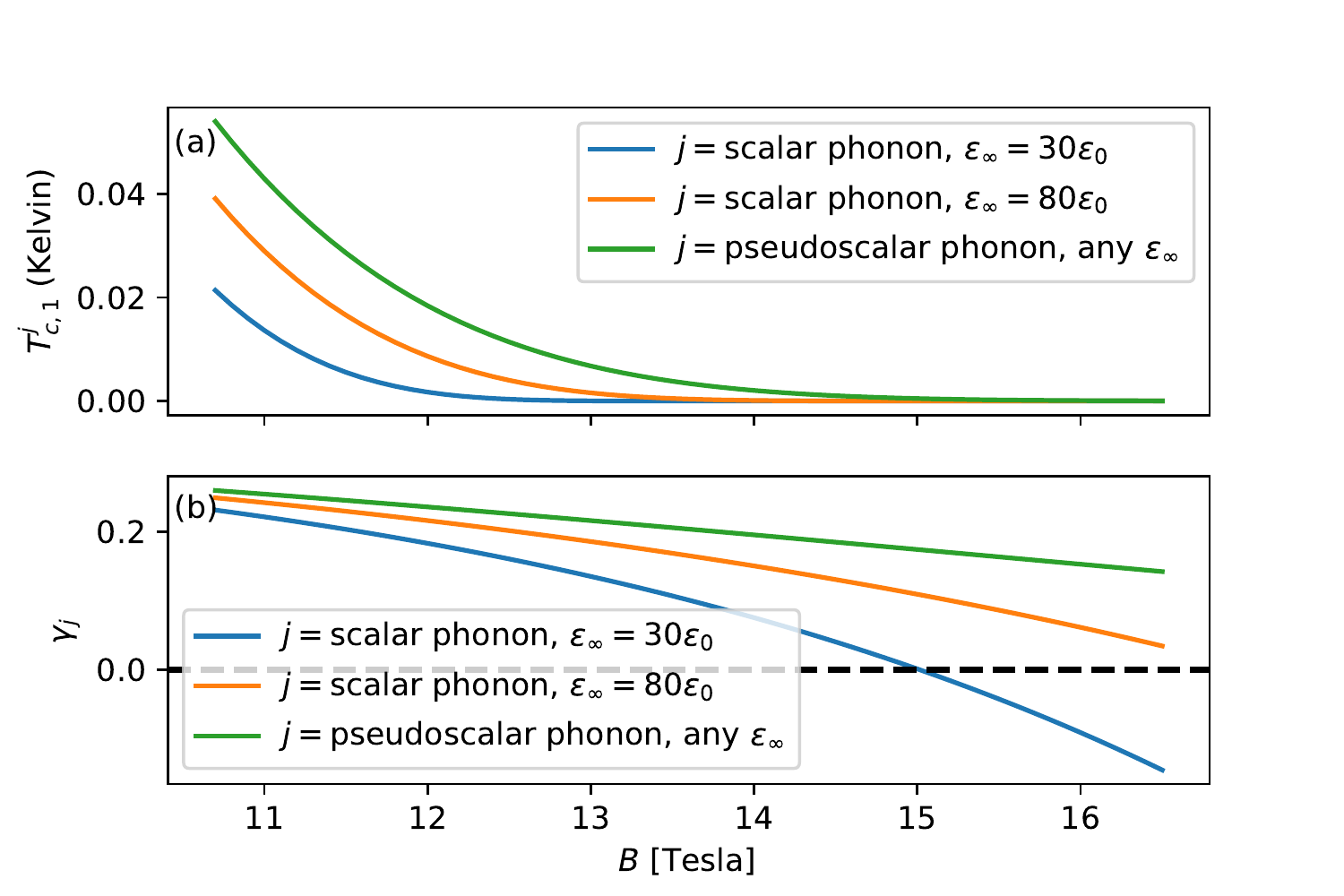}
\caption{(a) Peierls transition temperature for scalar and pseudoscalar phonons, at the nesting wave vector ${\bf q}_1$ (see Fig.~\ref{fig:nonc}).
The parameter values are the same as in Fig.~\ref{fig:Tc}.
Only the magnetic field range in which the Fermi level intersects the $n=0$ and $|n|=1$ Landau levels is shown. Also, we limit the data to $q_1 l_B > 1$.
For a given strength of electron-phonon interactions, the transition temperature for pseudoscalar phonons is enhanced with respect to that of scalar phonons; the two converge to one another when $\epsilon_\infty$ is large.
(b) The bare value of $\gamma_j$ as a function of the magnetic field. For the scalar phonon, $\gamma_s=g_2 - 2g_1$ changes sign at sufficiently high magnetic fields, eventually eliminating the Peierls instability. For the pseudoscalar phonon, $\gamma_{ps}= g_2$ remains always positive. We note that the numerical values for $\gamma_j$ are much smaller than in the quantum limit (Fig.~\ref{fig:Tc}) because $g_2$ is relatively reduced by the additional screening coming from the $|n|=1$ Landau levels. As a a result, the critical temperatures in this figure are much lower than the ones expected in Fig.~\ref{fig:Tc}.}
 \label{fig:Tc_nonchiral}
  \end{center}
\end{figure} 

Figure \ref{fig:Tc_nonchiral} provides a numerical estimate of $T_{c, 1}^s$ and $T_{c,1}^{ps}$.
For equal magnitude of electron-phonon coupling, the transition temperature for pseudoscalar phonons is higher than that of scalar phonons, because $\gamma_{ps}>\gamma_s$.
This difference can be very significant when $g_1/g_2$ is not small, i.e. at stronger magnetic fields, for reasons mentioned above.
Therefore, we conclude that pseudoscalar phonons are more prone than scalar phonons to Peierls instabilities.

For the same parameter values as in Sec.~\ref{sec:cll}, the transition temperature for the Peierls instability at wave vector ${\bf q}_1$ is very low  (sub Kelvin).
This is so for three reasons: (i) when the Fermi level intersects only $n=0$ and $|n|=1$ Landau levels, the cutoff $\Lambda_1$ can be no larger than $(2-\sqrt{2}) \hbar v/l_B$; (ii) $g_2$ is smaller than in the quantum limit, due to the additional screening coming from the partially occupied $|n|=1$ Landau levels (compare Eqs.~(\ref{eq:g2_nc}) and (\ref{g20})); 
(iii) as the magnetic field is made stronger, $\Lambda_1$ and $g_2$ both decrease, while $g_1/g_2$ increases.  

In view of the preceding paragraph, a Peierls transition that originates exclusively from $|n|=1$ Landau levels at wave vector $q_1$ appears difficult to observe.
For a given Fermi energy, the instability will be often dominated by the nesting wave vector connecting nodes of opposite chirality, mainly because 
$g_1$ is smaller for the larger (internode) nesting wave vectors.
An exception may take place if $g_1>g_2$ for both $n=0$ and $|n|=1$ bands.  In that case, for a weak enough electron-phonon interaction strength, there would be no instability originating from the chiral Landau levels (recall Eq.~(\ref{Tc0})). There could be, however, an instability originating from the $|n|=1$ levels for (and only for) pseudoscalar phonons, although at very low temperature.

We conclude this section by discussing the cases of broken mirror  symmetry and nonadiabaticity.
Let us suppose that the mirror symmetry is broken (which can be done by rotating the direction of the magnetic field).
In this case, the wave vector ${\bf q}_1$ is different (by $\delta k_F$) in the two nodes of opposite chirality. 
If the energy scale $\hbar v_F \delta k_F$ is small  compared to the critical temperature obtained from Eq.~(\ref{eq:Tc_nc}), then the effect of the mirror-breaking perturbation on the Peierls instability is unimportant.
If, to the contrary,  $\hbar v_F \delta k_F$ is large, only one node will contribute to the Peierls instability. 
In this case, the Peierls transition temperature will still be of exponential form, like in Eq.~(\ref{eq:Tc_nc}).
However, there is no longer a qualitative distinction between the softening of scalar and pseudoscalar phonons. 
In fact, a single phonon mode will couple to electrons via a combination of scalar and pseudoscalar coupling. 
The corresponding transition temperature is given by
\begin{align}
\label{eq:Tc_nc_bm}
T_{c} &=\Lambda_{1}\exp\left(-\frac{2}{\alpha_{1} g_{1}^{\prime2}+\gamma_{1}}\right), 
\end{align}
where $\gamma_{1}=\alpha_{1}(g_{2}-g_{1})$, and ($g_{2}$, $g_{1}$, $g_{1}^{\prime}$) refer to the electron-electron and electron-phonon couplings defined for a single nonchiral Landau level.
While Eq.~(\ref{eq:Tc_nc_bm})  has the same form as Eq.~(\ref{Tc0Delta2}), it is quantitatively much lower, mainly because $g_2$ is suppressed when the $|n|=1$ bands are populated and hence contribute to screening. 

Next, we briefly discuss the nonadiabatic regime of instabilities for nonchiral Landau levels. In the two-cutoff scaling scheme, when for some given magnetic field interval, $\Lambda_1(l) $ is scaled down to $\Lambda_1(l^*)=\hbar\omega_{0,j}< \Lambda_1$, the adiabatic softening condition (Eq.~(\ref{eq:phprop_nc})) no longer holds and  must be  replaced by the flow equations shown in Eq. (\ref{eq:rg_ee_nc}). In these, the conditions at $l^*$ for  non retarded couplings  become $g_{1,2}^* = g_{1,2}(l^*) + \mathfrak{g}_{1,2} $, where $ \mathfrak{g}_{1,2} $ are the attractive phonon-induced backward and forward scattering amplitudes. These can be obtained from Eqs.~(\ref{g1ph}-\ref{g2ph}) after the substitution of the appropriate electron-phonon coupling matrix element (Eqs.~(\ref{geph}) and (\ref{eq:gLR})), of the renormalized phonon propagator (Eq.~(\ref{eq:phprop_nc})) and of vertex parts $z_{j}(l^*)$ (Eq.~(\ref{eq:ep_nc_sol}))  at $l^*$. Note that when the applied magnetic field is sufficiently large and $\Lambda_1 < \hbar \omega_{0,j}$, the electron-phonon system is entirely nonadiabatic so that $l^*=0$.  

It follows from Eqs.~(\ref{eq:rg_ee_nc}) and (\ref{eq:rg_ep_nc}) that, for $\lambda_P\approx 1$, the combinations of couplings 
\begin{align}
\gamma_j(l)  = {\gamma_j^*\over 1 -{{\gamma}^*_j \over 2}(l-l^*)}
\end{align}
and the electron-phonon vertex parts  
\begin{align}
z_j(l)  = {z_j(l^*)\over 1 -{{\gamma}^*_j \over 2}(l-l^*)}
\end{align}
are both developing a simple pole structure. Here $ \gamma_{ps}^*=\alpha_{1} g_2^*$ and $\gamma_{s}^*= \alpha_{1} (g_2^* -2g_1^*)$, whereas the expressions for $z_j(l^*)$ are given by Eq.~(\ref{eq:ep_nc_sol}) at $l^*$. As a function of temperature, we obtain the characteristic scale
\begin{equation}
\label{Tc1nonadia}
T_{1,j}^{*} = {\Lambda_{1,j}\over k_B} e^{-2/\gamma_j^*}
\end{equation} 
for an instability of the electron system against a $2k_F\hat{\bf z}$ charge density wave, which is  distinct for the scalar  ($j=s$) and pseudoscalar ($j=ps$)   cases. Here the cutoff energy is $\Lambda_{1,j}= \ {\rm min}\{\hbar\omega_{0,j},\Lambda_1\}$ where $\Lambda_1$ is field-dependent. Therefore at  a given applied magnetic field nonadiabatic corrections, which  predominantly screen $g_1$ to the attractive sector, will enhance (reduce) the critical temperature (Eq. (\ref{eq:Tc_nc})) for scalar (pseudoscalar) phonons.  
The corresponding transition temperature for broken mirror symmetry between the two Weyl nodes, where we effectively consider a single nonchiral Landau level for our analysis, is given by \begin{equation}
\label{Tc1nonadia_bm}
T_{1}^{*} = \frac{\Lambda_{1}}{k_B} e^{-2/\gamma_{1}^*},
\end{equation} 
where $\gamma_{1}^*=\alpha_{1} (g_2^* -g_1^*)$.



\section{Application to real materials}
\label{sec:real}

Thus far, we have considered a toy model for two Weyl nodes of opposite chirality, with untilted Weyl cones. 
In this section, we extrapolate our results to more realistic situations. 
This extrapolation is nontrivial, due to at least three aspects. 

 First, in the absence of magnetic fields, the Fermi surfaces of several real WSM are non spherical around the Weyl nodes, and moreover host topologically trivial hole bands away from the nodes. This goes well beyond the simple model that we have considered at zero magnetic field. Fortunately, in the strong magnetic field regime that we are interested in, the topologically trivial hole pockets are pushed away from the Fermi energy and the bands at the Fermi level become simpler and rather universal: one-dimensional (though highly degenerate) chiral Landau and nonchiral Landau levels. 
Admittedly, our model neglects the curvature of the chiral Landau level. Nevertheless, the main effect of such curvature in our theory (namely a magnetic-field-dependence of the Fermi velocity in the chiral LL) is qualitatively unimportant.

Second, one common element of real Weyl semimetals, which we have ignored throughout the paper, is the tilt of the Weyl cones. It turns out that such a tilt does not change the main results of our theory because the Peierls scattering amplitude (and also the Cooper scattering amplitude, in the presence of mirror symmetry) retains its logarithmic divergence; this can be checked by an explicit calculation. 

Third, real WSM most often host multiple pairs of Weyl nodes. This has various implications for our theory, which are worth elaborating on for the remainder of this section.
For example, TaAs and related materials  (NbAs, TaP, NbP) display two symmetry-inequivalent multiplets of Weyl nodes (8 W1 nodes and 16 W2 nodes).
The eight symmetry-equivalent W1 nodes are located on the $k_z=0$ plane of the Brillouin zone (see Fig. \ref{fig:taas}). 
The sixteen symmetry-equivalent W2 nodes are distributed into two planes  ($k_z=\pm k_0$), which are related to one another by time-reversal.
Because the Fermi energy of the W1 nodes is significantly greater than that of the W2 nodes, the quantum limit is first attained in the latter.

\begin{figure}[h!]
\centering
\includegraphics[width=0.3\textwidth]{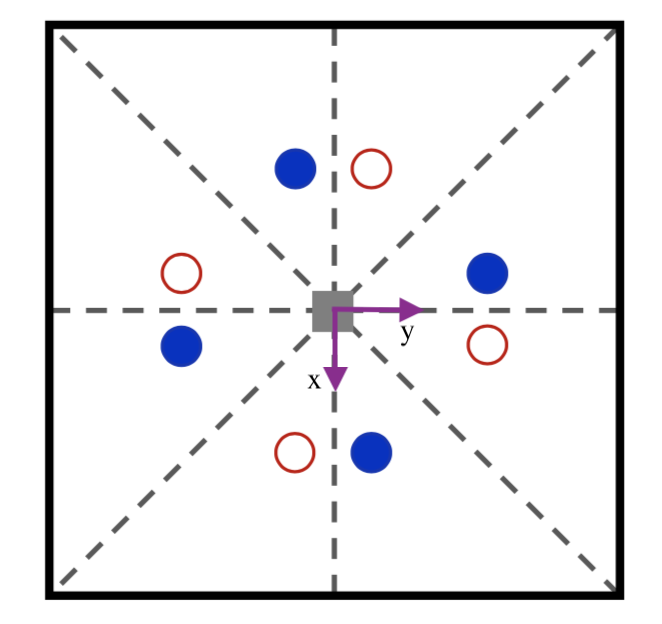}
\caption{$k_z=0$ plane of the Brillouin zone in TaAs. Eight Weyl nodes are shown (blue and red correspond to opposite chiralities). The grey dashed lines denote the mirror planes.}
\label{fig:taas}
\end{figure}

In the absence of a magnetic field, TaAs belongs to point group $C_{4 v}$.
Weyl nodes of opposite chirality in Fig. \ref{fig:taas} are related to one another by a mirror plane.
At zero field, none of the long-wavelength phonons in TaAs are pseudoscalar\cite{Aroyo2006a, Aroyo2006b}.

Let us suppose that the magnetic field is applied along $z$. Then, all mirror planes are broken but the rotational symmetry is preserved. 
The eight W1 nodes of Fig. \ref{fig:taas} can then be subdivided into four symmetry-related nodes of positive chirality, and another four symmetry-equivalent nodes of negative chirality.
The sixteen W2 nodes can be similarly subdivided. 

In the quantum limit, the possible phonon instabilities occur for wave vectors connecting Weyl nodes of opposite chirality.
Indeed, since chiral Landau levels of the same chirality have the same dispersion along the direction of $B$, it follows that the nesting condition necessary for the logarithmic singularity of the Peierls and Cooper channels is not satisfied for phonon wave vectors connecting two nodes of the same chirality.
For a wave vector in the $xy$ plane connecting two nodes of opposite chirality, two pairs of chiral Landau levels (in the case of W1 nodes) or four pairs of chiral Landau levels (in the case of W2 nodes)  contribute to the instability. For the remaining pairs, the phonon wave vector does not connect the two nodes of opposite chirality. 

If we neglect the Coulomb scattering between different pairs of nodes, then our theory of Sec. \ref{sec:cll_no_mirror} for the Peierls transition temperature can be applied, with some modifications, to the case of multiple pairs of nodes.
Specifically, we note the following points: (i)  since ${\bf B}||\hat{\bf z}$ breaks all mirrors, the Peierls transition temperature will be of the BCS form;
(ii) the coupling $g_2$ will be reduced by a factor of $N$, where $N$ is the number of pairs of Weyl nodes contributing to the screening of the Coulomb interaction;
(iii) the effective electron-phonon parameter $g_{x,j}^2$ will be enhanced by a factor $N'$, where $N'$ is the number of pairs of Weyl nodes that are connected by the same nesting wave vector.
While (ii) leads to a depletion of the transition temperature, (iii) implies an increase. Hence, the transition temperature for multiple pairs of Weyl nodes can be either higher or lower than the transition temperature for a single pair of Weyl nodes, depending on whether the dominant contribution comes from electron-electron or electron-phonon interactions.
It must also be mentioned that, in the case of multiple pairs of nodes, there could be additional types of instabilities that are not present in the single-pair model. 

One aspect of having the magnetic field along the $z$ axis is that the field is perpendicular to the separation between all the Weyl nodes in a constant $k_z$ plane.
Therefore, significant single-particle hybridization gaps between chiral Landau levels of opposite chirality are expected at high fields. 
When the size of these gaps exceeds the Peierls transition temperature computed for gapless chiral Landau levels, we expect that the instability will be suppressed.


Next, let us suppose that the magnetic field is applied along the $x$ direction and let us consider a nesting wave vector ${\bf q}$ along $x$ as well.
The single-particle hybridization between chiral Landau levels is relatively unimportant in this configuration, because the magnetic field is parallel to the vector separating the two Weyl nodes that contribute to the Peierls instability.
Such field reduces the symmetry of the crystal from $C_{4v}$ to $C_{1h}$.
As a result,\cite{dresselhaus2008} the $E$ phonons of TaAs, doubly degenerate in the absence of a magnetic field, split into nondegenerate scalar ($A'$) and pseudoscalar ($A''$) phonons.
According to Eq.~(\ref{eq:Tc_nc}), we would expect to see a more significant softening for the $A''$ phonon than for the $A'$ phonon, in the regime in which both $n=0$ and $|n|=1$ Landau levels intersect the Fermi energy.
One caveat is that the hybridization between $A'$ and $A''$ phonons, neglected in our theory, may be significant due to the near degeneracy of these modes.

In the quantum limit with ${\bf B}=B\hat{\bf x}$, the results from Sec.~\ref{sec:cll} indicate that the $B$-dependence of the Peierls transition temperature would be qualitatively different for ${\bf q}||\hat{\bf x}$ and ${\bf q}||\hat{\bf y}$ phonons (compare e.g. Eqs.~(\ref{Tc0}) and ~(\ref{Tc0Delta})), because the former phonons would couple two Weyl nodes related by a mirror plane and the latter phonons would couple Weyl nodes not related by a symmetry.

 To close this section, we note that our theory is also applicable to a Dirac semimetal, under the condition that the Zeeman splitting can transform it into a Weyl semimetal.

\section{Summary and conclusions}
\label{sec:conc}


\begin{table*}
\label{table}
\begin{centering}
\begin{tabular}{|c||c|c|c|c|}
\hline 
Magnetic field regime & Phonon regime & Mirror symmetry & Field-dependence of transition temperature\tabularnewline
\hline 
\hline 
\multirow{4}{*}{Quantum limit} & \multirow{2}{*}{Adiabatic} & Yes & Power-law [Eq. (\ref{Tc0})].\tabularnewline
\cline{3-4} \cline{4-4} 
 &  & No & Exponential [Eq. (\ref{Tc0Delta})]. \tabularnewline
\cline{2-4} \cline{3-4} \cline{4-4} 
  & \multirow{3}{*}{Nonadiabatic} & Yes & No instability (Luttinger liquid).\tabularnewline
\cline{3-4} \cline{4-4} 
&  & \multirow{2}{*}{No} & \multirow{2}{*}{Exponential [Eq. (\ref{Tc0nonadia}) for $\Delta < \hbar\omega_0$ and Eq. (\ref{Tc0nonadiadelta}) for $\Delta>\hbar\omega_0$].
}\tabularnewline
  &  &  & \tabularnewline
\hline 
 \multirow{4}{*}{Near-quantum limit} & \multirow{2}{*}{Adiabatic} & Yes & Exponential [Eq. (\ref{eq:Tc_nc})]. Pseudoscalar $>$ scalar.\tabularnewline
\cline{3-4} \cline{4-4} 
  &  & No & Exponential [Eq. (\ref{eq:Tc_nc_bm})].\tabularnewline
\cline{2-4} \cline{3-4} \cline{4-4} 
  & \multirow{2}{*}{Nonadiabatic} & Yes & Exponential [Eq. (\ref{Tc1nonadia})]. Pseudoscalar $>$ scalar.\tabularnewline
\cline{3-4} \cline{4-4} 
 &  & No & Exponential [Eq. (\ref{Tc1nonadia_bm})].\tabularnewline
\hline 
\end{tabular}
\par\end{centering}
\caption{Summary of the main results. {\em First column}: in the quantum limit, only the chiral Landau levels intersect the Fermi energy. In the "near-quantum" limit, the chiral and the first nonchiral  Landau levels intersect the Fermi energy. {\em Second column}: the adiabatic (nonadiabatic) phonon regime is realized when $k_B T_c \gg \hbar \omega_0$ ($k_B T_c \ll \hbar\omega_0$), where $T_c$ is  the critical temperature for the lattice instability and $\omega_0$ is the bare phonon frequency. 
 {\em Third column}: when present, mirror symmetry relates the two nodes of opposite chirality. Mirror symmetry may be broken intrinsically (due to lattice structure) or extrinsically (due to the application of a magnetic field that is not perpendicular to the preexisting mirror plane). The mirror symmetry-breaking  energy scale $\Delta$ is defined in Eq. (\ref{eq:Delta_main}). 
 {\em Fourth column}: the magnetic-field-dependence of $T_c$ for the lattice instability can be exponential or power-law. For the near-quantum regime, we consider instabilities at nesting wave vectors that connect electronic states of the same chirality. The power-law behavior in $T_c$ originates from the interference between the Peierls and Cooper channels. In the near-quantum regime with mirror symmetry, "pseudoscalar $>$ scalar" stands for the fact that the instability is qualitatively more likely to occur for pseudoscalar phonons than for scalar phonons, because the former  do not generate net charge fluctuations and are thus not screened by Coulomb interactions.} 
\label{table}
\end{table*}


To summarize, we have developed a theory of phonon instabilities for a simple model of Weyl semimetal
with two Weyl nodes of opposite chiralities, in the presence of a large applied
magnetic field and electron-electron interactions, 
using a Kadanoff-Wilson renormalization group (RG) approach.
In this section, we begin by reviewing the main results collected in Table \ref{table}.

We have focused on two different parameter regimes for the applied
magnetic field $B$: the quantum limit, where the Fermi energy intersects
only the chiral ($n=0$) Landau levels, and the "near-quantum" limit, 
where the Fermi level also crosses one nonchiral ($n=1$) Landau level on each node. 
We have assumed that the hybridization gap coming from the single-particle magnetic tunneling is smaller than the Fermi energy, so that the system is a metal in the noninteracting limit.

In the quantum limit, the instabilities take place for nesting wave vectors connecting $n=0$ Landau levels of opposite chirality.
We have first considered the situation where the two
Weyl nodes of opposite chirality are related by a mirror plane, and
the field is oriented perpendicular to this plane. In this case, the Peierls
transition temperature $T_c$ obeys a power-law dependence on $B$ (Eq. (\ref{Tc0})), which  
originates from the destructive interference between the Peierls and Cooper scattering channels. 
The Cooper channel is important even though the two Weyl nodes are not time-reversed partners of one another, provided that the two nodes have opposite momenta along the direction of the magnetic field. 
The Peierls transition in this situation depends crucially on the electron-phonon coupling, that is to say we find no instabilities in the absence of electron-phonon interactions. 
Moreover, the dependence of $T_c$ on $B$ can be nonmonotonic (Fig.~\ref{fig:Tc}).

Again in the quantum limit, when the two Weyl nodes of opposite chirality
are not related by symmetry, the 
Cooper channel is suppressed due to the asymmetry in the positions of the Weyl nodes in momentum space. 
On the other hand, the Peierls channel remains unaffected because of the perfect nesting.
This opens the possibility of having a purely electronic instability.
However, in the presence of a nonzero electron-phonon coupling, this instability is preempted by an instability of the lattice, with a BCS-like expression for $T_c$
(Eq. (\ref{Tc0Delta})).
In a material that is mirror-symmetric in the absence of a magnetic field, such symmetry can be broken in a controllable fashion by rotating the magnetic field away from the normal to the mirror plane. 
One can therefore gradually turn from Eq. (\ref{Tc0}) to Eq. (\ref{Tc0Delta}); the crossover takes place when the energy scale associated to mirror symmetry-breaking exceeds the $T_c$ in Eq. (\ref{Tc0}).

Still in the quantum limit,  we have investigated the nonadiabatic regime, where the bare phonon energy
exceeds the value of $T_c$ obtained in the adiabatic regime.  
In this regime, if the two Weyl nodes are related by mirror symmetry,
Luttinger liquid behavior is predicted, with no instability at finite
temperature. 
On the other hand, if mirror symmetry is broken,
the aforementioned Luttinger liquid state becomes
unstable against the formation of a charge-density wave order (with $T_c$ given by Eq. (\ref{Tc0nonadia}) or (\ref{Tc0nonadiadelta}), depending on whether the energy scale for mirror symmetry-breaking exceeds or not the bare phonon energy).

Next, we have considered lower magnetic fields. 
In contrast to the quantum limit, here lattice instabilities can also emerge at nesting wave vectors that connect two Fermi points of $n=1$ Landau levels within the same chirality. In the next two paragraphs, we focus on such instabilities emerging from $n=1$ Landau levels.

For the adiabatic phonon regime, we obtain a Peierls instability for $n=1$ Landau levels with a $T_c$ of the BCS form; $T_c$ decreases as $B$ increases (Fig. ~\ref{fig:Tc_nonchiral}).
Here, the Cooper channel is suppressed because the Fermi points connected by the nesting wave vector are not symmetric with respect to the $\Gamma$ point.
When mirror symmetry is present, a qualitative difference appears between the behaviors of pseudoscalar phonons (whose deformation potentials have equal magnitudes and opposite signs on Weyl nodes of opposite chirality) and scalar phonons (whose deformation potentials have equal magnitudes and equal signs on Weyl nodes of opposite chirality). 
Because pseudoscalar phonons do not generate net charge fluctuations, they are less screened than scalar phonons by Coulomb interactions and are accordingly more prone to undergo a Peierls instability.  

In the nonadiabatic regime with mirror symmetry, we obtain a characteristic scale for an instability of the electron system against a charge-density wave order, which is distinct for scalar and pseudoscalar phonons (Eq. (\ref{Tc1nonadia})). 
When the mirror symmetry relating the Weyl nodes is broken, there is no longer a qualitative difference between the softening of scalar and pseudoscalar phonons (Eq.~(\ref{eq:Tc_nc_bm})).

If we extrapolate our results to lower fields, where the Fermi level intersects multiple nonchiral Landau levels, we anticipate that the leading instability will occur at the nesting wave vector connecting chiral Landau levels of opposite chirality. Such instability will however take place at significantly lower temperatures than the corresponding one in the quantum limit, mainly because of the contribution of the nonchiral Landau levels  to the screening of the forward Coulomb scattering.

This completes the summary of our work. 
The main messages are the following: (1) it is important to take electron-phonon interactions into account, along with electronic interaction effects, for a complete understanding of interacting phases of matter of Weyl semimetals at high fields; (2) electron-phonon interactions tend to augment the charge-density wave fluctuations in the system and, whenever a Peierls transition is possible, lead to an enhancement of the corresponding transition temperature; 
(3) the Cooper channel plays an important role when the the two nodes are related by a mirror plane, and can prevent purely electronic instabilities while enabling non-BCS-like lattice instabilities; (4) breaking mirror symmetry increases $T_c$ for the Peierls transition by suppressing the Cooper channel; (5) $T_c$ is expected to be higher for instabilities emerging from chiral (as opposed to nonchiral) Landau levels.



For the remainder of this section, we compare our theory to the existing literature.
First, the importance of electron-phonon interactions in the putative charge-density wave ordered state in the Dirac semimetal ZrTe$_5$ at high magnetic field
has been suggested in a recent theoretical study\cite{Qin2020}. 
The authors of this paper use a mean-field approach to compare the threshold magnetic field for the order parameter, calculated using electron-phonon and electron-electron interactions alone, whereas we perform a detailed analysis of the Peierls transition resulting from the interplay of these two effects, to compare their relative importance for the transition. Moreover, Ref.~[\onlinecite{Qin2020}] neglects the Cooper channel contribution, which, as we have shown, can qualitatively change the behavior of the transition through its interference with the Peierls channel. 

Second, Ref.~[\onlinecite{Zhao2020}] has developed an RG  theory for a metal-insulator transition in the quantum limit of a Dirac material, treating electron-electron interactions, electron-phonon interactions and disorder on an equal footing. The authors of Ref.~[\onlinecite{Zhao2020}] do not consider forward scattering and vertex corrections, and do not take into account the Cooper channel contributions, all of which play a role in our final expression for the Peierls transition temperature. Moreover, Ref.~[\onlinecite{Zhao2020}]  considers phonon-induced electronic interactions as the starting point, which would correspond to the nonadiabatic regime in our analysis. We, in turn, do not consider the effect of disorder. These differences make it difficult to directly compare the results of Ref.~[\onlinecite{Zhao2020}] with ours.


Third, the lattice instabilities studied in our work differ in a number of ways from the ones that are well-known in quasi one-dimensional conductors\cite{Pouget2015}. 
The nonadiabatic regime is less likely to be relevant  in the latter, because they are soft materials with low-frequency phonons and a relatively high value for the tight-binding electron-phonon matrix element. Also, the cutoff band energy entering the Peierls transition temperature is significantly larger compared to the values found for WSM. 
Moreover, in quasi-1D conductors, the low-energy electron bands are symmetric with respect to the $\Gamma$ point. This is not the case for Weyl fermion bands in Weyl semimetals, which can affect the importance of the Cooper channel depending on the orientation of the magnetic field.
Finally, there is no spin degeneracy in WSM, contrary to quasi-1D conductors. In these, the possibility of a density-wave instability for electrons induced by an applied magnetic field 
is known to exist, but is restricted to the spin sector in the form of a field-induced spin-density wave state\cite{Lebed08}. 


Fourth, we compare our work to the literature on magnetic catalysis, which has been long-studied in high-energy physics and has recently received attention in the context of Weyl and Dirac semimetals\cite{Miransky2015}.
Magnetic catalysis consists of a dynamical generation of a Dirac mass in an interacting system of initially massless Dirac (or Weyl) fermions subjected to a strong magnetic field.
Our work can be regarded as a theory of magnetic catalysis in two-node Weyl semimetals, in the presence of both electron-electron and electron-phonon interactions. 

If we turn off electron-phonon interactions in our theory, we can compare our results to earlier works in the literature of magnetic catalysis. While Ref. [\onlinecite{Miransky2015}] reviews mainly the mean-field approach, consistent results have been reported in RG approaches that are closer in spirit to our work (see e.g. Refs. [\onlinecite{Hong1996}] and [\onlinecite{Hattori2017}]). 
In the latter RG approaches, an effective electron-electron interaction parameter $g$ flows to strong coupling provided that the bare value of $g$ is positive, thereby producing the dynamical generation of a Dirac mass. The dynamical mass is BCS-like. The flow equation for $g$ in those theories coincides with our flow equation (\ref{g2mg1}) for $g_2-g_1$ (i.e. difference between forward and backward Coulomb scattering), but only provided that the Cooper channel contribution is negligible.  

While the neglect of the Cooper channel may be justified in models with multiple fermion flavors, 
we have shown that the Cooper channel contribution is important in the quantum limit of two-node Weyl semimetals, if the nodes are related by a crystal symmetry. In that situation, the Cooper channel interferes destructively with the Peierls channel, thereby leading to an effective electron-electron coupling $g_2-g_1$ that no longer flows to strong coupling. In other words, the Cooper channel contribution precludes the emergence of magnetic catalysis in the absence of electron-phonon interactions. 
When electron-phonon interactions are included, we find the magnetic catalysis re-emerges in the form of a Peierls instability, but with a critical temperature that is no longer BCS-like. 

In our theory, the contribution from the Cooper channel becomes suppressed when the two Weyl nodes are not related by a crystal symmetry. In this case, only the Peierls channel remains intact and $g_2-g_1$ flows to strong coupling provided that the bare value of $g_2-g_1$ is positive, like predicted in the RG treatments of magnetic catalysis. Yet, even in this case, our theory offers a new result: the purely electronic instability resulting from the divergence of $g_2-g_1$ is preempted by a lattice instability, because the electron-phonon vertex diverges before $g_2-g_1$ does.


To conclude, we mention that possible avenues for future work include a generalization of our theory to semimetals with multiple pairs of nodes, and its application to study electronic and lattice instabilities in other interesting semimetals such as graphite, bismuth and chiral multifold fermion systems.

\acknowledgements
This paper has been financially supported by the Canada First Research Excellence Fund (CFREF) and by the Natural Sciences and Engineering Research Council of Canada (NSERC). C.B. thanks NSERC under Grant No. RGPIN- 2016-06017. I.G. thanks NSERC under Grant No. RGPIN- 2018-05385. S.K. acknowledges financial support from the Postdoctoral Fellowship from Institut Quantique, and from UF Project No. P0224175 - Dirac postdoc fellowship, sponsored by the Florida State University National High Magnetic Field Laboratory (NHMFL). I.G. thanks R. C\^ot\'e for a useful discussion. S.K. gratefully acknowledges useful discussions with Vikram Tripathi.


\begin{thebibliography}{88}%
\makeatletter
\providecommand \@ifxundefined [1]{%
 \@ifx{#1\undefined}
}%
\providecommand \@ifnum [1]{%
 \ifnum #1\expandafter \@firstoftwo
 \else \expandafter \@secondoftwo
 \fi
}%
\providecommand \@ifx [1]{%
 \ifx #1\expandafter \@firstoftwo
 \else \expandafter \@secondoftwo
 \fi
}%
\providecommand \natexlab [1]{#1}%
\providecommand \enquote  [1]{``#1''}%
\providecommand \bibnamefont  [1]{#1}%
\providecommand \bibfnamefont [1]{#1}%
\providecommand \citenamefont [1]{#1}%
\providecommand \href@noop [0]{\@secondoftwo}%
\providecommand \href [0]{\begingroup \@sanitize@url \@href}%
\providecommand \@href[1]{\@@startlink{#1}\@@href}%
\providecommand \@@href[1]{\endgroup#1\@@endlink}%
\providecommand \@sanitize@url [0]{\catcode `\\12\catcode `\$12\catcode
  `\&12\catcode `\#12\catcode `\^12\catcode `\_12\catcode `\%12\relax}%
\providecommand \@@startlink[1]{}%
\providecommand \@@endlink[0]{}%
\providecommand \url  [0]{\begingroup\@sanitize@url \@url }%
\providecommand \@url [1]{\endgroup\@href {#1}{\urlprefix }}%
\providecommand \urlprefix  [0]{URL }%
\providecommand \Eprint [0]{\href }%
\providecommand \doibase [0]{http://dx.doi.org/}%
\providecommand \selectlanguage [0]{\@gobble}%
\providecommand \bibinfo  [0]{\@secondoftwo}%
\providecommand \bibfield  [0]{\@secondoftwo}%
\providecommand \translation [1]{[#1]}%
\providecommand \BibitemOpen [0]{}%
\providecommand \bibitemStop [0]{}%
\providecommand \bibitemNoStop [0]{.\EOS\space}%
\providecommand \EOS [0]{\spacefactor3000\relax}%
\providecommand \BibitemShut  [1]{\csname bibitem#1\endcsname}%
\let\auto@bib@innerbib\@empty
\bibitem [{\citenamefont {Rau}\ \emph {et~al.}(2016)\citenamefont {Rau},
  \citenamefont {Lee},\ and\ \citenamefont
  {Kee}}]{doi:10.1146/annurev-conmatphys-031115-011319}%
  \BibitemOpen
  \bibfield  {author} {\bibinfo {author} {\bibfnamefont {J.~G.}\ \bibnamefont
  {Rau}}, \bibinfo {author} {\bibfnamefont {E.~K.-H.}\ \bibnamefont {Lee}}, \
  and\ \bibinfo {author} {\bibfnamefont {H.-Y.}\ \bibnamefont {Kee}},\ }\href
  {\doibase 10.1146/annurev-conmatphys-031115-011319} {\bibfield  {journal}
  {\bibinfo  {journal} {Annual Review of Condensed Matter Physics}\ }\textbf
  {\bibinfo {volume} {7}},\ \bibinfo {pages} {195} (\bibinfo {year}
  {2016})}\BibitemShut {NoStop}%
\bibitem [{\citenamefont {Go}\ \emph {et~al.}(2012)\citenamefont {Go},
  \citenamefont {Witczak-Krempa}, \citenamefont {Jeon}, \citenamefont {Park},\
  and\ \citenamefont {Kim}}]{PhysRevLett.109.066401}%
  \BibitemOpen
  \bibfield  {author} {\bibinfo {author} {\bibfnamefont {A.}~\bibnamefont
  {Go}}, \bibinfo {author} {\bibfnamefont {W.}~\bibnamefont {Witczak-Krempa}},
  \bibinfo {author} {\bibfnamefont {G.~S.}\ \bibnamefont {Jeon}}, \bibinfo
  {author} {\bibfnamefont {K.}~\bibnamefont {Park}}, \ and\ \bibinfo {author}
  {\bibfnamefont {Y.~B.}\ \bibnamefont {Kim}},\ }\href {\doibase
  10.1103/PhysRevLett.109.066401} {\bibfield  {journal} {\bibinfo  {journal}
  {Phys. Rev. Lett.}\ }\textbf {\bibinfo {volume} {109}},\ \bibinfo {pages}
  {066401} (\bibinfo {year} {2012})}\BibitemShut {NoStop}%
\bibitem [{\citenamefont {Schaffer}\ \emph {et~al.}(2016)\citenamefont
  {Schaffer}, \citenamefont {Lee}, \citenamefont {Yang},\ and\ \citenamefont
  {Kim}}]{Schaffer_2016}%
  \BibitemOpen
  \bibfield  {author} {\bibinfo {author} {\bibfnamefont {R.}~\bibnamefont
  {Schaffer}}, \bibinfo {author} {\bibfnamefont {E.~K.-H.}\ \bibnamefont
  {Lee}}, \bibinfo {author} {\bibfnamefont {B.-J.}\ \bibnamefont {Yang}}, \
  and\ \bibinfo {author} {\bibfnamefont {Y.~B.}\ \bibnamefont {Kim}},\ }\href
  {\doibase 10.1088/0034-4885/79/9/094504} {\bibfield  {journal} {\bibinfo
  {journal} {Reports on Progress in Physics}\ }\textbf {\bibinfo {volume}
  {79}},\ \bibinfo {pages} {094504} (\bibinfo {year} {2016})}\BibitemShut
  {NoStop}%
\bibitem [{\citenamefont {Ando}\ and\ \citenamefont
  {Fu}(2015)}]{doi:10.1146/annurev-conmatphys-031214-014501}%
  \BibitemOpen
  \bibfield  {author} {\bibinfo {author} {\bibfnamefont {Y.}~\bibnamefont
  {Ando}}\ and\ \bibinfo {author} {\bibfnamefont {L.}~\bibnamefont {Fu}},\
  }\href {\doibase 10.1146/annurev-conmatphys-031214-014501} {\bibfield
  {journal} {\bibinfo  {journal} {Annual Review of Condensed Matter Physics}\
  }\textbf {\bibinfo {volume} {6}},\ \bibinfo {pages} {361} (\bibinfo {year}
  {2015})}\BibitemShut {NoStop}%
\bibitem [{\citenamefont {Hasan}\ and\ \citenamefont
  {Moore}(2011)}]{doi:10.1146/annurev-conmatphys-062910-140432}%
  \BibitemOpen
  \bibfield  {author} {\bibinfo {author} {\bibfnamefont {M.~Z.}\ \bibnamefont
  {Hasan}}\ and\ \bibinfo {author} {\bibfnamefont {J.~E.}\ \bibnamefont
  {Moore}},\ }\href {\doibase 10.1146/annurev-conmatphys-062910-140432}
  {\bibfield  {journal} {\bibinfo  {journal} {Annual Review of Condensed Matter
  Physics}\ }\textbf {\bibinfo {volume} {2}},\ \bibinfo {pages} {55} (\bibinfo
  {year} {2011})}\BibitemShut {NoStop}%
\bibitem [{\citenamefont {Ando}(2013)}]{doi:10.7566/JPSJ.82.102001}%
  \BibitemOpen
  \bibfield  {author} {\bibinfo {author} {\bibfnamefont {Y.}~\bibnamefont
  {Ando}},\ }\href {\doibase 10.7566/JPSJ.82.102001} {\bibfield  {journal}
  {\bibinfo  {journal} {Journal of the Physical Society of Japan}\ }\textbf
  {\bibinfo {volume} {82}},\ \bibinfo {pages} {102001} (\bibinfo {year}
  {2013})}\BibitemShut {NoStop}%
\bibitem [{\citenamefont {Hasan}\ \emph {et~al.}(2015)\citenamefont {Hasan},
  \citenamefont {Xu},\ and\ \citenamefont {Bian}}]{Hasan_2015}%
  \BibitemOpen
  \bibfield  {author} {\bibinfo {author} {\bibfnamefont {M.~Z.}\ \bibnamefont
  {Hasan}}, \bibinfo {author} {\bibfnamefont {S.-Y.}\ \bibnamefont {Xu}}, \
  and\ \bibinfo {author} {\bibfnamefont {G.}~\bibnamefont {Bian}},\ }\href
  {\doibase 10.1088/0031-8949/2015/t164/014001} {\bibfield  {journal} {\bibinfo
   {journal} {Physica Scripta}\ }\textbf {\bibinfo {volume} {T164}},\ \bibinfo
  {pages} {014001} (\bibinfo {year} {2015})}\BibitemShut {NoStop}%
\bibitem [{\citenamefont {Hasan}\ and\ \citenamefont
  {Kane}(2010)}]{RevModPhys.82.3045}%
  \BibitemOpen
  \bibfield  {author} {\bibinfo {author} {\bibfnamefont {M.~Z.}\ \bibnamefont
  {Hasan}}\ and\ \bibinfo {author} {\bibfnamefont {C.~L.}\ \bibnamefont
  {Kane}},\ }\href {\doibase 10.1103/RevModPhys.82.3045} {\bibfield  {journal}
  {\bibinfo  {journal} {Rev. Mod. Phys.}\ }\textbf {\bibinfo {volume} {82}},\
  \bibinfo {pages} {3045} (\bibinfo {year} {2010})}\BibitemShut {NoStop}%
\bibitem [{\citenamefont {Qi}\ and\ \citenamefont
  {Zhang}(2011)}]{RevModPhys.83.1057}%
  \BibitemOpen
  \bibfield  {author} {\bibinfo {author} {\bibfnamefont {X.-L.}\ \bibnamefont
  {Qi}}\ and\ \bibinfo {author} {\bibfnamefont {S.-C.}\ \bibnamefont {Zhang}},\
  }\href {\doibase 10.1103/RevModPhys.83.1057} {\bibfield  {journal} {\bibinfo
  {journal} {Rev. Mod. Phys.}\ }\textbf {\bibinfo {volume} {83}},\ \bibinfo
  {pages} {1057} (\bibinfo {year} {2011})}\BibitemShut {NoStop}%
\bibitem [{\citenamefont {Yan}\ and\ \citenamefont {Zhang}(2012)}]{Yan_2012}%
  \BibitemOpen
  \bibfield  {author} {\bibinfo {author} {\bibfnamefont {B.}~\bibnamefont
  {Yan}}\ and\ \bibinfo {author} {\bibfnamefont {S.-C.}\ \bibnamefont
  {Zhang}},\ }\href {\doibase 10.1088/0034-4885/75/9/096501} {\bibfield
  {journal} {\bibinfo  {journal} {Reports on Progress in Physics}\ }\textbf
  {\bibinfo {volume} {75}},\ \bibinfo {pages} {096501} (\bibinfo {year}
  {2012})}\BibitemShut {NoStop}%
\bibitem [{\citenamefont {Yan}\ and\ \citenamefont
  {Felser}(2017)}]{doi:10.1146/annurev-conmatphys-031016-025458}%
  \BibitemOpen
  \bibfield  {author} {\bibinfo {author} {\bibfnamefont {B.}~\bibnamefont
  {Yan}}\ and\ \bibinfo {author} {\bibfnamefont {C.}~\bibnamefont {Felser}},\
  }\href {\doibase 10.1146/annurev-conmatphys-031016-025458} {\bibfield
  {journal} {\bibinfo  {journal} {Annual Review of Condensed Matter Physics}\
  }\textbf {\bibinfo {volume} {8}},\ \bibinfo {pages} {337} (\bibinfo {year}
  {2017})}\BibitemShut {NoStop}%
\bibitem [{\citenamefont {Vafek}\ and\ \citenamefont
  {Vishwanath}(2014)}]{doi:10.1146/annurev-conmatphys-031113-133841}%
  \BibitemOpen
  \bibfield  {author} {\bibinfo {author} {\bibfnamefont {O.}~\bibnamefont
  {Vafek}}\ and\ \bibinfo {author} {\bibfnamefont {A.}~\bibnamefont
  {Vishwanath}},\ }\href {\doibase 10.1146/annurev-conmatphys-031113-133841}
  {\bibfield  {journal} {\bibinfo  {journal} {Annual Review of Condensed Matter
  Physics}\ }\textbf {\bibinfo {volume} {5}},\ \bibinfo {pages} {83} (\bibinfo
  {year} {2014})}\BibitemShut {NoStop}%
\bibitem [{\citenamefont {Armitage}\ \emph {et~al.}(2018)\citenamefont
  {Armitage}, \citenamefont {Mele},\ and\ \citenamefont
  {Vishwanath}}]{RevModPhys.90.015001}%
  \BibitemOpen
  \bibfield  {author} {\bibinfo {author} {\bibfnamefont {N.~P.}\ \bibnamefont
  {Armitage}}, \bibinfo {author} {\bibfnamefont {E.~J.}\ \bibnamefont {Mele}},
  \ and\ \bibinfo {author} {\bibfnamefont {A.}~\bibnamefont {Vishwanath}},\
  }\href {\doibase 10.1103/RevModPhys.90.015001} {\bibfield  {journal}
  {\bibinfo  {journal} {Rev. Mod. Phys.}\ }\textbf {\bibinfo {volume} {90}},\
  \bibinfo {pages} {015001} (\bibinfo {year} {2018})}\BibitemShut {NoStop}%
\bibitem [{\citenamefont {Roy}\ \emph {et~al.}(2017)\citenamefont {Roy},
  \citenamefont {Goswami},\ and\ \citenamefont {Juricic}}]{PhysRevB.95.201102}%
  \BibitemOpen
  \bibfield  {author} {\bibinfo {author} {\bibfnamefont {B.}~\bibnamefont
  {Roy}}, \bibinfo {author} {\bibfnamefont {P.}~\bibnamefont {Goswami}}, \ and\
  \bibinfo {author} {\bibfnamefont {V.}~\bibnamefont {Juricic}},\ }\href
  {\doibase 10.1103/PhysRevB.95.201102} {\bibfield  {journal} {\bibinfo
  {journal} {Phys. Rev. B}\ }\textbf {\bibinfo {volume} {95}},\ \bibinfo
  {pages} {201102} (\bibinfo {year} {2017})}\BibitemShut {NoStop}%
\bibitem [{\citenamefont {Witczak-Krempa}\ \emph {et~al.}(2014)\citenamefont
  {Witczak-Krempa}, \citenamefont {Knap},\ and\ \citenamefont
  {Abanin}}]{PhysRevLett.113.136402}%
  \BibitemOpen
  \bibfield  {author} {\bibinfo {author} {\bibfnamefont {W.}~\bibnamefont
  {Witczak-Krempa}}, \bibinfo {author} {\bibfnamefont {M.}~\bibnamefont
  {Knap}}, \ and\ \bibinfo {author} {\bibfnamefont {D.}~\bibnamefont
  {Abanin}},\ }\href {\doibase 10.1103/PhysRevLett.113.136402} {\bibfield
  {journal} {\bibinfo  {journal} {Phys. Rev. Lett.}\ }\textbf {\bibinfo
  {volume} {113}},\ \bibinfo {pages} {136402} (\bibinfo {year}
  {2014})}\BibitemShut {NoStop}%
\bibitem [{\citenamefont {Wei}\ \emph {et~al.}(2012)\citenamefont {Wei},
  \citenamefont {Chao},\ and\ \citenamefont {Aji}}]{PhysRevLett.109.196403}%
  \BibitemOpen
  \bibfield  {author} {\bibinfo {author} {\bibfnamefont {H.}~\bibnamefont
  {Wei}}, \bibinfo {author} {\bibfnamefont {S.-P.}\ \bibnamefont {Chao}}, \
  and\ \bibinfo {author} {\bibfnamefont {V.}~\bibnamefont {Aji}},\ }\href
  {\doibase 10.1103/PhysRevLett.109.196403} {\bibfield  {journal} {\bibinfo
  {journal} {Phys. Rev. Lett.}\ }\textbf {\bibinfo {volume} {109}},\ \bibinfo
  {pages} {196403} (\bibinfo {year} {2012})}\BibitemShut {NoStop}%
\bibitem [{\citenamefont {Carlstr\"om}\ and\ \citenamefont
  {Bergholtz}(2018)}]{PhysRevB.98.241102}%
  \BibitemOpen
  \bibfield  {author} {\bibinfo {author} {\bibfnamefont {J.}~\bibnamefont
  {Carlstr\"om}}\ and\ \bibinfo {author} {\bibfnamefont {E.~J.}\ \bibnamefont
  {Bergholtz}},\ }\href {\doibase 10.1103/PhysRevB.98.241102} {\bibfield
  {journal} {\bibinfo  {journal} {Phys. Rev. B}\ }\textbf {\bibinfo {volume}
  {98}},\ \bibinfo {pages} {241102} (\bibinfo {year} {2018})}\BibitemShut
  {NoStop}%
\bibitem [{\citenamefont {Xue}\ and\ \citenamefont
  {Zhang}(2017)}]{PhysRevB.96.195160}%
  \BibitemOpen
  \bibfield  {author} {\bibinfo {author} {\bibfnamefont {F.}~\bibnamefont
  {Xue}}\ and\ \bibinfo {author} {\bibfnamefont {X.-X.}\ \bibnamefont
  {Zhang}},\ }\href {\doibase 10.1103/PhysRevB.96.195160} {\bibfield  {journal}
  {\bibinfo  {journal} {Phys. Rev. B}\ }\textbf {\bibinfo {volume} {96}},\
  \bibinfo {pages} {195160} (\bibinfo {year} {2017})}\BibitemShut {NoStop}%
\bibitem [{\citenamefont {Laubach}\ \emph {et~al.}(2016)\citenamefont
  {Laubach}, \citenamefont {Platt}, \citenamefont {Thomale}, \citenamefont
  {Neupert},\ and\ \citenamefont {Rachel}}]{PhysRevB.94.241102}%
  \BibitemOpen
  \bibfield  {author} {\bibinfo {author} {\bibfnamefont {M.}~\bibnamefont
  {Laubach}}, \bibinfo {author} {\bibfnamefont {C.}~\bibnamefont {Platt}},
  \bibinfo {author} {\bibfnamefont {R.}~\bibnamefont {Thomale}}, \bibinfo
  {author} {\bibfnamefont {T.}~\bibnamefont {Neupert}}, \ and\ \bibinfo
  {author} {\bibfnamefont {S.}~\bibnamefont {Rachel}},\ }\href {\doibase
  10.1103/PhysRevB.94.241102} {\bibfield  {journal} {\bibinfo  {journal} {Phys.
  Rev. B}\ }\textbf {\bibinfo {volume} {94}},\ \bibinfo {pages} {241102}
  (\bibinfo {year} {2016})}\BibitemShut {NoStop}%
\bibitem [{\citenamefont {Zhai}\ \emph {et~al.}(2016)\citenamefont {Zhai},
  \citenamefont {Chou},\ and\ \citenamefont {Mou}}]{PhysRevB.94.125135}%
  \BibitemOpen
  \bibfield  {author} {\bibinfo {author} {\bibfnamefont {L.-J.}\ \bibnamefont
  {Zhai}}, \bibinfo {author} {\bibfnamefont {P.-H.}\ \bibnamefont {Chou}}, \
  and\ \bibinfo {author} {\bibfnamefont {C.-Y.}\ \bibnamefont {Mou}},\ }\href
  {\doibase 10.1103/PhysRevB.94.125135} {\bibfield  {journal} {\bibinfo
  {journal} {Phys. Rev. B}\ }\textbf {\bibinfo {volume} {94}},\ \bibinfo
  {pages} {125135} (\bibinfo {year} {2016})}\BibitemShut {NoStop}%
\bibitem [{\citenamefont {Maciejko}\ and\ \citenamefont
  {Nandkishore}(2014)}]{PhysRevB.90.035126}%
  \BibitemOpen
  \bibfield  {author} {\bibinfo {author} {\bibfnamefont {J.}~\bibnamefont
  {Maciejko}}\ and\ \bibinfo {author} {\bibfnamefont {R.}~\bibnamefont
  {Nandkishore}},\ }\href {\doibase 10.1103/PhysRevB.90.035126} {\bibfield
  {journal} {\bibinfo  {journal} {Phys. Rev. B}\ }\textbf {\bibinfo {volume}
  {90}},\ \bibinfo {pages} {035126} (\bibinfo {year} {2014})}\BibitemShut
  {NoStop}%
\bibitem [{\citenamefont {Wei}\ \emph {et~al.}(2014{\natexlab{a}})\citenamefont
  {Wei}, \citenamefont {Chao},\ and\ \citenamefont {Aji}}]{PhysRevB.89.235109}%
  \BibitemOpen
  \bibfield  {author} {\bibinfo {author} {\bibfnamefont {H.}~\bibnamefont
  {Wei}}, \bibinfo {author} {\bibfnamefont {S.-P.}\ \bibnamefont {Chao}}, \
  and\ \bibinfo {author} {\bibfnamefont {V.}~\bibnamefont {Aji}},\ }\href
  {\doibase 10.1103/PhysRevB.89.235109} {\bibfield  {journal} {\bibinfo
  {journal} {Phys. Rev. B}\ }\textbf {\bibinfo {volume} {89}},\ \bibinfo
  {pages} {235109} (\bibinfo {year} {2014}{\natexlab{a}})}\BibitemShut
  {NoStop}%
\bibitem [{\citenamefont {Wei}\ \emph {et~al.}(2014{\natexlab{b}})\citenamefont
  {Wei}, \citenamefont {Chao},\ and\ \citenamefont {Aji}}]{PhysRevB.89.014506}%
  \BibitemOpen
  \bibfield  {author} {\bibinfo {author} {\bibfnamefont {H.}~\bibnamefont
  {Wei}}, \bibinfo {author} {\bibfnamefont {S.-P.}\ \bibnamefont {Chao}}, \
  and\ \bibinfo {author} {\bibfnamefont {V.}~\bibnamefont {Aji}},\ }\href
  {\doibase 10.1103/PhysRevB.89.014506} {\bibfield  {journal} {\bibinfo
  {journal} {Phys. Rev. B}\ }\textbf {\bibinfo {volume} {89}},\ \bibinfo
  {pages} {014506} (\bibinfo {year} {2014}{\natexlab{b}})}\BibitemShut
  {NoStop}%
\bibitem [{\citenamefont {Wang}\ \emph
  {et~al.}(2017{\natexlab{a}})\citenamefont {Wang}, \citenamefont {Li},\ and\
  \citenamefont {Bian}}]{PhysRevB.96.165203}%
  \BibitemOpen
  \bibfield  {author} {\bibinfo {author} {\bibfnamefont {Y.-X.}\ \bibnamefont
  {Wang}}, \bibinfo {author} {\bibfnamefont {F.}~\bibnamefont {Li}}, \ and\
  \bibinfo {author} {\bibfnamefont {B.}~\bibnamefont {Bian}},\ }\href {\doibase
  10.1103/PhysRevB.96.165203} {\bibfield  {journal} {\bibinfo  {journal} {Phys.
  Rev. B}\ }\textbf {\bibinfo {volume} {96}},\ \bibinfo {pages} {165203}
  (\bibinfo {year} {2017}{\natexlab{a}})}\BibitemShut {NoStop}%
\bibitem [{\citenamefont {Yang}(2019)}]{PhysRevB.100.245137}%
  \BibitemOpen
  \bibfield  {author} {\bibinfo {author} {\bibfnamefont {M.-F.}\ \bibnamefont
  {Yang}},\ }\href {\doibase 10.1103/PhysRevB.100.245137} {\bibfield  {journal}
  {\bibinfo  {journal} {Phys. Rev. B}\ }\textbf {\bibinfo {volume} {100}},\
  \bibinfo {pages} {245137} (\bibinfo {year} {2019})}\BibitemShut {NoStop}%
\bibitem [{\citenamefont {Yi}\ \emph {et~al.}(2017)\citenamefont {Yi},
  \citenamefont {Wang}, \citenamefont {Wang},\ and\ \citenamefont
  {Wang}}]{Yi2017possible}%
  \BibitemOpen
  \bibfield  {author} {\bibinfo {author} {\bibfnamefont {W.}~\bibnamefont
  {Yi}}, \bibinfo {author} {\bibfnamefont {Q.-S.}\ \bibnamefont {Wang}},
  \bibinfo {author} {\bibfnamefont {R.}~\bibnamefont {Wang}}, \ and\ \bibinfo
  {author} {\bibfnamefont {B.}~\bibnamefont {Wang}},\ }\href
  {https://arxiv.org/abs/1707.09576} {\bibfield  {journal} {\bibinfo  {journal}
  {arXiv preprint arXiv:1707.09576}\ } (\bibinfo {year} {2017})}\BibitemShut
  {NoStop}%
\bibitem [{\citenamefont {Arnold}\ \emph {et~al.}(2017)\citenamefont {Arnold},
  \citenamefont {Isidori}, \citenamefont {Kampert}, \citenamefont {Yager},
  \citenamefont {Eschrig},\ and\ \citenamefont {Saunders}}]{Arnold2017}%
  \BibitemOpen
  \bibfield  {author} {\bibinfo {author} {\bibfnamefont {F.}~\bibnamefont
  {Arnold}}, \bibinfo {author} {\bibfnamefont {A.}~\bibnamefont {Isidori}},
  \bibinfo {author} {\bibfnamefont {E.}~\bibnamefont {Kampert}}, \bibinfo
  {author} {\bibfnamefont {B.}~\bibnamefont {Yager}}, \bibinfo {author}
  {\bibfnamefont {M.}~\bibnamefont {Eschrig}}, \ and\ \bibinfo {author}
  {\bibfnamefont {J.}~\bibnamefont {Saunders}},\ }\href {\doibase
  10.1103/PhysRevLett.119.136601} {\bibfield  {journal} {\bibinfo  {journal}
  {Phys. Rev. Lett.}\ }\textbf {\bibinfo {volume} {119}},\ \bibinfo {pages}
  {136601} (\bibinfo {year} {2017})}\BibitemShut {NoStop}%
\bibitem [{\citenamefont {Yakovenko}(1993)}]{Yakovenko93}%
  \BibitemOpen
  \bibfield  {author} {\bibinfo {author} {\bibfnamefont {V.~M.}\ \bibnamefont
  {Yakovenko}},\ }\href {\doibase 10.1103/PhysRevB.47.8851} {\bibfield
  {journal} {\bibinfo  {journal} {Phys. Rev. B}\ }\textbf {\bibinfo {volume}
  {47}},\ \bibinfo {pages} {8851} (\bibinfo {year} {1993})}\BibitemShut
  {NoStop}%
\bibitem [{\citenamefont {Kumar}\ \emph {et~al.}(2010)\citenamefont {Kumar},
  \citenamefont {Poumirol}, \citenamefont {Escoffier}, \citenamefont {Goiran},
  \citenamefont {Raquet},\ and\ \citenamefont {Pivin}}]{Kumar_2010}%
  \BibitemOpen
  \bibfield  {author} {\bibinfo {author} {\bibfnamefont {A.}~\bibnamefont
  {Kumar}}, \bibinfo {author} {\bibfnamefont {J.-M.}\ \bibnamefont {Poumirol}},
  \bibinfo {author} {\bibfnamefont {W.}~\bibnamefont {Escoffier}}, \bibinfo
  {author} {\bibfnamefont {M.}~\bibnamefont {Goiran}}, \bibinfo {author}
  {\bibfnamefont {B.}~\bibnamefont {Raquet}}, \ and\ \bibinfo {author}
  {\bibfnamefont {J.~C.}\ \bibnamefont {Pivin}},\ }\href {\doibase
  10.1088/0953-8984/22/43/436004} {\bibfield  {journal} {\bibinfo  {journal}
  {Journal of Physics: Condensed Matter}\ }\textbf {\bibinfo {volume} {22}},\
  \bibinfo {pages} {436004} (\bibinfo {year} {2010})}\BibitemShut {NoStop}%
\bibitem [{\citenamefont {LeBoeuf}\ \emph {et~al.}(2017)\citenamefont
  {LeBoeuf}, \citenamefont {Rischau}, \citenamefont {Seyfarth}, \citenamefont
  {Küchler}, \citenamefont {Berben}, \citenamefont {Wiedmann}, \citenamefont
  {Tabis}, \citenamefont {Frachet}, \citenamefont {Behnia},\ and\ \citenamefont
  {Fauque}}]{LeBeouf2017}%
  \BibitemOpen
  \bibfield  {author} {\bibinfo {author} {\bibfnamefont {D.}~\bibnamefont
  {LeBoeuf}}, \bibinfo {author} {\bibfnamefont {C.~W.}\ \bibnamefont
  {Rischau}}, \bibinfo {author} {\bibfnamefont {G.}~\bibnamefont {Seyfarth}},
  \bibinfo {author} {\bibfnamefont {R.}~\bibnamefont {Küchler}}, \bibinfo
  {author} {\bibfnamefont {M.}~\bibnamefont {Berben}}, \bibinfo {author}
  {\bibfnamefont {S.}~\bibnamefont {Wiedmann}}, \bibinfo {author}
  {\bibfnamefont {W.}~\bibnamefont {Tabis}}, \bibinfo {author} {\bibfnamefont
  {M.}~\bibnamefont {Frachet}}, \bibinfo {author} {\bibfnamefont
  {K.}~\bibnamefont {Behnia}}, \ and\ \bibinfo {author} {\bibfnamefont
  {B.}~\bibnamefont {Fauque}},\ }\href {\doibase 10.1038/s41467-017-01394-7}
  {\bibfield  {journal} {\bibinfo  {journal} {Nat. Commun.}\ }\textbf {\bibinfo
  {volume} {8}},\ \bibinfo {pages} {1337} (\bibinfo {year} {2017})}\BibitemShut
  {NoStop}%
\bibitem [{\citenamefont {Alicea}\ and\ \citenamefont
  {Balents}(2009)}]{PhysRevB.79.241101}%
  \BibitemOpen
  \bibfield  {author} {\bibinfo {author} {\bibfnamefont {J.}~\bibnamefont
  {Alicea}}\ and\ \bibinfo {author} {\bibfnamefont {L.}~\bibnamefont
  {Balents}},\ }\href {\doibase 10.1103/PhysRevB.79.241101} {\bibfield
  {journal} {\bibinfo  {journal} {Phys. Rev. B}\ }\textbf {\bibinfo {volume}
  {79}},\ \bibinfo {pages} {241101} (\bibinfo {year} {2009})}\BibitemShut
  {NoStop}%
\bibitem [{\citenamefont {Fauqu\'e}\ \emph {et~al.}(2013)\citenamefont
  {Fauqu\'e}, \citenamefont {LeBoeuf}, \citenamefont {Vignolle}, \citenamefont
  {Nardone}, \citenamefont {Proust},\ and\ \citenamefont
  {Behnia}}]{PhysRevLett.110.266601}%
  \BibitemOpen
  \bibfield  {author} {\bibinfo {author} {\bibfnamefont {B.}~\bibnamefont
  {Fauqu\'e}}, \bibinfo {author} {\bibfnamefont {D.}~\bibnamefont {LeBoeuf}},
  \bibinfo {author} {\bibfnamefont {B.}~\bibnamefont {Vignolle}}, \bibinfo
  {author} {\bibfnamefont {M.}~\bibnamefont {Nardone}}, \bibinfo {author}
  {\bibfnamefont {C.}~\bibnamefont {Proust}}, \ and\ \bibinfo {author}
  {\bibfnamefont {K.}~\bibnamefont {Behnia}},\ }\href {\doibase
  10.1103/PhysRevLett.110.266601} {\bibfield  {journal} {\bibinfo  {journal}
  {Phys. Rev. Lett.}\ }\textbf {\bibinfo {volume} {110}},\ \bibinfo {pages}
  {266601} (\bibinfo {year} {2013})}\BibitemShut {NoStop}%
\bibitem [{\citenamefont {Iye}\ and\ \citenamefont
  {Dresselhaus}(1985)}]{PhysRevLett.54.1182}%
  \BibitemOpen
  \bibfield  {author} {\bibinfo {author} {\bibfnamefont {Y.}~\bibnamefont
  {Iye}}\ and\ \bibinfo {author} {\bibfnamefont {G.}~\bibnamefont
  {Dresselhaus}},\ }\href {\doibase 10.1103/PhysRevLett.54.1182} {\bibfield
  {journal} {\bibinfo  {journal} {Phys. Rev. Lett.}\ }\textbf {\bibinfo
  {volume} {54}},\ \bibinfo {pages} {1182} (\bibinfo {year}
  {1985})}\BibitemShut {NoStop}%
\bibitem [{\citenamefont {Takahashi}\ and\ \citenamefont
  {Takada}(1994)}]{TAKAHASHI1994384}%
  \BibitemOpen
  \bibfield  {author} {\bibinfo {author} {\bibfnamefont {K.}~\bibnamefont
  {Takahashi}}\ and\ \bibinfo {author} {\bibfnamefont {Y.}~\bibnamefont
  {Takada}},\ }\href {\doibase https://doi.org/10.1016/0921-4526(94)91120-7}
  {\bibfield  {journal} {\bibinfo  {journal} {Physica B: Condensed Matter}\
  }\textbf {\bibinfo {volume} {201}},\ \bibinfo {pages} {384} (\bibinfo {year}
  {1994})}\BibitemShut {NoStop}%
\bibitem [{\citenamefont {Yaguchi}\ and\ \citenamefont
  {Singleton}(2009)}]{Yaguchi_2009}%
  \BibitemOpen
  \bibfield  {author} {\bibinfo {author} {\bibfnamefont {H.}~\bibnamefont
  {Yaguchi}}\ and\ \bibinfo {author} {\bibfnamefont {J.}~\bibnamefont
  {Singleton}},\ }\href {\doibase 10.1088/0953-8984/21/34/344207} {\bibfield
  {journal} {\bibinfo  {journal} {Journal of Physics: Condensed Matter}\
  }\textbf {\bibinfo {volume} {21}},\ \bibinfo {pages} {344207} (\bibinfo
  {year} {2009})}\BibitemShut {NoStop}%
\bibitem [{\citenamefont {Zhang}\ \emph {et~al.}(2016)\citenamefont {Zhang},
  \citenamefont {Tong}, \citenamefont {Yuan}, \citenamefont {Lin},
  \citenamefont {Wang}, \citenamefont {Zhang}, \citenamefont {Xi},
  \citenamefont {Wang}, \citenamefont {Jia},\ and\ \citenamefont
  {Zhang}}]{Zhang2016}%
  \BibitemOpen
  \bibfield  {author} {\bibinfo {author} {\bibfnamefont {C.-L.}\ \bibnamefont
  {Zhang}}, \bibinfo {author} {\bibfnamefont {B.}~\bibnamefont {Tong}},
  \bibinfo {author} {\bibfnamefont {Z.}~\bibnamefont {Yuan}}, \bibinfo {author}
  {\bibfnamefont {Z.}~\bibnamefont {Lin}}, \bibinfo {author} {\bibfnamefont
  {J.}~\bibnamefont {Wang}}, \bibinfo {author} {\bibfnamefont {J.}~\bibnamefont
  {Zhang}}, \bibinfo {author} {\bibfnamefont {C.-Y.}\ \bibnamefont {Xi}},
  \bibinfo {author} {\bibfnamefont {Z.}~\bibnamefont {Wang}}, \bibinfo {author}
  {\bibfnamefont {S.}~\bibnamefont {Jia}}, \ and\ \bibinfo {author}
  {\bibfnamefont {C.}~\bibnamefont {Zhang}},\ }\href {\doibase
  10.1103/PhysRevB.94.205120} {\bibfield  {journal} {\bibinfo  {journal} {Phys.
  Rev. B}\ }\textbf {\bibinfo {volume} {94}},\ \bibinfo {pages} {205120}
  (\bibinfo {year} {2016})}\BibitemShut {NoStop}%
\bibitem [{\citenamefont {Yuan}\ \emph {et~al.}(2020)\citenamefont {Yuan},
  \citenamefont {Zhang}, \citenamefont {Zhang}, \citenamefont {Yan},
  \citenamefont {Lyu}, \citenamefont {Zhang}, \citenamefont {Li}, \citenamefont
  {Song}, \citenamefont {Zhao}, \citenamefont {Leng}, \citenamefont {Ozerov},
  \citenamefont {Chen}, \citenamefont {Wang}, \citenamefont {Shi},
  \citenamefont {Yan},\ and\ \citenamefont {Xiu}}]{Wang2020}%
  \BibitemOpen
  \bibfield  {author} {\bibinfo {author} {\bibfnamefont {X.}~\bibnamefont
  {Yuan}}, \bibinfo {author} {\bibfnamefont {C.}~\bibnamefont {Zhang}},
  \bibinfo {author} {\bibfnamefont {Y.}~\bibnamefont {Zhang}}, \bibinfo
  {author} {\bibfnamefont {Z.}~\bibnamefont {Yan}}, \bibinfo {author}
  {\bibfnamefont {T.}~\bibnamefont {Lyu}}, \bibinfo {author} {\bibfnamefont
  {M.}~\bibnamefont {Zhang}}, \bibinfo {author} {\bibfnamefont
  {Z.}~\bibnamefont {Li}}, \bibinfo {author} {\bibfnamefont {C.}~\bibnamefont
  {Song}}, \bibinfo {author} {\bibfnamefont {M.}~\bibnamefont {Zhao}}, \bibinfo
  {author} {\bibfnamefont {P.}~\bibnamefont {Leng}}, \bibinfo {author}
  {\bibfnamefont {M.}~\bibnamefont {Ozerov}}, \bibinfo {author} {\bibfnamefont
  {X.}~\bibnamefont {Chen}}, \bibinfo {author} {\bibfnamefont {N.}~\bibnamefont
  {Wang}}, \bibinfo {author} {\bibfnamefont {Y.}~\bibnamefont {Shi}}, \bibinfo
  {author} {\bibfnamefont {H.}~\bibnamefont {Yan}}, \ and\ \bibinfo {author}
  {\bibfnamefont {F.}~\bibnamefont {Xiu}},\ }\href {\doibase
  https://doi.org/10.1038/s41467-020-14749-4} {\bibfield  {journal} {\bibinfo
  {journal} {Nat. Commun.}\ }\textbf {\bibinfo {volume} {11}},\ \bibinfo
  {pages} {1259} (\bibinfo {year} {2020})}\BibitemShut {NoStop}%
\bibitem [{\citenamefont {Tang}\ \emph {et~al.}(2019)\citenamefont {Tang},
  \citenamefont {Ren}, \citenamefont {Wang}, \citenamefont {Zhong},
  \citenamefont {Schneeloch}, \citenamefont {Yang}, \citenamefont {Yang},
  \citenamefont {Lee}, \citenamefont {Gu}, \citenamefont {Qiao},\ and\
  \citenamefont {Zhang}}]{Tang2019}%
  \BibitemOpen
  \bibfield  {author} {\bibinfo {author} {\bibfnamefont {F.}~\bibnamefont
  {Tang}}, \bibinfo {author} {\bibfnamefont {Y.}~\bibnamefont {Ren}}, \bibinfo
  {author} {\bibfnamefont {P.}~\bibnamefont {Wang}}, \bibinfo {author}
  {\bibfnamefont {R.}~\bibnamefont {Zhong}}, \bibinfo {author} {\bibfnamefont
  {J.}~\bibnamefont {Schneeloch}}, \bibinfo {author} {\bibfnamefont {S.~A.}\
  \bibnamefont {Yang}}, \bibinfo {author} {\bibfnamefont {K.}~\bibnamefont
  {Yang}}, \bibinfo {author} {\bibfnamefont {P.~A.}\ \bibnamefont {Lee}},
  \bibinfo {author} {\bibfnamefont {G.}~\bibnamefont {Gu}}, \bibinfo {author}
  {\bibfnamefont {Z.}~\bibnamefont {Qiao}}, \ and\ \bibinfo {author}
  {\bibfnamefont {L.}~\bibnamefont {Zhang}},\ }\href {\doibase
  10.1038/s41586-019-1180-9} {\bibfield  {journal} {\bibinfo  {journal}
  {Nature}\ }\textbf {\bibinfo {volume} {569}},\ \bibinfo {pages} {537}
  (\bibinfo {year} {2019})}\BibitemShut {NoStop}%
\bibitem [{\citenamefont {Ramshaw}\ \emph {et~al.}(2018)\citenamefont
  {Ramshaw}, \citenamefont {Modic}, \citenamefont {Shekhter}, \citenamefont
  {Zhang}, \citenamefont {Kim}, \citenamefont {Moll}, \citenamefont {Bachmann},
  \citenamefont {Chan}, \citenamefont {Betts}, \citenamefont {Balakirev},
  \citenamefont {Migliori}, \citenamefont {Ghimire}, \citenamefont {Bauer},
  \citenamefont {Ronning},\ and\ \citenamefont {McDonald}}]{Ramshaw2018}%
  \BibitemOpen
  \bibfield  {author} {\bibinfo {author} {\bibfnamefont {B.~J.}\ \bibnamefont
  {Ramshaw}}, \bibinfo {author} {\bibfnamefont {K.~A.}\ \bibnamefont {Modic}},
  \bibinfo {author} {\bibfnamefont {A.}~\bibnamefont {Shekhter}}, \bibinfo
  {author} {\bibfnamefont {Y.}~\bibnamefont {Zhang}}, \bibinfo {author}
  {\bibfnamefont {E.-A.}\ \bibnamefont {Kim}}, \bibinfo {author} {\bibfnamefont
  {P.~J.~W.}\ \bibnamefont {Moll}}, \bibinfo {author} {\bibfnamefont {M.~D.}\
  \bibnamefont {Bachmann}}, \bibinfo {author} {\bibfnamefont {M.~K.}\
  \bibnamefont {Chan}}, \bibinfo {author} {\bibfnamefont {J.~B.}\ \bibnamefont
  {Betts}}, \bibinfo {author} {\bibfnamefont {F.}~\bibnamefont {Balakirev}},
  \bibinfo {author} {\bibfnamefont {A.}~\bibnamefont {Migliori}}, \bibinfo
  {author} {\bibfnamefont {N.~J.}\ \bibnamefont {Ghimire}}, \bibinfo {author}
  {\bibfnamefont {E.~D.}\ \bibnamefont {Bauer}}, \bibinfo {author}
  {\bibfnamefont {F.}~\bibnamefont {Ronning}}, \ and\ \bibinfo {author}
  {\bibfnamefont {R.~D.}\ \bibnamefont {McDonald}},\ }\href {\doibase
  10.1038/s41467-018-04542-9} {\bibfield  {journal} {\bibinfo  {journal} {Nat.
  Commun.}\ }\textbf {\bibinfo {volume} {9}},\ \bibinfo {pages} {2217}
  (\bibinfo {year} {2018})}\BibitemShut {NoStop}%
\bibitem [{\citenamefont {Liu}\ \emph {et~al.}(2016)\citenamefont {Liu},
  \citenamefont {Yuan}, \citenamefont {Zhang}, \citenamefont {Jin},
  \citenamefont {Narayan}, \citenamefont {Luo}, \citenamefont {Chen},
  \citenamefont {Yang}, \citenamefont {Zou}, \citenamefont {Wu}, \citenamefont
  {Sanvito}, \citenamefont {Xia}, \citenamefont {Li}, \citenamefont {Wang},\
  and\ \citenamefont {Xiu}}]{Liu2016}%
  \BibitemOpen
  \bibfield  {author} {\bibinfo {author} {\bibfnamefont {Y.}~\bibnamefont
  {Liu}}, \bibinfo {author} {\bibfnamefont {X.}~\bibnamefont {Yuan}}, \bibinfo
  {author} {\bibfnamefont {C.}~\bibnamefont {Zhang}}, \bibinfo {author}
  {\bibfnamefont {Z.}~\bibnamefont {Jin}}, \bibinfo {author} {\bibfnamefont
  {A.}~\bibnamefont {Narayan}}, \bibinfo {author} {\bibfnamefont
  {C.}~\bibnamefont {Luo}}, \bibinfo {author} {\bibfnamefont {Z.}~\bibnamefont
  {Chen}}, \bibinfo {author} {\bibfnamefont {L.}~\bibnamefont {Yang}}, \bibinfo
  {author} {\bibfnamefont {J.}~\bibnamefont {Zou}}, \bibinfo {author}
  {\bibfnamefont {X.}~\bibnamefont {Wu}}, \bibinfo {author} {\bibfnamefont
  {S.}~\bibnamefont {Sanvito}}, \bibinfo {author} {\bibfnamefont
  {Z.}~\bibnamefont {Xia}}, \bibinfo {author} {\bibfnamefont {L.}~\bibnamefont
  {Li}}, \bibinfo {author} {\bibfnamefont {Z.}~\bibnamefont {Wang}}, \ and\
  \bibinfo {author} {\bibfnamefont {F.}~\bibnamefont {Xiu}},\ }\href {\doibase
  10.1038/ncomms12516} {\bibfield  {journal} {\bibinfo  {journal} {Nat.
  Commun.}\ }\textbf {\bibinfo {volume} {7}},\ \bibinfo {pages} {12516}
  (\bibinfo {year} {2016})}\BibitemShut {NoStop}%
\bibitem [{\citenamefont {Hui}\ \emph {et~al.}(2019)\citenamefont {Hui},
  \citenamefont {Zhang},\ and\ \citenamefont {Kim}}]{Hui2019}%
  \BibitemOpen
  \bibfield  {author} {\bibinfo {author} {\bibfnamefont {A.}~\bibnamefont
  {Hui}}, \bibinfo {author} {\bibfnamefont {Y.}~\bibnamefont {Zhang}}, \ and\
  \bibinfo {author} {\bibfnamefont {E.-A.}\ \bibnamefont {Kim}},\ }\href
  {\doibase 10.1103/PhysRevB.100.085144} {\bibfield  {journal} {\bibinfo
  {journal} {Phys. Rev. B}\ }\textbf {\bibinfo {volume} {100}},\ \bibinfo
  {pages} {085144} (\bibinfo {year} {2019})}\BibitemShut {NoStop}%
\bibitem [{\citenamefont {Kim}\ \emph {et~al.}(2017)\citenamefont {Kim},
  \citenamefont {Ryoo},\ and\ \citenamefont {Park}}]{Kim2017}%
  \BibitemOpen
  \bibfield  {author} {\bibinfo {author} {\bibfnamefont {P.}~\bibnamefont
  {Kim}}, \bibinfo {author} {\bibfnamefont {J.~H.}\ \bibnamefont {Ryoo}}, \
  and\ \bibinfo {author} {\bibfnamefont {C.-H.}\ \bibnamefont {Park}},\ }\href
  {\doibase 10.1103/PhysRevLett.119.266401} {\bibfield  {journal} {\bibinfo
  {journal} {Phys. Rev. Lett.}\ }\textbf {\bibinfo {volume} {119}},\ \bibinfo
  {pages} {266401} (\bibinfo {year} {2017})}\BibitemShut {NoStop}%
\bibitem [{\citenamefont {Rinkel}\ \emph {et~al.}(2019)\citenamefont {Rinkel},
  \citenamefont {Lopes},\ and\ \citenamefont {Garate}}]{rinkel2019}%
  \BibitemOpen
  \bibfield  {author} {\bibinfo {author} {\bibfnamefont {P.}~\bibnamefont
  {Rinkel}}, \bibinfo {author} {\bibfnamefont {P.~L.~S.}\ \bibnamefont
  {Lopes}}, \ and\ \bibinfo {author} {\bibfnamefont {I.}~\bibnamefont
  {Garate}},\ }\href {\doibase 10.1103/PhysRevB.99.144301} {\bibfield
  {journal} {\bibinfo  {journal} {Phys. Rev. B}\ }\textbf {\bibinfo {volume}
  {99}},\ \bibinfo {pages} {144301} (\bibinfo {year} {2019})}\BibitemShut
  {NoStop}%
\bibitem [{\citenamefont {Zyuzin}\ and\ \citenamefont
  {Burkov}(2012)}]{PhysRevB.86.115133}%
  \BibitemOpen
  \bibfield  {author} {\bibinfo {author} {\bibfnamefont {A.~A.}\ \bibnamefont
  {Zyuzin}}\ and\ \bibinfo {author} {\bibfnamefont {A.~A.}\ \bibnamefont
  {Burkov}},\ }\href {\doibase 10.1103/PhysRevB.86.115133} {\bibfield
  {journal} {\bibinfo  {journal} {Phys. Rev. B}\ }\textbf {\bibinfo {volume}
  {86}},\ \bibinfo {pages} {115133} (\bibinfo {year} {2012})}\BibitemShut
  {NoStop}%
\bibitem [{\citenamefont {Parameswaran}\ \emph {et~al.}(2014)\citenamefont
  {Parameswaran}, \citenamefont {Grover}, \citenamefont {Abanin}, \citenamefont
  {Pesin},\ and\ \citenamefont {Vishwanath}}]{PhysRevX.4.031035}%
  \BibitemOpen
  \bibfield  {author} {\bibinfo {author} {\bibfnamefont {S.~A.}\ \bibnamefont
  {Parameswaran}}, \bibinfo {author} {\bibfnamefont {T.}~\bibnamefont
  {Grover}}, \bibinfo {author} {\bibfnamefont {D.~A.}\ \bibnamefont {Abanin}},
  \bibinfo {author} {\bibfnamefont {D.~A.}\ \bibnamefont {Pesin}}, \ and\
  \bibinfo {author} {\bibfnamefont {A.}~\bibnamefont {Vishwanath}},\ }\href
  {\doibase 10.1103/PhysRevX.4.031035} {\bibfield  {journal} {\bibinfo
  {journal} {Phys. Rev. X}\ }\textbf {\bibinfo {volume} {4}},\ \bibinfo {pages}
  {031035} (\bibinfo {year} {2014})}\BibitemShut {NoStop}%
\bibitem [{\citenamefont {Zhang}\ \emph {et~al.}(2019)\citenamefont {Zhang},
  \citenamefont {Zeng}, \citenamefont {Chiu}, \citenamefont {Sch\"onemann},
  \citenamefont {Memaran}, \citenamefont {Zheng}, \citenamefont {Rhodes},
  \citenamefont {Chen}, \citenamefont {Besara}, \citenamefont {Sankar},
  \citenamefont {Chou}, \citenamefont {McCandless}, \citenamefont {Chan},
  \citenamefont {Alidoust}, \citenamefont {Xu}, \citenamefont {Belopolski},
  \citenamefont {Hasan}, \citenamefont {Balakirev},\ and\ \citenamefont
  {Balicas}}]{Zhang2019}%
  \BibitemOpen
  \bibfield  {author} {\bibinfo {author} {\bibfnamefont {Q.~R.}\ \bibnamefont
  {Zhang}}, \bibinfo {author} {\bibfnamefont {B.}~\bibnamefont {Zeng}},
  \bibinfo {author} {\bibfnamefont {Y.~C.}\ \bibnamefont {Chiu}}, \bibinfo
  {author} {\bibfnamefont {R.}~\bibnamefont {Sch\"onemann}}, \bibinfo {author}
  {\bibfnamefont {S.}~\bibnamefont {Memaran}}, \bibinfo {author} {\bibfnamefont
  {W.}~\bibnamefont {Zheng}}, \bibinfo {author} {\bibfnamefont
  {D.}~\bibnamefont {Rhodes}}, \bibinfo {author} {\bibfnamefont {K.-W.}\
  \bibnamefont {Chen}}, \bibinfo {author} {\bibfnamefont {T.}~\bibnamefont
  {Besara}}, \bibinfo {author} {\bibfnamefont {R.}~\bibnamefont {Sankar}},
  \bibinfo {author} {\bibfnamefont {F.}~\bibnamefont {Chou}}, \bibinfo {author}
  {\bibfnamefont {G.~T.}\ \bibnamefont {McCandless}}, \bibinfo {author}
  {\bibfnamefont {J.~Y.}\ \bibnamefont {Chan}}, \bibinfo {author}
  {\bibfnamefont {N.}~\bibnamefont {Alidoust}}, \bibinfo {author}
  {\bibfnamefont {S.-Y.}\ \bibnamefont {Xu}}, \bibinfo {author} {\bibfnamefont
  {I.}~\bibnamefont {Belopolski}}, \bibinfo {author} {\bibfnamefont {M.~Z.}\
  \bibnamefont {Hasan}}, \bibinfo {author} {\bibfnamefont {F.~F.}\ \bibnamefont
  {Balakirev}}, \ and\ \bibinfo {author} {\bibfnamefont {L.}~\bibnamefont
  {Balicas}},\ }\href {\doibase 10.1103/PhysRevB.100.115138} {\bibfield
  {journal} {\bibinfo  {journal} {Phys. Rev. B}\ }\textbf {\bibinfo {volume}
  {100}},\ \bibinfo {pages} {115138} (\bibinfo {year} {2019})}\BibitemShut
  {NoStop}%
\bibitem [{\citenamefont {Roy}\ and\ \citenamefont {Sau}(2015)}]{Roy2015}%
  \BibitemOpen
  \bibfield  {author} {\bibinfo {author} {\bibfnamefont {B.}~\bibnamefont
  {Roy}}\ and\ \bibinfo {author} {\bibfnamefont {J.~D.}\ \bibnamefont {Sau}},\
  }\href {\doibase 10.1103/PhysRevB.92.125141} {\bibfield  {journal} {\bibinfo
  {journal} {Phys. Rev. B}\ }\textbf {\bibinfo {volume} {92}},\ \bibinfo
  {pages} {125141} (\bibinfo {year} {2015})}\BibitemShut {NoStop}%
\bibitem [{\citenamefont {Miransky}\ and\ \citenamefont
  {Shovkovy}(2015)}]{Miransky2015}%
  \BibitemOpen
  \bibfield  {author} {\bibinfo {author} {\bibfnamefont {V.~A.}\ \bibnamefont
  {Miransky}}\ and\ \bibinfo {author} {\bibfnamefont {I.~A.}\ \bibnamefont
  {Shovkovy}},\ }\href {\doibase https://doi.org/10.1016/j.physrep.2015.02.003}
  {\bibfield  {journal} {\bibinfo  {journal} {Physics Reports}\ }\textbf
  {\bibinfo {volume} {576}},\ \bibinfo {pages} {1} (\bibinfo {year} {2015})},\
  \bibinfo {note} {quantum field theory in a magnetic field: From quantum
  chromodynamics to graphene and Dirac semimetals}\BibitemShut {NoStop}%
\bibitem [{\citenamefont {Pan}\ and\ \citenamefont
  {Shindou}(2019)}]{Shindou2019}%
  \BibitemOpen
  \bibfield  {author} {\bibinfo {author} {\bibfnamefont {Z.}~\bibnamefont
  {Pan}}\ and\ \bibinfo {author} {\bibfnamefont {R.}~\bibnamefont {Shindou}},\
  }\href {\doibase 10.1103/PhysRevB.100.165124} {\bibfield  {journal} {\bibinfo
   {journal} {Phys. Rev. B}\ }\textbf {\bibinfo {volume} {100}},\ \bibinfo
  {pages} {165124} (\bibinfo {year} {2019})}\BibitemShut {NoStop}%
\bibitem [{\citenamefont {Song}\ \emph {et~al.}(2017)\citenamefont {Song},
  \citenamefont {Fang},\ and\ \citenamefont {Dai}}]{Song2017}%
  \BibitemOpen
  \bibfield  {author} {\bibinfo {author} {\bibfnamefont {Z.}~\bibnamefont
  {Song}}, \bibinfo {author} {\bibfnamefont {Z.}~\bibnamefont {Fang}}, \ and\
  \bibinfo {author} {\bibfnamefont {X.}~\bibnamefont {Dai}},\ }\href {\doibase
  10.1103/PhysRevB.96.235104} {\bibfield  {journal} {\bibinfo  {journal} {Phys.
  Rev. B}\ }\textbf {\bibinfo {volume} {96}},\ \bibinfo {pages} {235104}
  (\bibinfo {year} {2017})}\BibitemShut {NoStop}%
\bibitem [{\citenamefont {Trescher}\ \emph {et~al.}(2017)\citenamefont
  {Trescher}, \citenamefont {Bergholtz}, \citenamefont {Udagawa},\ and\
  \citenamefont {Knolle}}]{Trescher2017}%
  \BibitemOpen
  \bibfield  {author} {\bibinfo {author} {\bibfnamefont {M.}~\bibnamefont
  {Trescher}}, \bibinfo {author} {\bibfnamefont {E.~J.}\ \bibnamefont
  {Bergholtz}}, \bibinfo {author} {\bibfnamefont {M.}~\bibnamefont {Udagawa}},
  \ and\ \bibinfo {author} {\bibfnamefont {J.}~\bibnamefont {Knolle}},\ }\href
  {\doibase 10.1103/PhysRevB.96.201101} {\bibfield  {journal} {\bibinfo
  {journal} {Phys. Rev. B}\ }\textbf {\bibinfo {volume} {96}},\ \bibinfo
  {pages} {201101} (\bibinfo {year} {2017})}\BibitemShut {NoStop}%
\bibitem [{\citenamefont {Li}\ \emph {et~al.}(2016)\citenamefont {Li},
  \citenamefont {Roy},\ and\ \citenamefont {Das~Sarma}}]{Xiao2016}%
  \BibitemOpen
  \bibfield  {author} {\bibinfo {author} {\bibfnamefont {X.}~\bibnamefont
  {Li}}, \bibinfo {author} {\bibfnamefont {B.}~\bibnamefont {Roy}}, \ and\
  \bibinfo {author} {\bibfnamefont {S.}~\bibnamefont {Das~Sarma}},\ }\href
  {\doibase 10.1103/PhysRevB.94.195144} {\bibfield  {journal} {\bibinfo
  {journal} {Phys. Rev. B}\ }\textbf {\bibinfo {volume} {94}},\ \bibinfo
  {pages} {195144} (\bibinfo {year} {2016})}\BibitemShut {NoStop}%
\bibitem [{\citenamefont {Yang}\ \emph {et~al.}(2011)\citenamefont {Yang},
  \citenamefont {Lu},\ and\ \citenamefont {Ran}}]{Yang2011}%
  \BibitemOpen
  \bibfield  {author} {\bibinfo {author} {\bibfnamefont {K.-Y.}\ \bibnamefont
  {Yang}}, \bibinfo {author} {\bibfnamefont {Y.-M.}\ \bibnamefont {Lu}}, \ and\
  \bibinfo {author} {\bibfnamefont {Y.}~\bibnamefont {Ran}},\ }\href {\doibase
  10.1103/PhysRevB.84.075129} {\bibfield  {journal} {\bibinfo  {journal} {Phys.
  Rev. B}\ }\textbf {\bibinfo {volume} {84}},\ \bibinfo {pages} {075129}
  (\bibinfo {year} {2011})}\BibitemShut {NoStop}%
\bibitem [{\citenamefont {Song}\ \emph {et~al.}(2016)\citenamefont {Song},
  \citenamefont {Zhao}, \citenamefont {Fang},\ and\ \citenamefont
  {Dai}}]{Song2016}%
  \BibitemOpen
  \bibfield  {author} {\bibinfo {author} {\bibfnamefont {Z.}~\bibnamefont
  {Song}}, \bibinfo {author} {\bibfnamefont {J.}~\bibnamefont {Zhao}}, \bibinfo
  {author} {\bibfnamefont {Z.}~\bibnamefont {Fang}}, \ and\ \bibinfo {author}
  {\bibfnamefont {X.}~\bibnamefont {Dai}},\ }\href {\doibase
  10.1103/PhysRevB.94.214306} {\bibfield  {journal} {\bibinfo  {journal} {Phys.
  Rev. B}\ }\textbf {\bibinfo {volume} {94}},\ \bibinfo {pages} {214306}
  (\bibinfo {year} {2016})}\BibitemShut {NoStop}%
\bibitem [{\citenamefont {M\"oller}\ \emph {et~al.}(2017)\citenamefont
  {M\"oller}, \citenamefont {Sawatzky}, \citenamefont {Franz},\ and\
  \citenamefont {Berciu}}]{Moller2017}%
  \BibitemOpen
  \bibfield  {author} {\bibinfo {author} {\bibfnamefont {M.~M.}\ \bibnamefont
  {M\"oller}}, \bibinfo {author} {\bibfnamefont {G.~A.}\ \bibnamefont
  {Sawatzky}}, \bibinfo {author} {\bibfnamefont {M.}~\bibnamefont {Franz}}, \
  and\ \bibinfo {author} {\bibfnamefont {M.}~\bibnamefont {Berciu}},\ }\href
  {\doibase 10.1038/s41467-017-02442-y} {\bibfield  {journal} {\bibinfo
  {journal} {Nat. Commun.}\ }\textbf {\bibinfo {volume} {8}},\ \bibinfo {pages}
  {2267} (\bibinfo {year} {2017})}\BibitemShut {NoStop}%
\bibitem [{\citenamefont {Nguyen}\ \emph {et~al.}(2020)\citenamefont {Nguyen},
  \citenamefont {Han}, \citenamefont {Andrejevic}, \citenamefont {Pablo-Pedro},
  \citenamefont {Apte}, \citenamefont {Tsurimaki}, \citenamefont {Ding},
  \citenamefont {Zhang}, \citenamefont {Alatas}, \citenamefont {Alp},
  \citenamefont {Chi}, \citenamefont {Fernandez-Baca}, \citenamefont {Matsuda},
  \citenamefont {Tennant}, \citenamefont {Zhao}, \citenamefont {Xu},
  \citenamefont {Lynn}, \citenamefont {Huang},\ and\ \citenamefont
  {Li}}]{nguyen2020}%
  \BibitemOpen
  \bibfield  {author} {\bibinfo {author} {\bibfnamefont {T.}~\bibnamefont
  {Nguyen}}, \bibinfo {author} {\bibfnamefont {F.}~\bibnamefont {Han}},
  \bibinfo {author} {\bibfnamefont {N.}~\bibnamefont {Andrejevic}}, \bibinfo
  {author} {\bibfnamefont {R.}~\bibnamefont {Pablo-Pedro}}, \bibinfo {author}
  {\bibfnamefont {A.}~\bibnamefont {Apte}}, \bibinfo {author} {\bibfnamefont
  {Y.}~\bibnamefont {Tsurimaki}}, \bibinfo {author} {\bibfnamefont
  {Z.}~\bibnamefont {Ding}}, \bibinfo {author} {\bibfnamefont {K.}~\bibnamefont
  {Zhang}}, \bibinfo {author} {\bibfnamefont {A.}~\bibnamefont {Alatas}},
  \bibinfo {author} {\bibfnamefont {E.~E.}\ \bibnamefont {Alp}}, \bibinfo
  {author} {\bibfnamefont {S.}~\bibnamefont {Chi}}, \bibinfo {author}
  {\bibfnamefont {J.}~\bibnamefont {Fernandez-Baca}}, \bibinfo {author}
  {\bibfnamefont {M.}~\bibnamefont {Matsuda}}, \bibinfo {author} {\bibfnamefont
  {D.~A.}\ \bibnamefont {Tennant}}, \bibinfo {author} {\bibfnamefont
  {Y.}~\bibnamefont {Zhao}}, \bibinfo {author} {\bibfnamefont {Z.}~\bibnamefont
  {Xu}}, \bibinfo {author} {\bibfnamefont {J.~W.}\ \bibnamefont {Lynn}},
  \bibinfo {author} {\bibfnamefont {S.}~\bibnamefont {Huang}}, \ and\ \bibinfo
  {author} {\bibfnamefont {M.}~\bibnamefont {Li}},\ }\href {\doibase
  10.1103/PhysRevLett.124.236401} {\bibfield  {journal} {\bibinfo  {journal}
  {Phys. Rev. Lett.}\ }\textbf {\bibinfo {volume} {124}},\ \bibinfo {pages}
  {236401} (\bibinfo {year} {2020})}\BibitemShut {NoStop}%
\bibitem [{\citenamefont {Zhang}\ and\ \citenamefont
  {Zhou}(2020)}]{PhysRevB.101.085202}%
  \BibitemOpen
  \bibfield  {author} {\bibinfo {author} {\bibfnamefont {S.-B.}\ \bibnamefont
  {Zhang}}\ and\ \bibinfo {author} {\bibfnamefont {J.}~\bibnamefont {Zhou}},\
  }\href {\doibase 10.1103/PhysRevB.101.085202} {\bibfield  {journal} {\bibinfo
   {journal} {Phys. Rev. B}\ }\textbf {\bibinfo {volume} {101}},\ \bibinfo
  {pages} {085202} (\bibinfo {year} {2020})}\BibitemShut {NoStop}%
\bibitem [{\citenamefont {Wang}\ \emph
  {et~al.}(2017{\natexlab{b}})\citenamefont {Wang}, \citenamefont {Jo},
  \citenamefont {Wu}, \citenamefont {Wu}, \citenamefont {Kaminski},
  \citenamefont {Canfield},\ and\ \citenamefont
  {Johnson}}]{PhysRevB.95.165114}%
  \BibitemOpen
  \bibfield  {author} {\bibinfo {author} {\bibfnamefont {L.-L.}\ \bibnamefont
  {Wang}}, \bibinfo {author} {\bibfnamefont {N.~H.}\ \bibnamefont {Jo}},
  \bibinfo {author} {\bibfnamefont {Y.}~\bibnamefont {Wu}}, \bibinfo {author}
  {\bibfnamefont {Q.}~\bibnamefont {Wu}}, \bibinfo {author} {\bibfnamefont
  {A.}~\bibnamefont {Kaminski}}, \bibinfo {author} {\bibfnamefont {P.~C.}\
  \bibnamefont {Canfield}}, \ and\ \bibinfo {author} {\bibfnamefont {D.~D.}\
  \bibnamefont {Johnson}},\ }\href {\doibase 10.1103/PhysRevB.95.165114}
  {\bibfield  {journal} {\bibinfo  {journal} {Phys. Rev. B}\ }\textbf {\bibinfo
  {volume} {95}},\ \bibinfo {pages} {165114} (\bibinfo {year}
  {2017}{\natexlab{b}})}\BibitemShut {NoStop}%
\bibitem [{\citenamefont {Gordon}\ and\ \citenamefont
  {Kee}(2018)}]{PhysRevB.97.195106}%
  \BibitemOpen
  \bibfield  {author} {\bibinfo {author} {\bibfnamefont {J.~S.}\ \bibnamefont
  {Gordon}}\ and\ \bibinfo {author} {\bibfnamefont {H.-Y.}\ \bibnamefont
  {Kee}},\ }\href {\doibase 10.1103/PhysRevB.97.195106} {\bibfield  {journal}
  {\bibinfo  {journal} {Phys. Rev. B}\ }\textbf {\bibinfo {volume} {97}},\
  \bibinfo {pages} {195106} (\bibinfo {year} {2018})}\BibitemShut {NoStop}%
\bibitem [{\citenamefont {Garate}(2013)}]{PhysRevLett.110.046402}%
  \BibitemOpen
  \bibfield  {author} {\bibinfo {author} {\bibfnamefont {I.}~\bibnamefont
  {Garate}},\ }\href {\doibase 10.1103/PhysRevLett.110.046402} {\bibfield
  {journal} {\bibinfo  {journal} {Phys. Rev. Lett.}\ }\textbf {\bibinfo
  {volume} {110}},\ \bibinfo {pages} {046402} (\bibinfo {year}
  {2013})}\BibitemShut {NoStop}%
\bibitem [{\citenamefont {Rinkel}\ \emph {et~al.}(2017)\citenamefont {Rinkel},
  \citenamefont {Lopes},\ and\ \citenamefont
  {Garate}}]{PhysRevLett.119.107401}%
  \BibitemOpen
  \bibfield  {author} {\bibinfo {author} {\bibfnamefont {P.}~\bibnamefont
  {Rinkel}}, \bibinfo {author} {\bibfnamefont {P.~L.~S.}\ \bibnamefont
  {Lopes}}, \ and\ \bibinfo {author} {\bibfnamefont {I.}~\bibnamefont
  {Garate}},\ }\href {\doibase 10.1103/PhysRevLett.119.107401} {\bibfield
  {journal} {\bibinfo  {journal} {Phys. Rev. Lett.}\ }\textbf {\bibinfo
  {volume} {119}},\ \bibinfo {pages} {107401} (\bibinfo {year}
  {2017})}\BibitemShut {NoStop}%
\bibitem [{\citenamefont {Sengupta}\ \emph {et~al.}(2020)\citenamefont
  {Sengupta}, \citenamefont {Lhachemi},\ and\ \citenamefont
  {Garate}}]{PhysRevLett.125.146402}%
  \BibitemOpen
  \bibfield  {author} {\bibinfo {author} {\bibfnamefont {S.}~\bibnamefont
  {Sengupta}}, \bibinfo {author} {\bibfnamefont {M.~N.~Y.}\ \bibnamefont
  {Lhachemi}}, \ and\ \bibinfo {author} {\bibfnamefont {I.}~\bibnamefont
  {Garate}},\ }\href {\doibase 10.1103/PhysRevLett.125.146402} {\bibfield
  {journal} {\bibinfo  {journal} {Phys. Rev. Lett.}\ }\textbf {\bibinfo
  {volume} {125}},\ \bibinfo {pages} {146402} (\bibinfo {year}
  {2020})}\BibitemShut {NoStop}%
\bibitem [{\citenamefont {Zhao}\ \emph {et~al.}(2021)\citenamefont {Zhao},
  \citenamefont {Lu},\ and\ \citenamefont {Xie}}]{Zhao2020}%
  \BibitemOpen
  \bibfield  {author} {\bibinfo {author} {\bibfnamefont {P.-L.}\ \bibnamefont
  {Zhao}}, \bibinfo {author} {\bibfnamefont {H.-Z.}\ \bibnamefont {Lu}}, \ and\
  \bibinfo {author} {\bibfnamefont {X.~C.}\ \bibnamefont {Xie}},\ }\href
  {\doibase 10.1103/PhysRevLett.127.046602} {\bibfield  {journal} {\bibinfo
  {journal} {Phys. Rev. Lett.}\ }\textbf {\bibinfo {volume} {127}},\ \bibinfo
  {pages} {046602} (\bibinfo {year} {2021})}\BibitemShut {NoStop}%
\bibitem [{\citenamefont {Qin}\ \emph {et~al.}(2020)\citenamefont {Qin},
  \citenamefont {Li}, \citenamefont {Du}, \citenamefont {Wang}, \citenamefont
  {Zhang}, \citenamefont {Yu}, \citenamefont {Lu},\ and\ \citenamefont
  {Xie}}]{Qin2020}%
  \BibitemOpen
  \bibfield  {author} {\bibinfo {author} {\bibfnamefont {F.}~\bibnamefont
  {Qin}}, \bibinfo {author} {\bibfnamefont {S.}~\bibnamefont {Li}}, \bibinfo
  {author} {\bibfnamefont {Z.~Z.}\ \bibnamefont {Du}}, \bibinfo {author}
  {\bibfnamefont {C.~M.}\ \bibnamefont {Wang}}, \bibinfo {author}
  {\bibfnamefont {W.}~\bibnamefont {Zhang}}, \bibinfo {author} {\bibfnamefont
  {D.}~\bibnamefont {Yu}}, \bibinfo {author} {\bibfnamefont {H.-Z.}\
  \bibnamefont {Lu}}, \ and\ \bibinfo {author} {\bibfnamefont {X.~C.}\
  \bibnamefont {Xie}},\ }\href {\doibase 10.1103/PhysRevLett.125.206601}
  {\bibfield  {journal} {\bibinfo  {journal} {Phys. Rev. Lett.}\ }\textbf
  {\bibinfo {volume} {125}},\ \bibinfo {pages} {206601} (\bibinfo {year}
  {2020})}\BibitemShut {NoStop}%
\bibitem [{\citenamefont {Okada}\ \emph {et~al.}(1982)\citenamefont {Okada},
  \citenamefont {Sambongi}, \citenamefont {Ido}, \citenamefont {Tazuke},
  \citenamefont {Aoki},\ and\ \citenamefont {Fujita}}]{Okada1982}%
  \BibitemOpen
  \bibfield  {author} {\bibinfo {author} {\bibfnamefont {S.}~\bibnamefont
  {Okada}}, \bibinfo {author} {\bibfnamefont {T.}~\bibnamefont {Sambongi}},
  \bibinfo {author} {\bibfnamefont {M.}~\bibnamefont {Ido}}, \bibinfo {author}
  {\bibfnamefont {Y.}~\bibnamefont {Tazuke}}, \bibinfo {author} {\bibfnamefont
  {R.}~\bibnamefont {Aoki}}, \ and\ \bibinfo {author} {\bibfnamefont
  {O.}~\bibnamefont {Fujita}},\ }\href {\doibase 10.1143/JPSJ.51.460}
  {\bibfield  {journal} {\bibinfo  {journal} {Journal of the Physical Society
  of Japan}\ }\textbf {\bibinfo {volume} {51}},\ \bibinfo {pages} {460}
  (\bibinfo {year} {1982})},\ \Eprint
  {http://arxiv.org/abs/https://doi.org/10.1143/JPSJ.51.460}
  {https://doi.org/10.1143/JPSJ.51.460} \BibitemShut {NoStop}%
\bibitem [{\citenamefont {Tian}\ \emph {et~al.}(2021)\citenamefont {Tian},
  \citenamefont {Ghassemi},\ and\ \citenamefont {Ross}}]{Tian2021}%
  \BibitemOpen
  \bibfield  {author} {\bibinfo {author} {\bibfnamefont {Y.}~\bibnamefont
  {Tian}}, \bibinfo {author} {\bibfnamefont {N.}~\bibnamefont {Ghassemi}}, \
  and\ \bibinfo {author} {\bibfnamefont {J.~H.}\ \bibnamefont {Ross}},\ }\href
  {\doibase 10.1103/PhysRevLett.126.236401} {\bibfield  {journal} {\bibinfo
  {journal} {Phys. Rev. Lett.}\ }\textbf {\bibinfo {volume} {126}},\ \bibinfo
  {pages} {236401} (\bibinfo {year} {2021})}\BibitemShut {NoStop}%
\bibitem [{\citenamefont {Galeski}\ \emph {et~al.}(2021)\citenamefont
  {Galeski}, \citenamefont {Ehmcke}, \citenamefont {Wawrzynczak}, \citenamefont
  {Lozano}, \citenamefont {Cho}, \citenamefont {Sharma}, \citenamefont {Das},
  \citenamefont {Kuster}, \citenamefont {Sessi}, \citenamefont {Brando},
  \citenamefont {Kuchler}, \citenamefont {Markou}, \citenamefont {Konig},
  \citenamefont {Swekis}, \citenamefont {Felser}, \citenamefont {Sassa},
  \citenamefont {Li}, \citenamefont {Gu}, \citenamefont {Zimmermann},
  \citenamefont {Ivashko}, \citenamefont {Gorbunov}, \citenamefont
  {Zherlitsyn}, \citenamefont {Forster}, \citenamefont {Parkin}, \citenamefont
  {Wosnitza}, \citenamefont {Meng},\ and\ \citenamefont {Gooth}}]{Galeski2021}%
  \BibitemOpen
  \bibfield  {author} {\bibinfo {author} {\bibfnamefont {S.}~\bibnamefont
  {Galeski}}, \bibinfo {author} {\bibfnamefont {T.}~\bibnamefont {Ehmcke}},
  \bibinfo {author} {\bibfnamefont {R.}~\bibnamefont {Wawrzynczak}}, \bibinfo
  {author} {\bibfnamefont {P.~M.}\ \bibnamefont {Lozano}}, \bibinfo {author}
  {\bibfnamefont {K.}~\bibnamefont {Cho}}, \bibinfo {author} {\bibfnamefont
  {A.}~\bibnamefont {Sharma}}, \bibinfo {author} {\bibfnamefont
  {S.}~\bibnamefont {Das}}, \bibinfo {author} {\bibfnamefont {F.}~\bibnamefont
  {Kuster}}, \bibinfo {author} {\bibfnamefont {P.}~\bibnamefont {Sessi}},
  \bibinfo {author} {\bibfnamefont {M.}~\bibnamefont {Brando}}, \bibinfo
  {author} {\bibfnamefont {R.}~\bibnamefont {Kuchler}}, \bibinfo {author}
  {\bibfnamefont {A.}~\bibnamefont {Markou}}, \bibinfo {author} {\bibfnamefont
  {M.}~\bibnamefont {Konig}}, \bibinfo {author} {\bibfnamefont
  {P.}~\bibnamefont {Swekis}}, \bibinfo {author} {\bibfnamefont
  {C.}~\bibnamefont {Felser}}, \bibinfo {author} {\bibfnamefont
  {Y.}~\bibnamefont {Sassa}}, \bibinfo {author} {\bibfnamefont
  {Q.}~\bibnamefont {Li}}, \bibinfo {author} {\bibfnamefont {G.}~\bibnamefont
  {Gu}}, \bibinfo {author} {\bibfnamefont {M.~V.}\ \bibnamefont {Zimmermann}},
  \bibinfo {author} {\bibfnamefont {O.}~\bibnamefont {Ivashko}}, \bibinfo
  {author} {\bibfnamefont {D.~I.}\ \bibnamefont {Gorbunov}}, \bibinfo {author}
  {\bibfnamefont {S.}~\bibnamefont {Zherlitsyn}}, \bibinfo {author}
  {\bibfnamefont {T.}~\bibnamefont {Forster}}, \bibinfo {author} {\bibfnamefont
  {S.~S.~P.}\ \bibnamefont {Parkin}}, \bibinfo {author} {\bibfnamefont
  {J.}~\bibnamefont {Wosnitza}}, \bibinfo {author} {\bibfnamefont
  {T.}~\bibnamefont {Meng}}, \ and\ \bibinfo {author} {\bibfnamefont
  {J.}~\bibnamefont {Gooth}},\ }\href {\doibase
  https://doi.org/10.1038/s41467-021-23435-y} {\bibfield  {journal} {\bibinfo
  {journal} {Nat. Commun.}\ }\textbf {\bibinfo {volume} {12}},\ \bibinfo
  {pages} {3197} (\bibinfo {year} {2021})}\BibitemShut {NoStop}%
\bibitem [{\citenamefont {Ehmcke}\ \emph {et~al.}(2021)\citenamefont {Ehmcke},
  \citenamefont {Galeski}, \citenamefont {Gorbunov}, \citenamefont
  {Zherlitsyn}, \citenamefont {Wosnitza}, \citenamefont {Gooth},\ and\
  \citenamefont {Meng}}]{Ehmcke2021}%
  \BibitemOpen
  \bibfield  {author} {\bibinfo {author} {\bibfnamefont {T.}~\bibnamefont
  {Ehmcke}}, \bibinfo {author} {\bibfnamefont {S.}~\bibnamefont {Galeski}},
  \bibinfo {author} {\bibfnamefont {D.}~\bibnamefont {Gorbunov}}, \bibinfo
  {author} {\bibfnamefont {S.}~\bibnamefont {Zherlitsyn}}, \bibinfo {author}
  {\bibfnamefont {J.}~\bibnamefont {Wosnitza}}, \bibinfo {author}
  {\bibfnamefont {J.}~\bibnamefont {Gooth}}, \ and\ \bibinfo {author}
  {\bibfnamefont {T.}~\bibnamefont {Meng}},\ }\href@noop {} {\enquote {\bibinfo
  {title} {Propagation of longitudinal acoustic phonons in zrte$_5$ exposed to
  a quantizing magnetic field},}\ } (\bibinfo {year} {2021}),\ \Eprint
  {http://arxiv.org/abs/2109.03738} {arXiv:2109.03738 [cond-mat.str-el]}
  \BibitemShut {NoStop}%
\bibitem [{\citenamefont {Bourbonnais}\ and\ \citenamefont
  {Caron}(1989)}]{Bourbon89}%
  \BibitemOpen
  \bibfield  {author} {\bibinfo {author} {\bibfnamefont {C.}~\bibnamefont
  {Bourbonnais}}\ and\ \bibinfo {author} {\bibfnamefont {L.~G.}\ \bibnamefont
  {Caron}},\ }\href {\doibase 10.1051/jphys:0198900500180275100} {\bibfield
  {journal} {\bibinfo  {journal} {J. Phys. France}\ }\textbf {\bibinfo {volume}
  {50}},\ \bibinfo {pages} {2751} (\bibinfo {year} {1989})}\BibitemShut
  {NoStop}%
\bibitem [{\citenamefont {Bourbonnais}\ and\ \citenamefont
  {Caron}(1991)}]{Bourbon91}%
  \BibitemOpen
  \bibfield  {author} {\bibinfo {author} {\bibfnamefont {C.}~\bibnamefont
  {Bourbonnais}}\ and\ \bibinfo {author} {\bibfnamefont {L.~G.}\ \bibnamefont
  {Caron}},\ }\href {\doibase 10.1142/S0217979291000547} {\bibfield  {journal}
  {\bibinfo  {journal} {Int. J. Mod. Phys. B}\ }\textbf {\bibinfo {volume}
  {05}},\ \bibinfo {pages} {1033} (\bibinfo {year} {1991})}\BibitemShut
  {NoStop}%
\bibitem [{\citenamefont {Zhang}\ and\ \citenamefont
  {Shindou}(2017)}]{Zhang2017}%
  \BibitemOpen
  \bibfield  {author} {\bibinfo {author} {\bibfnamefont {X.-T.}\ \bibnamefont
  {Zhang}}\ and\ \bibinfo {author} {\bibfnamefont {R.}~\bibnamefont
  {Shindou}},\ }\href {\doibase 10.1103/PhysRevB.95.205108} {\bibfield
  {journal} {\bibinfo  {journal} {Phys. Rev. B}\ }\textbf {\bibinfo {volume}
  {95}},\ \bibinfo {pages} {205108} (\bibinfo {year} {2017})}\BibitemShut
  {NoStop}%
\bibitem [{\citenamefont {Chan}\ and\ \citenamefont {Lee}(2017)}]{Chan2017}%
  \BibitemOpen
  \bibfield  {author} {\bibinfo {author} {\bibfnamefont {C.-K.}\ \bibnamefont
  {Chan}}\ and\ \bibinfo {author} {\bibfnamefont {P.~A.}\ \bibnamefont {Lee}},\
  }\href {\doibase 10.1103/PhysRevB.96.195143} {\bibfield  {journal} {\bibinfo
  {journal} {Phys. Rev. B}\ }\textbf {\bibinfo {volume} {96}},\ \bibinfo
  {pages} {195143} (\bibinfo {year} {2017})}\BibitemShut {NoStop}%
\bibitem [{\citenamefont {Saykin}\ \emph {et~al.}(2018)\citenamefont {Saykin},
  \citenamefont {Tikhonov},\ and\ \citenamefont {Rodionov}}]{Saykin2018}%
  \BibitemOpen
  \bibfield  {author} {\bibinfo {author} {\bibfnamefont {D.~R.}\ \bibnamefont
  {Saykin}}, \bibinfo {author} {\bibfnamefont {K.~S.}\ \bibnamefont
  {Tikhonov}}, \ and\ \bibinfo {author} {\bibfnamefont {Y.~I.}\ \bibnamefont
  {Rodionov}},\ }\href {\doibase 10.1103/PhysRevB.97.041202} {\bibfield
  {journal} {\bibinfo  {journal} {Phys. Rev. B}\ }\textbf {\bibinfo {volume}
  {97}},\ \bibinfo {pages} {041202} (\bibinfo {year} {2018})}\BibitemShut
  {NoStop}%
\bibitem [{\citenamefont {Kaasbjerg}\ \emph {et~al.}(2012)\citenamefont
  {Kaasbjerg}, \citenamefont {Thygesen},\ and\ \citenamefont
  {Jacobsen}}]{Kaasbjerg2012}%
  \BibitemOpen
  \bibfield  {author} {\bibinfo {author} {\bibfnamefont {K.}~\bibnamefont
  {Kaasbjerg}}, \bibinfo {author} {\bibfnamefont {K.~S.}\ \bibnamefont
  {Thygesen}}, \ and\ \bibinfo {author} {\bibfnamefont {K.~W.}\ \bibnamefont
  {Jacobsen}},\ }\href {\doibase 10.1103/PhysRevB.85.115317} {\bibfield
  {journal} {\bibinfo  {journal} {Phys. Rev. B}\ }\textbf {\bibinfo {volume}
  {85}},\ \bibinfo {pages} {115317} (\bibinfo {year} {2012})}\BibitemShut
  {NoStop}%
\bibitem [{\citenamefont {Mahan}(2000)}]{Mahan2000b}%
  \BibitemOpen
  \bibfield  {author} {\bibinfo {author} {\bibfnamefont {G.~D.}\ \bibnamefont
  {Mahan}},\ }\href {"https://www.springer.com/gp/book/9780306463389"} {\emph
  {\bibinfo {title} {Many-Particle Physics}}}\ (\bibinfo  {publisher}
  {Springer},\ \bibinfo {address} {New York},\ \bibinfo {year}
  {2000})\BibitemShut {NoStop}%
\bibitem [{\citenamefont {MacDonald}\ \emph {et~al.}(1986)\citenamefont
  {MacDonald}, \citenamefont {Oji},\ and\ \citenamefont {Liu}}]{MacDonald1986}%
  \BibitemOpen
  \bibfield  {author} {\bibinfo {author} {\bibfnamefont {A.~H.}\ \bibnamefont
  {MacDonald}}, \bibinfo {author} {\bibfnamefont {H.~C.~A.}\ \bibnamefont
  {Oji}}, \ and\ \bibinfo {author} {\bibfnamefont {K.~L.}\ \bibnamefont
  {Liu}},\ }\href {\doibase 10.1103/PhysRevB.34.2681} {\bibfield  {journal}
  {\bibinfo  {journal} {Phys. Rev. B}\ }\textbf {\bibinfo {volume} {34}},\
  \bibinfo {pages} {2681} (\bibinfo {year} {1986})}\BibitemShut {NoStop}%
\bibitem [{\citenamefont {Goerbig}(2011)}]{Goerbig2011}%
  \BibitemOpen
  \bibfield  {author} {\bibinfo {author} {\bibfnamefont {M.~O.}\ \bibnamefont
  {Goerbig}},\ }\href {\doibase 10.1103/RevModPhys.83.1193} {\bibfield
  {journal} {\bibinfo  {journal} {Rev. Mod. Phys.}\ }\textbf {\bibinfo {volume}
  {83}},\ \bibinfo {pages} {1193} (\bibinfo {year} {2011})}\BibitemShut
  {NoStop}%
\bibitem [{\citenamefont {Fowler}(1976)}]{Fowler76}%
  \BibitemOpen
  \bibfield  {author} {\bibinfo {author} {\bibfnamefont {M.}~\bibnamefont
  {Fowler}},\ }\href {\doibase https://doi.org/10.1016/0038-1098(76)91462-9}
  {\bibfield  {journal} {\bibinfo  {journal} {Solid State Communications}\
  }\textbf {\bibinfo {volume} {18}},\ \bibinfo {pages} {241} (\bibinfo {year}
  {1976})}\BibitemShut {NoStop}%
\bibitem [{\citenamefont {Barisi\'{c}}(1983)}]{barisic83}%
  \BibitemOpen
  \bibfield  {author} {\bibinfo {author} {\bibfnamefont {S.}~\bibnamefont
  {Barisi\'{c}}},\ }\href {\doibase 0.1051/jphys:01983004402018500} {\bibfield
  {journal} {\bibinfo  {journal} {J. Phys. France}\ }\textbf {\bibinfo {volume}
  {44}},\ \bibinfo {pages} {185} (\bibinfo {year} {1983})}\BibitemShut
  {NoStop}%
\bibitem [{\citenamefont {Caron}\ and\ \citenamefont
  {Bourbonnais}(1984)}]{Caron84}%
  \BibitemOpen
  \bibfield  {author} {\bibinfo {author} {\bibfnamefont {L.~G.}\ \bibnamefont
  {Caron}}\ and\ \bibinfo {author} {\bibfnamefont {C.}~\bibnamefont
  {Bourbonnais}},\ }\href {\doibase 10.1103/PhysRevB.29.4230} {\bibfield
  {journal} {\bibinfo  {journal} {Phys. Rev. B}\ }\textbf {\bibinfo {volume}
  {29}},\ \bibinfo {pages} {4230} (\bibinfo {year} {1984})}\BibitemShut
  {NoStop}%
\bibitem [{\citenamefont {Hirsch}\ and\ \citenamefont
  {Fradkin}(1983)}]{Hirsch83}%
  \BibitemOpen
  \bibfield  {author} {\bibinfo {author} {\bibfnamefont {J.~E.}\ \bibnamefont
  {Hirsch}}\ and\ \bibinfo {author} {\bibfnamefont {E.}~\bibnamefont
  {Fradkin}},\ }\href {\doibase 10.1103/PhysRevB.27.4302} {\bibfield  {journal}
  {\bibinfo  {journal} {Phys. Rev. B}\ }\textbf {\bibinfo {volume} {27}},\
  \bibinfo {pages} {4302} (\bibinfo {year} {1983})}\BibitemShut {NoStop}%
\bibitem [{\citenamefont {Aroyo}\ \emph
  {et~al.}(2006{\natexlab{a}})\citenamefont {Aroyo}, \citenamefont
  {Perez-Mato}, \citenamefont {Capillas}, \citenamefont {Kroumova},
  \citenamefont {Ivantchev}, \citenamefont {Madariaga}, \citenamefont {Kirov},\
  and\ \citenamefont {Wondratschek}}]{Aroyo2006a}%
  \BibitemOpen
  \bibfield  {author} {\bibinfo {author} {\bibfnamefont {M.~I.}\ \bibnamefont
  {Aroyo}}, \bibinfo {author} {\bibfnamefont {J.~M.}\ \bibnamefont
  {Perez-Mato}}, \bibinfo {author} {\bibfnamefont {C.}~\bibnamefont
  {Capillas}}, \bibinfo {author} {\bibfnamefont {E.}~\bibnamefont {Kroumova}},
  \bibinfo {author} {\bibfnamefont {S.}~\bibnamefont {Ivantchev}}, \bibinfo
  {author} {\bibfnamefont {G.}~\bibnamefont {Madariaga}}, \bibinfo {author}
  {\bibfnamefont {A.}~\bibnamefont {Kirov}}, \ and\ \bibinfo {author}
  {\bibfnamefont {H.}~\bibnamefont {Wondratschek}},\ }\href {\doibase
  doi:10.1524/zkri.2006.221.1.15} {\bibfield  {journal} {\bibinfo  {journal}
  {Zeitschrift fur Kristallographie - Crystalline Materials}\ }\textbf
  {\bibinfo {volume} {221}},\ \bibinfo {pages} {15} (\bibinfo {year}
  {2006}{\natexlab{a}})}\BibitemShut {NoStop}%
\bibitem [{\citenamefont {Aroyo}\ \emph
  {et~al.}(2006{\natexlab{b}})\citenamefont {Aroyo}, \citenamefont {Kirov},
  \citenamefont {Capillas}, \citenamefont {Perez-Mato},\ and\ \citenamefont
  {Wondratschek}}]{Aroyo2006b}%
  \BibitemOpen
  \bibfield  {author} {\bibinfo {author} {\bibfnamefont {M.~I.}\ \bibnamefont
  {Aroyo}}, \bibinfo {author} {\bibfnamefont {A.}~\bibnamefont {Kirov}},
  \bibinfo {author} {\bibfnamefont {C.}~\bibnamefont {Capillas}}, \bibinfo
  {author} {\bibfnamefont {J.~M.}\ \bibnamefont {Perez-Mato}}, \ and\ \bibinfo
  {author} {\bibfnamefont {H.}~\bibnamefont {Wondratschek}},\ }\href {\doibase
  10.1107/S0108767305040286} {\bibfield  {journal} {\bibinfo  {journal} {Acta
  Crystallographica Section A}\ }\textbf {\bibinfo {volume} {62}},\ \bibinfo
  {pages} {115} (\bibinfo {year} {2006}{\natexlab{b}})}\BibitemShut {NoStop}%
\bibitem [{\citenamefont {Dresselhaus}\ \emph {et~al.}(2008)\citenamefont
  {Dresselhaus}, \citenamefont {Dresselhaus},\ and\ \citenamefont
  {Jorio}}]{dresselhaus2008}%
  \BibitemOpen
  \bibfield  {author} {\bibinfo {author} {\bibfnamefont {M.~D.}\ \bibnamefont
  {Dresselhaus}}, \bibinfo {author} {\bibfnamefont {G.}~\bibnamefont
  {Dresselhaus}}, \ and\ \bibinfo {author} {\bibfnamefont {A.}~\bibnamefont
  {Jorio}},\ }\href {https://www.springer.com/gp/book/9783540328971} {\emph
  {\bibinfo {title} {Group Theory: Application to the Physics of Condensed
  Matter}}}\ (\bibinfo  {publisher} {Springer-Verlag},\ \bibinfo {address}
  {Berlin},\ \bibinfo {year} {2008})\BibitemShut {NoStop}%
\bibitem [{\citenamefont {Pouget}(2016)}]{Pouget2015}%
  \BibitemOpen
  \bibfield  {author} {\bibinfo {author} {\bibfnamefont {J.-P.}\ \bibnamefont
  {Pouget}},\ }\href {\doibase https://doi.org/10.1016/j.crhy.2015.11.008}
  {\bibfield  {journal} {\bibinfo  {journal} {Comptes Rendus Physique}\
  }\textbf {\bibinfo {volume} {17}},\ \bibinfo {pages} {332} (\bibinfo {year}
  {2016})}\BibitemShut {NoStop}%
\bibitem [{\citenamefont {Lebed}(2008)}]{Lebed08}%
  \BibitemOpen
  \bibinfo {editor} {\bibfnamefont {A.}~\bibnamefont {Lebed}},\ ed.,\ \href
  {\doibase 10.1007/978-3-540-76672-8} {\emph {\bibinfo {title} {The Physics of
  Organic Superconductors and Conductors}}},\ Vol.\ \bibinfo {volume} {110}\
  (\bibinfo  {publisher} {Springer Series in Materials Science},\ \bibinfo
  {address} {Heidelberg},\ \bibinfo {year} {2008})\BibitemShut {NoStop}%
\bibitem [{\citenamefont {Hong}\ \emph {et~al.}(1996)\citenamefont {Hong},
  \citenamefont {Kim},\ and\ \citenamefont {Sin}}]{Hong1996}%
  \BibitemOpen
  \bibfield  {author} {\bibinfo {author} {\bibfnamefont {D.~K.}\ \bibnamefont
  {Hong}}, \bibinfo {author} {\bibfnamefont {Y.}~\bibnamefont {Kim}}, \ and\
  \bibinfo {author} {\bibfnamefont {S.-J.}\ \bibnamefont {Sin}},\ }\href
  {\doibase 10.1103/PhysRevD.54.7879} {\bibfield  {journal} {\bibinfo
  {journal} {Phys. Rev. D}\ }\textbf {\bibinfo {volume} {54}},\ \bibinfo
  {pages} {7879} (\bibinfo {year} {1996})}\BibitemShut {NoStop}%
\bibitem [{\citenamefont {Hattori}\ \emph {et~al.}(2017)\citenamefont
  {Hattori}, \citenamefont {Itakura},\ and\ \citenamefont
  {Ozaki}}]{Hattori2017}%
  \BibitemOpen
  \bibfield  {author} {\bibinfo {author} {\bibfnamefont {K.}~\bibnamefont
  {Hattori}}, \bibinfo {author} {\bibfnamefont {K.}~\bibnamefont {Itakura}}, \
  and\ \bibinfo {author} {\bibfnamefont {S.}~\bibnamefont {Ozaki}},\ }\href
  {\doibase https://doi.org/10.1016/j.physletb.2017.11.004} {\bibfield
  {journal} {\bibinfo  {journal} {Physics Letters B}\ }\textbf {\bibinfo
  {volume} {775}},\ \bibinfo {pages} {283} (\bibinfo {year}
  {2017})}\BibitemShut {NoStop}%
\end{thebibliography}
%


\appendix
\begin{widetext}
\appendix

\section{Derivation of renormalization group equations in the quantum limit}

In this section, we derive the recursion relations for the phonon self-energy,
the electron-phonon coupling and the electron-electron
couplings, up to one-loop order. 
We focus on the case in which the Fermi level crosses only the chiral Landau levels (in Appendix B, we treat the case in which the Fermi level intersects with one nonchiral Landau level).
We work under the assumption that the couplings are approximately local in the transverse
directions, within the magnetic length $l_{B}$, above which they
are exponentially suppressed. In practice, this translates into
the approximation $f(q_x l_B, q_y l_B)\exp[-q_{\bot}^{2}l_{B}^{2}/2]\simeq 0$ for $q_{\bot}l_{B}\gtrsim1$,
and $f(q_x l_B, q_y l_B) \exp[-q_{\bot}^{2}l_{B}^{2}/2]\simeq f(0,0)$ for $q_{\bot}l_{B}\lesssim1$, where $f$ is some function of the transverse momentum ${\bf q}_\perp=(q_x,q_y)$.

With the aforementioned approximation, the action for the (internode) electron-phonon interaction is given by 
\begin{equation}
S_{ep}[\psi^{\dagger},\psi,\phi]\simeq-\sqrt{\frac{\pi v_{F}}{\beta\mathcal{V}}}\sum_{\omega_{n},\omega_{m}}\sum_{k,\mathbf{q},j,X,\tau}z_{\tau\overline{\tau},j}g_{\tau\overline{\tau},j}^{ep}(\mathbf{q})\exp[iq_{x}X]\psi_{X,k+q-b_{\tau},\tau}^{\dagger}(\omega_{n}+\omega_{m})\psi_{X+q_{y}l_{B}^{2},k,\overline{\tau}}(\omega_{n})\phi_{j}(\mathbf{q},\omega_{m}),\label{eq:19-1}
\end{equation}
where $b_{\tau}=k_{\tau}-k_{\overline{\tau}}$ refers to the momentum separation between the two Weyl nodes of opposite chiralities, $z_{\tau\overline{\tau},j}$
stands as a renormalization factor for the internode electron-phonon coupling
for phonon mode $j$, $\tau$ refers to the Weyl node ($\tau=\pm 1$), $\psi$ and
$\psi^{\dagger}$ are Grassmann fields for electrons, and $\phi$ is the phonon
displacement field for mode $j$ with wave vector $\mathbf{q}$. 

The action of the electron-electron interactions is divided into two parts, corresponding to backward Coulomb scattering (of amplitude $g_1$)  and forward Coulomb scattering (of amplitude $g_2$):
\begin{align}
S_{ee}[\psi^{\dagger},\psi] &\simeq
-g_1\frac{\pi v_{F}}{2 \beta\mathcal{V}}\sum_{X,Y,\{\omega_{n}\}}\sum_{k_{1},k_{2},\mathbf{q}',\tau}\exp[iq_{x}'X-iq_{x}'Y]  \nonumber \\
&~~~~~~~~~~~~~~\times \psi_{X,k_{1}+q',\tau}^{\dagger}(\omega_{n_{1}})\psi_{Y,k_{2}-q',\overline{\tau}}^{\dagger}(\omega_{n_{2}})\psi_{Y-q'_{y}l_{B}^{2},k_{2},\tau}(\omega_{n_{3}})\psi_{X+q'_{y}l_{B}^{2},k_{1},\overline{\tau}}(\omega_{n_{1}}+\omega_{n_{2}}-\omega_{n_{3}})\, \nonumber\\
&-g_2\frac{\pi v_{F}}{2\beta\mathcal{V}}\sum_{X,Y,\{\omega_{n}\}}\sum_{k_{1},k_{2},\mathbf{q}',\tau}\exp[iq'_{x}X-iq'_{x}Y] \nonumber \\
&~~~~~~~~~~~~~~\times \psi_{X,k_{1}+q',\tau}^{\dagger}(\omega_{n_{1}})\psi_{Y,k_{2}-q',\overline{\tau}}^{\dagger}(\omega_{n_{2}})\psi_{Y-q'_{y}l_{B}^{2},k_{2},\overline{\tau}}(\omega_{n_{3}})\psi_{X+q'_{y}l_{B}^{2},k_{1},\tau}(\omega_{n_{1}}+\omega_{n_{2}}-\omega_{n_{3}}),
\end{align}
where, in the first term, $|q'|\simeq|k_{F+}-k_{F-}|-|k_{+}-k_{-}|\equiv 2k_{F}^{\prime}$ is a small momentum, $k_{\pm}$ being the positions of the two nodes and $k_{F\pm}$ being the corresponding Fermi momenta (see Fig. \ref{fig:ql}).
In the expressions that follow, we simplify the notation by often omitting the fermionic Matsubara frequencies associated with the Grassmann and phonon fields.

\subsection{One-loop corrections to the electron-phonon vertex}

Let us derive the RG transformation for the internode electron-phonon
vertex $z_{\tau\bar{\tau}, j}$, resulting from
forward and backward scattering amplitudes 
$g_{2}$ and $g_{1}$.
The one-loop correction to $z_{\tau\bar{\tau}, j}$ follows from $\langle S_{ep}S_{ee}\rangle_{dl}$ (see Fig. \ref{fig_RG1} and Eq. (\ref{zeph0}) in the main text).
The latter can be subdivided into contributions coming from $g_1$ and $g_2$ alone:
\begin{equation}
\langle S_{ep}S_{ee}\rangle_{dl} \equiv  \langle S_{ep}S_{ee}\rangle_{dl} (g_1) +  \langle S_{ep}S_{ee}\rangle_{dl} (g_2).
\end{equation}

First, we focus on
\begin{align}
\label{eq:sep0}
\langle S_{ep}S_{ee}\rangle_{dl} (g_2)&\simeq 
\left(\frac{\pi v_{F}}{\beta\mathcal{V}}\right)^{3/2}\sum_{\tau,X,X^{\prime},Y^{\prime}}\sum_{k,\mathbf{q},j}\sum_{k_{1}^{\prime},k_{2}^{\prime},\mathbf{q}^{\prime}} z_{\tau\overline{\tau},j}g_{\tau\overline{\tau},j}^{ep}g_{2}\exp[iq_{x}X]\exp[iq_{x}^{\prime}X^{\prime}-iq_{x}^{\prime}Y^{\prime}]   \nonumber\\
&~~~~~~~~~~~~~~~~~~~~~\times \langle\overline{\psi}_{X,k+q-b_{\tau},\tau}^{\dagger}\overline{\psi}_{X^{\prime}+q_{y}^{\prime}l_{B}^{2},k_{1}^{\prime},\tau}\rangle\langle\overline{\psi}_{Y^{\prime},k_{2}^{\prime}-q^{\prime},\overline{\tau}}^{\dagger}\overline{\psi}_{X+q_{y}l_{B}^{2},k,\overline{\tau}}\rangle\psi_{X^{\prime},k_{1}^{\prime}+q^{\prime},\tau}^{\dagger}\psi_{Y^{\prime}-q_{y}^{\prime}l_{B}^{2},k_{2}^{\prime},\overline{\tau}}\phi_{j}(\mathbf{q})  \nonumber\\
&=\left(\frac{\pi v_{F}}{\beta\mathcal{V}}\right)^{3/2}\sum_{\tau,X^{\prime}}\sum_{\mathbf{q},j}\sum_{q_{x}^{\prime},q_{y}^{\prime}}\sum_{k^{\prime},\omega_{m}} z_{\tau\overline{\tau},j}g_{\tau\overline{\tau},j}^{ep}g_{2}\exp[iq_{x}X^{\prime}] \nonumber\\
&~~~~~~~~~~~~~~~~\times  \sum_{\omega_{n}} \sumslashD_{k} \frac{1}{2}\left[G^{0}_{\tau}(k+q-b_{\tau},\omega_{n}+\omega_{m})G^{0}_{\overline{\tau}}(k,\omega_{n})+G^{0}_{\tau}(k,\omega_{n})G^{0}_{\overline{\tau}}(k-q+b_{\tau},\omega_{n}-\omega_{m})\right]\nonumber\\
&~~~~~~~~~~~~~~~~\times \psi_{X^{\prime},k^{\prime}+q-b_{\tau},\tau}^{\dagger}\psi_{X^{\prime}+q_{y}l_{B}^{2},k^{\prime},\overline{\tau}}\phi_{j}(\mathbf{q},\omega_{m}),
\end{align}
where $|q-b_{\tau}| \simeq 2k_{F}^{\prime}$ for phonons connecting the two Fermi points or nearby, 
\begin{equation}
G^{0}_{\tau}(k,\omega_{n})=(i\hbar \omega_{n}-\hbar v_{\tau} \tau (k-k_{F \tau}^{\prime}))^{-1}
\end{equation}
 is the Green's function for  a noninteracting fermion in the chiral Landau level, $k$ is the momentum measured with respect to the node position $k_{\tau}$ (see Eqs. (\ref{eq:LL_energies}) and (\ref{eq:freefer})), $k_{F \tau}^{\prime}=k_{F \tau}-k_{\tau}$ is the Fermi momentum measured from the node, $\overline{\psi}$ labels
 electronic degrees of freedom in the outer energy shells $[-\Lambda_{0}(l),-\Lambda_{0}(l+dl)]$ and $[\Lambda_{0}(l+dl),\Lambda_{0}(l)]$ above and below (respectively)  of the Fermi level (see Sec.\ref{subsec:method} in the main text), and $\sumslash_{k}$ denotes a summation over momenta in those energy shells.
Below, we convert the momentum sum into an energy integral via $(1/L_z)\sumslash_{k}\rightarrow (2\pi \hbar v_{F})^{-1} \intslash d\epsilon$.

In the second equality of Eq.~(\ref{eq:sep0}), we used a symmetric product of propagators in order to allow at least one fermion line in the outer energy shell at  non zero $\Delta$ and $\delta v$,  the other being tied in these conditions to a previous outer shell contraction. 

Then,  Eq.~(\ref{eq:sep0}) can be rewritten as
\begin{align}
\label{eq:sep1}
\langle S_{ep}S_{ee}\rangle_{dl} (g_2)= & -\sqrt{\frac{{\pi v_{F}}}{{\beta\mathcal{V}}}}\sum_{\tau,X^{\prime}}\sum_{\mathbf{q},j}\sum_{q_{x}^{\prime},q_{y}^{\prime}}\sum_{k^{\prime},\omega_{m}} z_{\tau\overline{\tau},j}g_{\tau\overline{\tau},j}^{ep}g_{2}\exp[iq_{x}X^{\prime}] I_{P}(dl) \psi_{X^{\prime},k^{\prime}+q-b_{\tau},\tau}^{\dagger}\psi_{X^{\prime}+q_{y}l_{B}^{2},k^{\prime},\overline{\tau}}\phi_{j}(\mathbf{q},\omega_{m})
\end{align}
with
\begin{align}
I_{P}(dl) &=-\frac{\pi v_{F}}{2 \beta L_z}\sum_{\omega_{n}} \sumslashD_{k}(G_{+}^{0}(k+2k_{F}^{\prime},\omega_{n})G_{-}^{0}(k,\omega_{n})+G_{+}^{0}(k,\omega_{n})G_{-}^{0}(k-2k_{F}^{\prime},\omega_{n})) \nonumber\\
&=\frac{dl}{2+\delta v/v_{F}}\frac{1}{4}\left\{2\tanh\left[\frac{\beta \Lambda_{0}(l)}{2}\right] +\tanh\left[\frac{\beta \Lambda_{0}(l)(1+\delta v/v_{F})}{2}\right]+\tanh\left[\frac{\beta \Lambda_{0}(l)}{2 (1+\delta v/v_{F})}\right]\right\} \nonumber\\
&\equiv \frac{dl}{2+\delta v/v_{F}} \lambda_{P}(l,T,\delta v),  \label{IPS}
\end{align}
evaluated in the static limit ($\omega_{m}=0$) and at the nesting wave vector $q=\pm 2k_{F}^{\prime}$ in order to obtain the Peierls transition temperature. 
As mentioned in the main text, $\delta v$ is the difference in the magnitude of group velocities between the two chiral Landau levels of opposite chirality, which is nonzero when the mirror symmetry is broken.
The above expression introduces $\lambda_{P}(l,T,\delta v)$ as the cutoff function of the Peierls loop ($0\leq\lambda_{P}(l,T,\delta v)\leq1$), which  is independent of the  energy  scale $\Delta$ for mirror symmetry breaking. 
This function is mainly governed by the temperature, such that $\lambda_P\approx 1$ when $l \ll l_T=\ln (\Lambda_0/k_BT)$ and $\lambda_P\approx 0$ for $l\gg l_T$.
The dependence of $\lambda_P$ on $\delta v$ is relatively weak.

Next, in Eq.~(\ref{eq:sep1}), 
we follow the prescription from the beginning of the section in order to approximate 
\begin{equation}
\label{eq:app}
\sum_{q_{x}^{\prime},q_{y}^{\prime}} 
\simeq  L_{x}L_{y} \int_{-1/l_B}^{1/l_B} \frac {dq_x dq_y}{(2\pi)^2} = \frac{ L_{x}L_{y}}{\pi^{2}l_{B}^{2}}.
\end{equation}
Clearly, the right hand side of this equation has some arbitrariness of order one in its prefactor (for example, one could do the integration in polar coordinates instead of in cartesian coordinates, or one could change the $1/l_B$ cutoff by a factor of order one, or one could simply do the convergent integral exactly).
Indeed, from our calculations, we find the coefficients of various one-loop correction terms to be proportional to $L_{x}L_{y}/(\pi^2 l_{B}^{2})$ up to slightly different prefactors. 
Below, we approximate the right hand side of Eq.~(\ref{eq:app}) to be equal to $L_{x}L_{y}/(2 \pi l_{B}^{2})$, so as to obtain the same coefficient as in Eq. (\ref{eq:2}). This choice is justified on physical grounds, to ensure at $\Delta=0$ that $g_{1}-g_{2}$ is an invariant of the RG flow in order to preserve electron-hole symmetry  for scattering processes that conserve the number of particles on each node. 
Henceforth, we introduce the factor 
\begin{equation}
\alpha_{0}=\frac{1}{2\pi l_{B}^{2}}\frac{1}{1+\delta v/2v_{F}},
\end{equation}
so that
\begin{equation}
\langle S_{ep}S_{ee}\rangle_{dl}(g_2)\simeq-\alpha_0 g_{2}\sqrt{\frac{\pi v_{F}}{\beta\mathcal{V}}}\sum_{\tau,X^{\prime}}\sum_{k^{\prime},\mathbf{q},j} z_{\tau\overline{\tau},j}g_{\tau\overline{\tau},j}^{ep}\frac{\lambda_P dl}{2}\exp[iq_{x}X^{\prime}] \psi_{X^{\prime},k^{\prime}+q-b_{\tau},\tau}^{\dagger}\psi_{X^{\prime}+q_{y}l_{B}^{2},k^{\prime},\overline{\tau}}\phi_{j}(\mathbf{q}).\label{eq:sep_g2}
\end{equation}


We can similarly calculate the one-loop corrections to the electron-phonon
action $S_{ep}$ from the $g_{1}$ term,
\begin{align}
\langle S_{ep}S_{ee}\rangle_{dl}(g_1)
&=-\left(\frac{\pi v_{F}}{\beta\mathcal{V}}\right)^{3/2}\sum_{X,X^{\prime},Y^{\prime}}\sum_{k,\mathbf{q},j,\tau}\sum_{k_{1}^{\prime},k_{2}^{\prime},\mathbf{q}^{\prime}}z_{\tau\overline{\tau},j}g_{\tau\overline{\tau},j}^{ep}g_{1}\exp[iq_{x}X]\exp[iq_{x}^{\prime}X^{\prime}-iq_{x}^{\prime}Y^{\prime}]\nonumber \\
&\langle\overline{\psi}_{X,k+q-b_{\tau},\tau}^{\dagger}\overline{\psi}_{Y^{\prime}-q_{y}^{\prime}l_{B}^{2},k_{2}^{\prime},\tau}\rangle\langle\overline{\psi}_{Y^{\prime},k_{2}^{\prime}-q^{\prime}+b_{\tau},\overline{\tau}}^{\dagger}\overline{\psi}_{X+q_{y}l_{B}^{2},k,\overline{\tau}}\rangle\psi_{X^{\prime},k_{1}^{\prime}+q^{\prime}-b_{\tau},\tau}^{\dagger}\psi_{X^{\prime}+q_{y}^{\prime}l_{B}^{2},k_{1}^{\prime},\overline{\tau}}\phi_{j}(\mathbf{q})\nonumber \\
&=\sqrt{\frac{\pi v_{F}}{\beta\mathcal{V}}}\frac{1}{L_{x}L_{y}}\sum_{X^{\prime}}\sum_{k_{1}^{\prime},\mathbf{q},j,\tau} z_{\tau\overline{\tau},j}g_{\tau\overline{\tau},j}^{ep}g_{1}\sum_{X}\exp[iq_{x}X]\sum_{q_{x}^{\prime}}\exp[iq_{x}^{\prime}(X^{\prime}-X)] \nonumber\\
&I_{P}(dl)\psi_{X^{\prime},k_{1}^{\prime}+q-b_{\tau},\tau}^{\dagger}\psi_{X^{\prime}+q_{y}l_{B}^{2},k_{1}^{\prime},\overline{\tau}}\phi_{j}(\mathbf{q}), 
\end{align}
following the same prescription as before.
We convert the sum over $q_{x}^{\prime}$ into
an integral from $-1/l_{B}$
to $1/l_{B}$, and obtain 
\begin{equation}
\frac{L_{x}}{2\pi}\int_{-\frac{1}{l_{B}}}^{\frac{1}{l_{B}}}dq_{x}^{\prime}\exp[iq_{x}^{\prime}(X^{\prime}-X)]=\frac{L_{x}}{2\pi}\frac{2\sin(\frac{X-X^{\prime}}{l_{B}})}{X-X^{\prime}}.
\end{equation}
Thereafter, we perform the sum over the guiding center $X$,
\begin{equation}
\sum_{X}\exp[iq_{x}X]\frac{2\sin(\frac{X-X^{\prime}}{l_{B}})}{X-X^{\prime}}=\frac{L_{y}B}{\phi_{0}}\int_{-\infty}^{\infty}dX\exp[iq_{x}X]\frac{2\sin(\frac{X-X^{\prime}}{l_{B}})}{X-X^{\prime}}\approx\frac{2\pi L_{y} B}{\phi_{0}}\exp[iq_{x}X^{\prime}].
\end{equation}
Thus we obtain
\begin{equation}
\langle S_{ep}S_{ee}\rangle_{dl} (g_1)=\sqrt{\frac{\pi v_{F}}{\beta\mathcal{V}}}\alpha_{0}g_{1}\sum_{X^{\prime},\tau}\sum_{k_{1}^{\prime},\mathbf{q},j} z_{\tau\overline{\tau},j}g_{\tau\overline{\tau},j}^{ep}\exp[iq_{x}X^{\prime}]  \lambda_{P} \frac{dl}{2} \psi_{X^{\prime},k_{1}^{\prime}+q-b_{\tau},\tau}^{\dagger}\psi_{X^{\prime}+q_{y}l_{B}^{2},k_{1}^{\prime},\overline{\tau}}\phi_{j}(\mathbf{q}).\label{eq:2}
\end{equation}
Combining  Eqs. (\ref{eq:sep_g2}) and (\ref{eq:2}) with Eq. (\ref{zeph0}), we arrive at 
\begin{equation}
\frac{1}{z_{\tau\overline{\tau},j}}\frac{dz_{\tau\overline{\tau},j}}{dl}=\frac{\alpha_{0}(g_{2}-g_{1})\lambda_{P}}{2}, \label{zep}
\end{equation}
which corresponds to  Eq. (\ref{zpm}) of the main text. 

\subsection{One-loop corrections to electron-electron interactions}

The renormalization of the backward and forward Coulomb scattering amplitudes, $g_{1}$ and $g_{2}$,
at the one-loop level are obtained from the outer-shell averages $\frac{1}{2}\langle S_{ee}^{2}\rangle_{dl}$ (see Eq. (\ref{g1g2}) in the main text).
Schematically, 
\begin{align}
&S_{ee}[\psi^{\dagger},\psi,\overline{\psi}^{\dagger},\overline{\psi}]=S_{ee}^{C}+S_{ee}^{P}+S_{ee}^{L}\nonumber\\
&~~~~~~~~~~~~~~~~~~~~~\Leftrightarrow(\overline{\psi}_{+}^{\dagger}\overline{\psi}_{-}^{\dagger}\psi_{-}\psi_{+}+\psi_{+}^{\dagger}\psi_{-}^{\dagger}\overline{\psi}_{-} \overline{\psi}_{+})+(\overline{\psi}_{+}^{\dagger}\psi_{-}^{\dagger}\overline{\psi}_{-}\psi_{+}+\psi_{+}^{\dagger}\overline{\psi}_{-}^{\dagger}\psi_{-}\overline{\psi}_{+}) \nonumber\\
&~~~~~~~~~~~~~~~~~~~~~~~~+(\overline{\psi}_{+}^{\dagger}\psi_{-}^{\dagger}\psi_{-}\overline{\psi}_{+}+\psi_{+}^{\dagger}\overline{\psi}_{-}^{\dagger}\overline{\psi}_{-}\psi_{+})
\end{align}
which in order  correspond to contributions of the Cooper channel (two particles or two holes
in the outer momentum shell), the Peierls channel (one particle and
one hole on opposite branches) and the Landau channel (a particle
and a hole on the same branch). 
The  Landau channel  does not lead to
logarithmic contributions at the one-loop level and will be hereafter ignored. 
For the
chiral Landau levels, we consider both the Peierls and the Cooper
channel contributions, since the Weyl nodes are located
at equal and opposite momenta with respect to the $\Gamma$ point in the presence of mirror symmetry,
giving rise to logarithmic divergences from both the channels.
Thus,
\begin{equation}
\langle S_{ee}^2\rangle_{dl} /2\simeq \langle (S_{ee}^{P})^2\rangle_{dl} /2+\langle (S_{ee}^{C})^2\rangle_{dl} /2.
\end{equation}
Cross terms of the form $\langle S_{ee}^{P} S_{ee}^C\rangle_{dl} /2$ 
vanish in outer shell averaging.
In the following, we study $\langle (S_{ee}^{P})^2\rangle_{dl} /2$ and $\langle (S_{ee}^{C})^2\rangle_{dl} /2$ separately.

\subsubsection{Peierls channel}

The contribution from the $g_{2}$ term to $\langle (S_{ee}^{P})^2\rangle_{dl} /2$ is given by 
\begin{align}
\label{eq:Peierls_g2}
&\frac{g_{2}^{2}}{2}(\frac{\pi v_{F}}{2\beta\mathcal{V}})^{2}\sum_{k_{1},k_{2},\mathbf{q},\tau,X,Y}\sum_{k_{1}^{\prime},k_{2}^{\prime},\mathbf{q^{\prime}},X^{\prime},Y^{\prime}}\exp[iq_{x}X-iq_{x}Y]\exp[iq_{x}^{\prime}X^{\prime}-iq_{x}^{\prime}Y^{\prime}]\nonumber \\
&(\langle\overline{\psi}_{X,k_{1}+q,\tau}^{\dagger}\overline{\psi}_{X^{\prime}+q_{y}^{\prime}l_{B}^{2},k_{1}^{\prime},\tau}\rangle\langle\overline{\psi}_{Y^{\prime},k_{2}^{\prime}-q^{\prime},\overline{\tau}}^{\dagger}\overline{\psi}_{Y-q_{y}l_{B}^{2},k_{2},\overline{\tau}}\rangle \psi_{X^{\prime},k_{1}^{\prime}+q^{\prime},\tau}^{\dagger}\psi_{Y,k_{2}-q,\overline{\tau}}^{\dagger}\psi_{Y^{\prime}-q_{y}^{\prime}l_{B}^{2},k_{2}^{\prime},\overline{\tau}}\psi_{X+q_{y}l_{B}^{2},k_{1},\tau} \nonumber\\
&+\langle\overline{\psi}_{X^{\prime},k_{1}^{\prime}+q^{\prime},\tau}^{\dagger}\overline{\psi}_{X+q_{y} l_{B}^{2},k_{1},\tau}\rangle\langle\overline{\psi}_{Y,k_{2}-q,\overline{\tau}}^{\dagger}\overline{\psi}_{Y^{\prime}-q_{y}^{\prime} l_{B}^{2},k_{2}^{\prime},\overline{\tau}}\rangle \psi_{X,k_{1}+q,\tau}^{\dagger}\psi_{Y^{\prime},k_{2}^{\prime}-q^{\prime},\overline{\tau}}^{\dagger}\psi_{Y-q_{y} l_{B}^{2},k_{2},\overline{\tau}}\psi_{X^{\prime}+q_{y}^{\prime} l_{B}^{2},k_{1}^{\prime},\tau}) \nonumber\\
&=-g_{2}^{2}(\frac{\pi v_{F}}{2\beta\mathcal{V}})\sum_{\mathbf{\widetilde{q}},\tau,Y}\sum_{k_{1},k_{2}^{\prime},\mathbf{\widetilde{q}^{\prime}},X^{\prime}}\exp[i\widetilde{q_{x}}X^{\prime}-i\widetilde{q_{x}}Y] I_{P}(dl)  \psi_{X^{\prime},k_{1}+\widetilde{q},\tau}^{\dagger}\psi_{Y,k_{2}^{\prime}-\widetilde{q},\overline{\tau}}^{\dagger}\psi_{Y-\widetilde{q_{y}}l_{B}^{2},k_{2}^{\prime},\overline{\tau}}\psi_{X^{\prime}+\widetilde{q_{y}}l_{B}^{2},k_{1},\tau}
\end{align}
where the couplings have once again been assumed to be approximately local in the transverse direction, and in the second line,
we have expressed
the correction in terms of a new set of variables, $\widetilde{\mathbf{q}}=\mathbf{q+q^{\prime}}$,
$\mathbf{\widetilde{q}^{\prime}=q-q^{\prime}}$, $k_{2}^{\prime}=k_{2}+q^{\prime}$, and $I_{P}(dl)$ is the symmetrized product of outer-shell averages defined in (\ref{IPS}) of the previous section. 

Summing over the transverse momenta using similar approximations as in the previous section, Eq.~(\ref{eq:Peierls_g2}) becomes
\begin{equation}
-\alpha_{0} \frac{g_{2}^{2}}{2}(\frac{\pi v_{F}}{\beta\mathcal{V}})\sum_{\mathbf{\widetilde{q}},\tau,Y}\sum_{k_{1},k_{2}^{\prime},X^{\prime}}\exp[i\widetilde{q_{x}}X^{\prime}-i\widetilde{q_{x}}Y] \lambda_{P} \frac{dl}{2} \psi_{X^{\prime},k_{1}+\widetilde{q},\tau}^{\dagger}\psi_{Y,k_{2}^{\prime}-\widetilde{q},\overline{\tau}}^{\dagger}\psi_{Y-\widetilde{q_{y}}l_{B}^{2},k_{2}^{\prime},\overline{\tau}}\psi_{X^{\prime}+\widetilde{q_{y}}l_{B}^{2},k_{1},\tau}
\end{equation}
which has the form of a $g_2$ term and thus contributes to the RG flow equation for $g_{2}$.


We can similarly calculate the contributions to the renormalization  of $g_1$ term from $\langle (S_{ee}^{P})^2\rangle_{dl} /2$, which comprises  terms proportional to  $g_1^2$ and $g_1g_2$, 
respectively. This makes explicit all contributions to the first diagram of the flow equation of Fig.~\ref{fig_RG1}~(c).
The outcome gives the  Peierls part of RG equations:
\begin{align}
\label{eq:g1g2P}
\left(\frac{dg_{2}}{dl}\right)_{P} &=\frac{\alpha_{0}\lambda_{P}g_{2}^{2}}{2}\nonumber\\
\left(\frac{dg_{1}}{dl}\right)_{P} &=-\frac{\alpha_{0}\lambda_{P}}{2}(g_{1}^{2}-2g_{1}g_{2}).
\end{align}

\subsubsection{Cooper channel}

We now consider the contribution coming from the Cooper scattering channel. At the one-loop level the contribution reads  
\begin{align}
\label{eq:SC2}
\frac{1}{2}\langle (S_{ee}^C)^{2}\rangle_{dl} &=-\frac{\left(g_1^2+g_{2}^{2}\right)}{2\pi l_{B}^{2}}(\frac{\pi v_{F}}{2\beta\mathcal{V}})\sum_{k^{\prime},\widetilde{\mathbf{q}},\tau,X,Y}\sum_{k_{2}^{\prime}}\exp[i\widetilde{q_{x}}X-i\widetilde{q_{x}}Y]I_{C}(dl) \psi_{X,k^{\prime}+\widetilde{q},\tau}^{\dagger}\psi_{Y,k_{2}^{\prime}-\widetilde{q},\overline{\tau}}^{\dagger}\psi_{Y-\widetilde{q_{y}}l_{B}^{2},k_{2}^{\prime},\overline{\tau}}\psi_{X+\widetilde{q_{y}}l_{B}^{2},k^{\prime},\tau}\nonumber\\
&-\frac{2 g_{1}g_{2}}{2\pi l_{B}^{2}}(\frac{\pi v_{F}}{2\beta\mathcal{V}})\sum_{\widetilde{q},\widetilde{\mathbf{q_{\bot}}},\tau,X,Y}\sum_{k_{1}^{\prime},k^{\prime}}\exp[i\widetilde{q_{x}}X-i\widetilde{q_{x}}Y]I_{C}(dl) \psi_{X,k_{1}^{\prime}+\widetilde{q},\tau}^{\dagger}\psi_{Y,k^{\prime}-\widetilde{q},\overline{\tau}}^{\dagger}\psi_{Y-\widetilde{q_{y}}l_{B}^{2},k^{\prime},\tau}\psi_{X+\widetilde{q_{y}}l_{B}^{2},k_{1}^{\prime},\overline{\tau}},
\end{align}
 where the first and last lines  contribute to the RG flow of $g_2$ and  $g_1$, respectively. The  expression for the outer shell Cooper loop at zero pair momentum and frequency is given by  
\begin{align}
I_{C}(dl)= &-\frac{\pi v_{F}}{2 \beta L}\sum_{\omega_{n}}\sumslashD_{k_{2}}(G^0_{+}(-(k_{+}+k_{-})-k_{2},-\omega_{n})G^0_{-}(k_{2},\omega_{n})+G^0_{+}(k_{2},\omega_{n})G^0_{-}(-(k_{+}+k_{-})-k_{2},-\omega_{n}))\cr
=& \ \frac{\pi v_{F}}{2L}\sumslashD_{k}(\frac{f[\epsilon_{+}(k)]-f[-\epsilon_{-}(-(k_{+}+k_{-})-k)]}{\epsilon_{+}(k)+\epsilon_{-}(-(k_{+}+k_{-})-k)}+\frac{f[\epsilon_{-}(k)]-f[-\epsilon_{+}(-(k_{+}+k_{-})-k)]}{\epsilon_{-}(k)+\epsilon_{+}(-(k_{+}+k_{-})-k)}),
\end{align}
where $\epsilon_{\tau}(k)=\hbar v_{\tau} \tau (k-k_{F \tau}^{\prime})$ is the energy of the Weyl fermions with respect to the Fermi energy and $f\left[z\right]$ is the Fermi function. In the case where both the velocities and positions of the two Weyl
nodes differ from one another, the corresponding dispersions satisfy
the relations
\begin{align}
\epsilon_{-}(-(k_{+}+k_{-})-k) &=\left(1+\frac{\delta v}{v_{F}}\right)\epsilon_{+}(k)+(1+\frac{\delta v}{v_{F}})\Delta\nonumber\\
\epsilon_{+}(-(k_{+}+k_{-})-k) &=\left(1+\frac{\delta v}{v_{F}}\right)^{-1}\epsilon_{-}(k)-\Delta,
\end{align}
with $\Delta=\hbar v_F (k_{F+}+k_{F-})$.
Thus, when $\Delta\neq 0$, the Fermi points are no longer symmetric with respect to the $\Gamma$ point, thereby removing the logarithmic divergence in the Cooper channel ($\Delta$ acts as a pair-breaking perturbation). 
 By carrying out the outer shell energy integration, it follows that
\begin{align}
 I_{C}(dl)= &-\frac{dl}{2+\delta v/v_{F}} {1\over 4}\left\{\frac{\tanh\left[\frac{\beta\Lambda_{0}(l)}{2}\right]}{1+\Delta\gamma/(\gamma^{\prime} \Lambda_{0}(l))}+\frac{\tanh\left[\frac{\beta\Lambda_{0}(l)}{2}\right]}{1-\Delta\gamma/(\gamma^{\prime} \Lambda_{0}(l))} -\frac{f[(\Lambda_{0}(l)+\Delta\gamma)/\gamma]-f[(-\Lambda_{0}(l)-\Delta)\gamma]}{1+\Delta\gamma/(\gamma^{\prime} \Lambda_{0}(l))}\right. \nonumber \\
&~~~~~~~~~~~~~~~~~~~~~~~ -\left.\frac{f[(\Lambda_{0}(l)-\Delta)\gamma]-f[(-\Lambda_{0}(l)+\Delta\gamma)/\gamma]}{1-\Delta\gamma/(\gamma^{\prime} \Lambda_{0}(l))}\right\},\cr 
\equiv&\ \frac{dl}{2+\delta v/v_{F}}\lambda_{C}(l,T,\Delta,\delta v),
 \label{Cooper}
\end{align}
where $\gamma=1+\frac{\delta v}{v_{F}}$ and $\gamma^{\prime}=1+\frac{\delta v}{2v_{F}}$.
The above expression defines the cutoff function $\lambda_{C}(l,T,\Delta,\delta v)$ of the Cooper loop ($-1 \le \lambda_{C}(l,T,\Delta,\delta v)\le 0$). One may easily verify the following limiting cases of $\lambda_C$. 
First, when $\Delta \gg k_BT$ and $l\ll l_\Delta =\ln (\Lambda_0/\Delta)$, the influence of $\Delta$ is small and  $\lambda_C\approx -\lambda_P\approx -1$;  at $l\gg l_\Delta $, one rather has $\lambda_C\approx 0$  which  suppresses the logarithmic singularity of the Cooper channel. Second, when  $\Delta \ll k_BT$, $\Delta$ is essentially irrelevant so that the Cooper loop is only cut off by the temperature, where $\lambda_C\approx -1$ for $l\ll l_T$ and  $\lambda_C\approx0$ for $l\gg l_T$. This 
 allows to define a sharp cut-off  procedure for $\lambda_C$ as a function of $l$, which is used in the main text to determine the different limits of the flow equations.  Note that for $\Delta=0$, we
have the expression 
\begin{equation}
\lambda_C =  -{1\over 4}\left\{2\tanh[\frac{\beta\Lambda_{0}(l)}{2}]+\tanh[\frac{\beta\Lambda_{0}(l)}{2\gamma}]+\tanh[\frac{\beta\Lambda_{0}(l)\gamma}{2}]\right\}=-\lambda_P,
\end{equation}
whose amplitude coincides with the cutoff function determined previously for the Peierls channel.




The corresponding Cooper contributions to the RG equations are given
by 
\begin{align}
\label{eq:g1g2C}
\left(\frac{dg_{2}}{dl}\right)_{C} &=\frac{\alpha_{0}}{2}(g_{2}^{2}+g_{1}^{2})\lambda_{C}\nonumber\\
\left(\frac{dg_{1}}{dl}\right)_{C} &=\alpha_{0}g_{1}g_{2}\lambda_{C}.
\end{align}

\subsubsection{Total contribution}

Summing Eqs. (\ref{eq:g1g2P}) and (\ref{eq:g1g2C}), we have 
\begin{align}
\frac{dg_{2}}{dl} &=\left(\frac{dg_{2}}{dl}\right)_{P}+\left(\frac{dg_{2}}{dl}\right)_{C} =\frac{g_{1}^{2}\alpha_{0}\lambda_{C}}{2}+\frac{g_{2}^{2}\alpha_{0}(\lambda_{P}+\lambda_{C})}{2}\nonumber\\
\frac{dg_{1}}{dl} &=\left(\frac{dg_{1}}{dl}\right)_{P}+\left(\frac{dg_{1}}{dl}\right)_{C} =-\frac{g_{1}^{2}\alpha_{0}\lambda_{P}}{2}+g_{1}g_{2}\alpha_{0}(\lambda_{C}+\lambda_{P}), \label{chiralrg}
\end{align}
in agreement with Eqs. (\ref{g1g2ChiralA}) and  (\ref{g1g2ChiralB}) 
of the main text. 

\subsection{One-loop corrections to the phonon propagator}

The self-energy correction for the phonon mode $j$ is (see Fig. \ref{fig_RG1} and Eq. \ref{Sp0} of the main text)
\begin{align}
\frac{1}{2}\langle S_{ep}^{2}\rangle_{dl} &=-\frac{1}{2}(\frac{\pi v_{F}}{\beta\mathcal{V}})\sum_{\tau,X,X^{\prime},j, j^{\prime}}\sum_{k,k^{\prime},\mathbf{q},\mathbf{q^{\prime}}}g_{\tau\overline{\tau}}^{2}z_{\tau,\overline{\tau}}z_{\overline{\tau},\tau}\exp[iq_{x}X]\exp[iq_{x}^{\prime}X^{\prime}] \langle\overline{\psi}_{X,k+q-b_{\tau},\tau}^{\dagger}\overline{\psi}_{X^{\prime}+q_{y}^{\prime}l_{B}^{2},k^{\prime},\tau}\rangle
\nonumber \\
&~~~~~~ \langle\overline{\psi}_{X^{\prime},k^{\prime}+q^{\prime}-b_{\overline{\tau}},\overline{\tau}}^{\dagger}\overline{\psi}_{X+q_{y}l_{B}^{2},k,\overline{\tau}}\rangle\phi_{j}(\mathbf{q})\phi_{j^{\prime}}(\mathbf{q^{\prime})} \nonumber \\
&=\alpha_{0}\sum_{\mathbf{q},\tau,j}g_{\tau\overline{\tau}}^{2} z_{\tau,\overline{\tau}}^{2}\frac{\lambda_{P}dl}{2}|\phi_{j}(\mathbf{q})|^{2},
\end{align}
where, in the second equality, we have only kept terms corresponding to $j=j^{\prime}$. This amounts to neglect the hybridization between different phonon modes, induced by electron-phonon coupling. This approximation, which we adopt for simplicity, is justified if the bare phonon frequencies for different $j$ are well separated (compared to the off-diagonal part of the electron-phonon self energy). 
Accordingly, the inverse of the phonon propagator is 
\begin{equation}
 \mathscr{D}_{j,l}^{-1}({\bf q},\omega_m)=\omega_{m}^{2}+\omega_{0,j}^{2}(\mathbf{q})\left(1-\frac{\alpha_{0}g_{x,j}^{\prime2}}{2}\int_{0}^{l}z_{+-,j}(l^{\prime})^{2}\lambda_{P}dl^{\prime}\right),
\end{equation}
where $g_{x,j}^{\prime}=g_{x,j}/\omega_{0,j}$, in agreement with Eq.~(\ref{Kohn_chiral}) of the main text. 



\section{Derivation of renormalization group equations in the near-quantum limit}

In this section, we derive the recursion equations for the phonon self-energy, the electron-phonon vertex and the electron-electron couplings, up to one-loop order, when the Fermi level intersects both the $n=0$ (chiral) and $|n|=1$ (nonchiral) Landau levels.
We tailor our theory to the situation in which the phonon wave vector of interest connects the two Fermi points of a nonchiral Landau level in a given node.
Moreover, we restrict our analysis to the case in which the two Weyl nodes are related by an improper (e.g. mirror) symmetry.
The case for broken mirror symmetry will be briefly discussed in App. C.

In the mirror-symmetric situation, the electronic dispersion for the $|n|=1$ Landau level on a given node (cf. Eq.~(\ref{eq:LL_energies})) can be linearized around the Fermi points $\pm k_F$, where $k_F$ is the Fermi momentum measured from the location of the node:
\begin{equation}
\epsilon_{n k \tau,\mu}\simeq \tau \, {\rm sign}(n)\left(\hbar v_{F}\sqrt{k_{F}^{2}+\frac{2}{l_{B}^{2}}}+\frac{\hbar v_{F}k_{F}}{\sqrt{k_{F}^{2}+\frac{2}{l_{B}^{2}}}}(\mu k-k_{F})\right),
\end{equation}
where we consider $\tau=n=1$ and $\tau=n=-1$ (see Fig.~\ref{fig:nonc}), 
and $\mu=\pm1$ is an index referring to the two branches on a given node. 
The fermion fields can be labeled by $\psi_{\mu,n,X,k,\tau}$ but, because $n=\tau$ for the bands of interest, the label $n$ will be omitted henceforth. 
Also, while we can generally consider multiple phonon modes $j$, here we limit the analysis to the scalar and pseudoscalar phonon modes only. 
Finally, like in App. A, we once again work under the assumption that the couplings are approximately local in the transverse directions, within the magnetic length $l_{B}$, above which they are exponentially suppressed.


The electron-phonon part of the action is then given by
\begin{align}
S_{ep}[\psi^{\dagger},\psi,\phi] &=-\sqrt{\frac{\pi v_{F}}{\beta\mathcal{V}}}\sum_{k,\mathbf{q},\mu,X,\omega_{n},\omega_{m}} e^{iq_{x}X}\left(z_{s}g_{s} \phi_{s}(\mathbf{q},\omega_{m}) \sum_\tau\psi_{\mu,X,k+q,\tau}^{\dagger}(\omega_{n}+\omega_{m})\psi_{\overline{\mu},X+q_{y}l_{B}^{2},k,\tau}(\omega_{n}) \right.\nonumber \\
&~~~~~~~~~~~~~~~~~~~~~~~~~~~~~~~~~~~~~~~~\left.+z_{ps}g_{ps} \phi_{ps}(\mathbf{q},\omega_{m}) \sum_\tau \tau \psi_{\mu,X,k+q,\tau}^{\dagger}(\omega_{n}+\omega_{m})\psi_{\overline{\mu},X+q_{y}l_{B}^{2},k,\tau}(\omega_{n}) \right),
\end{align}
where $\phi_{s}(\mathbf{q})$ and $\phi_{ps}(\mathbf{q})$ refer to the scalar and
pseudoscalar phonon modes, and $g_s z_s$ and $g_{ps} z_{ps}$ denote their respective couplings to electrons. 
Motivated by the fact that the nesting vector of interest is much smaller than the size of the Brillouin zone, we have adopted the long-wavelength approximation for phonons. 
In this approximation, scalar (pseudoscalar) phonons couple with the same magnitude and the same (opposite) sign to electrons of the two nodes, at every step of the RG flow. 
A factor $\lambda=1/\sqrt{(k_{F}l_{B})^{2}/2+1}$, originating from the pseudospin textures of the nonchiral Landau levels (see Sec.~\ref{sec:noncll}), has been
absorbed in the definition of $g_{s,ps}$.

The electron-electron part of the action can be written as
\begin{align}
S_{ee}[\psi^{\dagger},\psi] \simeq-\frac{\pi v_{F}}{2\beta\mathcal{V}}\sum_{\tau,\tau^{\prime}}\sum_{X,Y}\sum_{k_{1},k_{2},\mathbf{q}}\sum_{\mu}\exp[iq_{x}X-iq_{x}Y] \nonumber \\
(g_{1} \psi_{\mu,X,k_{1}+q,\tau}^{\dagger}\psi_{\overline{\mu},Y,k_{2}-q,\tau^{\prime}}^{\dagger}\psi_{\mu,Y-q_{y}l_{B}^{2},k_{2},\tau^{\prime}}\psi_{\overline{\mu},X+q_{y}l_{B}^{2},k_{1},\tau} + \nonumber \\
 g_{2}
\psi_{\mu,X,k_{1}+q,\tau}^{\dagger}\psi_{\overline{\mu},Y,k_{2}-q,\tau^{\prime}}^{\dagger}\psi_{\overline{\mu},Y-q_{y}l_{B}^{2},k_{2},\tau^{\prime}}\psi_{\mu,X+q_{y}l_{B}^{2},k_{1},\tau}) \label{eq:21-1-1}
\end{align}
corresponding to the sum of the backscattering $g_{1}$ and forward-scattering $g_{2}$ terms. The labels $\tau,\tau^{\prime}$
are used to distinguish the $g_{1}$ and $g_{2}$ scattering terms
within a given node from those between different nodes.
Since the nodes are related by mirror symmetry, we take the amplitudes of the scattering terms $g_{1}$ and $g_{2}$ to be the same for the two nodes, with no additional $\tau,\tau^{\prime}$ labels.

\subsection{One-loop corrections to the electron-phonon vertex}

Here, we calculate the one-loop corrections to the electron-phonon
vertex, from the $g_{1}$ and $g_{2}$ terms, for both the scalar and pseudoscalar phonon modes. As an illustrative example, we write down the various corrections for the electron-phonon coupling $z_{s}$ for the scalar phonons. 
The one-loop correction to the electron-phonon action from the $g_{1}$ and $g_{2}$ terms is given by 

\begin{align}
&\langle S_{ep}S_{ee}\rangle_{dl} = \frac{1}{2\pi l_{B}^{2}} \left(\frac{\pi v_{F}}{\beta\mathcal{V}}\right)^{3/2}(\sum_{\mu,\tau}\sum_{k^{\prime},\mathbf{q}}\sum_{X^{\prime}} (z_{s}g_{s} g_{2} \exp[iq_{x}X^{\prime}]\sumslashD_{k} G^{0}_{\mu,\tau}(k+q)G^{0}_{\overline{\mu},\tau}(k) \nonumber \\
&~~~~~~~~~~~~~~~~~~~~~~~~~~~~~~~~~~~~~\psi_{\mu,X^{\prime},k^{\prime}+q,\tau}^{\dagger}\psi_{\overline{\mu},X^{\prime}+q_{y}l_{B}^{2},k^{\prime},\tau})\phi_{s} (\mathbf{q}) \nonumber \\
&~~~~~~~~~~~~~~~~~~~~~~~~~~~~~~~~~~~~~-\sum_{\tau,\tau^{\prime}}\sum_{\mu}\sum_{k_{2}^{\prime},\mathbf{q}}\sum_{Y^{\prime}}(z_{s}g_{s} g_{1}\exp[iq_{x}Y^{\prime}]\sumslashD_{k} G^{0}_{\mu,\tau}(k+q)G^{0}_{\overline{\mu},\tau}(k)  \nonumber \\
&~~~~~~~~~~~~~~~~~~~~~~~~~~~~~~~~~~~~~\psi_{\mu,Y^{\prime},k_{2}^{\prime}+q,\tau^{\prime}}^{\dagger}\psi_{\overline{\mu},Y^{\prime}+q_{y}l_{B}^{2},k_{2}^{\prime},\tau^{\prime}})\phi_{s}(\mathbf{q})) \nonumber \\
&~~~~~~~~~~~~\simeq -\frac{1}{2\pi l_{B}^{2}} \sqrt{\frac{\pi v_{F}}{\beta\mathcal{V}}}(\sum_{\mu,\tau}\sum_{k^{\prime},\mathbf{q}}\sum_{X^{\prime}} (z_{s}g_{s} g_{2} \exp[iq_{x}X^{\prime}]I_{P}(dl) \psi_{\mu,X^{\prime},k^{\prime}+q,\tau}^{\dagger}\psi_{\overline{\mu},X^{\prime}+q_{y}l_{B}^{2},k^{\prime},\tau})\phi_{s} (\mathbf{q}) \nonumber \\
&~~~~~~~~~~~~~~~~~~~~~~~~~+\sum_{\tau,\tau^{\prime}}\sum_{\mu,\nu}\sum_{k_{2}^{\prime},\mathbf{q}}\sum_{Y^{\prime}}(z_{s}g_{s} g_{1}\exp[iq_{x}Y^{\prime}]I_{P}(dl) \psi_{\mu,Y^{\prime},k_{2}^{\prime}+q,\tau^{\prime}}^{\dagger}\psi_{\overline{\mu},Y^{\prime}+q_{y}l_{B}^{2},k_{2}^{\prime},\tau^{\prime}})\phi_{s}(\mathbf{q})).
\label{eq:sepnc}
\end{align}
where 
\begin{equation}
I_{P}(dl) =\frac{dl}{2}\tanh\left[\frac{\beta \Lambda_{1}(l)}{2}\right] 
\equiv \frac{dl}{2} \lambda_{P}(l,T)  \label{eq:ipnc}
\end{equation}
and $\Lambda_{1}$ refers to the cutoff energy scale for the nonchiral Landau levels (See Fig.~\ref{fig:nonc}).
Henceforth, we define $\alpha_{1}=1/(2\pi l_{B}^{2})$. 
It is easy to see that the  second part of the correction in Eq. (\ref{eq:sepnc}) above involves additive contributions from both nodes because scalar phonons couple to both nodes with the same magnitude and sign. This results in a multiplicative prefactor of $-2g_{1}$ for scalar phonons.
On the other hand, the first part of the correction, resulting from the $g_{2}$ coupling, 
only involves contributions to the electron-phonon coupling defined on a given node, from the electronic  interactions within the same node.
The corrections to the electron-phonon couplings corresponding to
the scalar phonon modes are then given by 
\begin{equation}
\label{eq:8}
\frac{dz_{s}}{dl} =z_{s}(g_{2}-2g_{1})\frac{\lambda_{P}\alpha_{1}}{2}.
\end{equation}

A similar analysis can be carried out for pseudoscalar phonons. Because such phonons couple to both nodes with the same magnitude but opposite sign, the contribution of $g_1$ terms to the electron-phonon vertex cancels out.
 This results in 
\begin{equation}
\label{eq:7}
\frac{dz_{ps}}{dl} =z_{ps}g_{2}\frac{\lambda_{P}\alpha_{1}}{2}.
\end{equation}
Equations (\ref{eq:8}) and (\ref{eq:7})  can be more concisely rewritten as 
\begin{equation}
\frac{dz_{j}}{dl}=z_{j}\frac{\gamma_{j}}{2}\lambda_{P},
\end{equation}
where $\gamma_{j}(l)=\alpha_{1}(g_{2}(l)-2g_{1}(l))$ for $j=s$ and $\gamma_{j}(l)=\alpha_{1}g_{2}(l)$ for $j=ps$, in agreement with Eq.~(\ref{eq:rg_ep_nc}) of the main text. 

\subsection{One-loop corrections to the electron-electron interactions}

Here, we calculate the renormalization of the electron-electron interaction
couplings to one-loop order.
Since our analysis only involves interbranch and intrabranch electronic interactions within a given node, where the two branches are not located at equal and opposite momenta with respect to the $\Gamma$ point,  
 we no longer consider
contributions from the Cooper channel, limiting instead ourselves
to the Peierls channel contributions. These are given by 
\begin{align}
\frac{1}{2}\langle S_{ee}^{2}\rangle_{dl}=-\frac{1}{2\pi l_{B}^{2}}\frac{\pi v_{F}}{2\beta\mathcal{V}}&\left(\sum_{\tau,\tau^{\prime}}\sum_{X^{\prime},Y}\sum_{\mu}\sum_{\widetilde{q},\widetilde{q_{\bot}}}\sum_{k_{1},k_{2}^{\prime}}g_{2}^{2} \exp[i\widetilde{q_{x}} X^{\prime}-i\widetilde{q_{x}} Y]I_{P}(dl)\right. \nonumber \\
&\psi_{\mu,X^{\prime},k_{1}+\widetilde{q},\tau}^{\dagger}\psi_{\overline{\mu},Y,k_{2}^{\prime}-\widetilde{q},\tau^{\prime}}^{\dagger}\psi_{\overline{\mu},Y-\widetilde{q_{y}}l_{B}^{2},k_{2}^{\prime},\tau^{\prime}}\psi_{\mu,X^{\prime}+\widetilde{q_{y}}l_{B}^{2},k_{1},\tau} \nonumber \\
&- \sum_{\tau,\tau^{\prime}}\sum_{X,Y^{\prime}}\sum_{\mu}\sum_{k_{1},k_{2}^{\prime},\mathbf{q}}g_{1}^{2} \exp[iq_{x}(X-Y^{\prime})]I_{P}(dl) \nonumber \\
&\psi_{\mu,X,k_{1}+q,\tau}^{\dagger}\psi_{\overline{\mu},Y^{\prime},k_{2}^{\prime}-q,\tau^{\prime}}^{\dagger}\psi{}_{\mu,Y^{\prime}-q_{y} l_{B}^{2},k_{2}^{\prime},\tau^{\prime}}\psi_{\overline{\mu},X+q_{y}l_{B}^{2},k_{1},\tau} \nonumber \\
&+\sum_{\tau,\tau^{\prime}}\sum_{X,X^{\prime}}\sum_{\mu}\sum_{k_{1},k_{2}^{\prime},\mathbf{q}}2 g_{1} g_{2} \exp[iq_{x}(X-X^{\prime})]I_{P}(dl) \nonumber \\
&\left. \psi_{\mu,X,k_{1}+q,\tau}^{\dagger}\psi_{\overline{\mu},X^{\prime},k_{2}^{\prime}-q,\tau^{\prime}}^{\dagger}\psi{}_{\mu,X^{\prime}-q_{y} l_{B}^{2},k_{2}^{\prime},\tau^{\prime}}\psi_{\overline{\mu},X+q_{y}l_{B}^{2},k_{1},\tau}\right), \label{10}
\end{align}
where the first term contributes to the RG flow of $g_{2}$, while the latter two terms contribute to the RG flow of $g_{1}$.
Collecting the above contributions, we get  the RG equations 
\begin{align}
\frac{dg_{1}}{dl} &=(-g_{1}^{2}+g_{1}g_{2})\alpha_{1}\lambda_{P}\label{eq:1-1}\\
\frac{dg_{2}}{dl} &=\frac{g_{2}^{2}}{2}\alpha_{1}\lambda_{P},\label{eq:2-1}
\end{align}
as given in Eq. (\ref{eq:rg_ee_nc}) of the main text. We note that while the RG equation for the forward-scattering term $g_{2}$ only involves contributions from a given node, the corrections to the backscattering term $g_{1}$ involve additive contributions from both nodes.

\subsection{One-loop corrections to the phonon action }

The one-loop correction to the bare phonon action is given by
\begin{equation}
\frac{1}{2}\langle S_{ep}^{2}\rangle_{dl}=\alpha_{1}\sum_{j\in\{s,ps\}}\sum_{\mathbf{q}}(z_{j}g_{j})^{2}\lambda_{P}dl|\phi_{j}(\mathbf{q})|^{2},
\end{equation}
which leads to  
\begin{equation}
\mathscr{D}_{j,l}^{-1}({\bf q},\omega_m)=\omega_{m}^{2}+\omega_{0,j}^{2}({\bf q})\left(1-\alpha_{1}\frac{g_{j}^2}{\omega_{0,j}^2({\bf q})}\int_{0}^{l}z_{j}^{2}(l^{\prime})\lambda_{P}(l')dl^{\prime}\right)
\end{equation}
for the inverse propagator for the scalar ($j=s$) and pseudoscalar
($j=ps$) phonon modes at step $l$ of the RG flow. 
This agrees with Eq.~(\ref{eq:phprop_nc}) of the main text.

\section{RG equations for the near-quantum limit in the absence of mirror symmetry}

To conclude, we briefly generalize the analysis of App. B to the case in which the mirror (or inversion) symmetry relating the two Weyl nodes of opposite chirality is broken.
If the electronic energy scale associated with the mirror symmetry breaking, $\Delta=\hbar v_{F} (k_{F,+}+k_{F,-})$, is small compared to the Peierls transition temperature $T_c$ computed for the mirror-symmetric case, the results from App. B are still applicable.
However, if the energy scale $\Delta$ exceeds $T_{c}$, the effect of mirror symmetry-breaking in the Peierls transition is significant. 
In the case in which the mirror symmetry is strongly broken, the intranodal nesting wave vector (${\bf q}_1$ in Fig.~\ref{fig:nonc}) becomes largely different for the two nodes of opposite chirality. 
Accordingly, only one of the nodes contributes to the Peierls instability and the transition temperature can be approximated by considering a model of a single nonchiral Landau level with interbranch electron-phonon coupling.
For such a model, the Cooper channel contribution is once again negligible, since the two Fermi points in a nonchiral Landau level of a fixed chirality are not symmetric with respect to the $\Gamma$ point. Then, the RG equations for the electron-phonon couplings become identical in form to the corresponding couplings defined between the chiral Landau levels in Eq. (\ref{zep}), and the equations for the electron-electron interactions become identical to those in Eq.  (\ref{chiralrg}), with $\lambda_{C}=0$. 

\end{widetext}

\end{document}